\documentclass[11pt,letterpaper]{article}

\usepackage{geometry}
\usepackage{bbm}
\usepackage[toc,page]{appendix}
\usepackage{amsmath, amsthm, amsfonts, amssymb}
\usepackage{mathrsfs} % for calligraphic C
\usepackage{hyperref}
\usepackage{cleveref}
\usepackage{slashed}
\usepackage{dsfont}
\usepackage{wrapfig}
\usepackage{minibox}
\usepackage{graphicx}
\usepackage{graphics}
\usepackage{epsfig}
\usepackage{xcolor}
\usepackage{array}
\usepackage{tikz}
\usepackage{tikz-cd}
\usetikzlibrary{arrows,calc}
\usepackage{relsize}
\usetikzlibrary{decorations.pathmorphing,calc}
\usepackage{authblk}
\usepackage[labelfont=bf]{caption}
\usepackage{caption}
\usepackage{subcaption}
\usepackage{placeins}

\def\b{\beta}
\def\g{\gamma}
\def\G{\Gamma}
\def\d{\delta}
\def\ve{\varepsilon}
\def\m{\mu}

\def\l{\lambda}
\def\L{\Lambda}

\def\s{\sigma}
\def\thintablerule{\hrule height0.4pt}
\def\wt{\widetilde}
\def\bu{\bullet}

\newcommand{\be}{\begin{equation}}
\newcommand{\ee}{\end{equation}}
\newcommand{\bea}{\begin{eqnarray}}
\newcommand{\eea}{\end{eqnarray}}
\newcommand{\bal}{\begin{aligned}}
\newcommand{\eal}{\end{aligned}}
\newcommand{\eq}[1]{Eq.~(\ref{#1})}
\newcommand{\fig}[1]{Fig.~\ref{#1}}

\numberwithin{equation}{section}

\allowdisplaybreaks

\begin{document}

%---------------------------------------------------------------------------------------------------------------------------------------------------
\tikzset{
    photon/.style={decorate, decoration={snake}, draw=black},
    electron/.style={draw=black, postaction={decorate},
        decoration={markings,mark=at position .55 with {\arrow[draw=black]{>}}}},
    gluon/.style={decorate, draw=black,
        decoration={coil,amplitude=4pt, segment length=5pt}} 
}
%---------------------------------------------------------------------------------------------------------------------------------------------------

\centerline{\LARGE On RG flows in Generalized Effective Field Theory}
\vskip .5cm

\vskip 2 cm
\centerline{\large Nikos Irges and Fotis Koutroulis}
\vskip 1cm
\centerline{\it Department of Physics}
\centerline{\it National Technical University of Athens}
\centerline{\it Zografou Campus, GR-15780 Athens, Greece}
\centerline{\it e-mail: irges@mail.ntua.gr, fkoutroulis@central.ntua.gr}

\vskip 2.2 true cm
\thintablerule
\vskip 2.0ex

\centerline{\bf Abstract}
Generalized Effective Field Theory (GEFT) is the non-renormalizable extension of an Effective Field Theory where the Wilson 
coefficients are endowed by their own, independent scale dependence. Such an effective theory can be constructed by
quantizing a Lagrangian in the presence of all internal symmetry respecting operators of any mass dimension. The resulting theory may be
practically useful in regimes of its phase diagram where the perturbative expansion and a truncation of the infinite tower of 
Higher Dimensional Operators (HDO) are valid.
The massless limit of GEFT is especially interesting as the spontaneous breaking of internal symmetry and of scale invariance are correlated.

\vskip 1.0ex\noindent
\vskip 2.0ex
\thintablerule

\newpage

\tableofcontents

\pagebreak

%---------------------------------------------------------------------------------------------------------------------------------------------------

%---------------------------------------------------------------------------------------------------------------------------------------------------
\section{Introduction}\label{Intro}
%---------------------------------------------------------------------------------------------------------------------------------------------------

There are various methods to compute the path integral in a field theory.
In the continuum among approximate methods perhaps the most popular one is the loop expansion. 
The practical recipe starts by considering a classical Lagrangian consisting of a kinetic term and then
add to it interactions. For a Lorentz invariant scalar field theory in $d=4$ dimensions, invariant under an internal, global 
$\mathbb Z_2$ symmetry $\phi \longrightarrow -\phi$, we have
(the 0 subscript is used to denote bare quantities; for simplicity we use the same notation for bare and renormalized fields)
\be\label{L24}
{\cal L} = -\frac{1}{2} \phi \Box \phi - \frac{1}{2} m_0^2 \phi^2 - \frac{\l_0}{4!} \phi^4 + \cdots = {\cal L}^{(2)} + {\cal L}_{\rm int}\, ,
\ee
with ${\cal L}^{(2)} = {\cal L}_{\rm kin} + {\cal L}_{\rm mass}$, ${\cal L}_{\rm kin} = - \frac{1}{2} \phi \Box \phi$ and ${\cal L}_{\rm mass} = - \frac{1}{2} m_0^2 \phi^2$.
The interaction part typically contains the marginal operators, for example in the basic, "renormalizable" context
\be
{\cal L}_{\rm int} \equiv {\cal L}^{(4)} = - \frac{\l_0}{4!} \phi^4\, ,
\ee
but we should point out that there is no fundamental reason why ${\cal L}_{\rm mass}$ can not be part of ${\cal L}_{\rm int}$ too.
The quantum theory is defined through the path integral
\be
{\cal Z} = \int D\phi e^{i \int d^4x {\cal L} + J \phi }
\ee
and its functional derivatives. We do not know how to compute $\cal Z$ exactly so we resort to various expansions.
The expansion in small $\l_0$ results in the loop expansion. Renormalization absorbs infinities in counter-terms, eliminates the bare couplings $m^2_0$ and $\l_0$
and returns finite, but scale dependent renormalized couplings $m^2$ and $\l$. The renormalized Lagrangian has the same form as the bare Lagrangian,
with the bare couplings and fields replaced by renormalized ones.
The computation of all terms in the loop expansion reconstructs entire non-perturbative sectors of the quantum theory. 
Let us now fix the loop order and ask if there are processes at 1-loop that generate "effective" terms that
do not appear in ${\cal L}$. Indeed, the Triangle diagram 
%-------------------------------------
\vskip .5cm
\begin{center}
\begin{tikzpicture}[scale=0.5]
\draw [dashed] (-1,0) -- (1,1);
\draw [dashed] (-1,0) -- (1,-1);
\draw [dashed] (1,1) -- (1,-1);
%%%%%%
\draw [dashed] (1,1) -- (2.5,1);%p3
\draw [dashed] (1,-1) -- (2.5,-1);%p4
%%%%%%
\draw [dashed] (1,1) -- (1,2.5);
\draw [dashed] (1,-1) -- (1,-2.5);
%%%%%%
\draw [dashed] (-2.5,-1.2)--(-1,0);
\draw [dashed] (-2.5,1.2)--(-1,0);
\end{tikzpicture}
\end{center}
%-------------------------------------
proportional to $\l^3$ generates the new interaction $\phi^6$. It is 
fixed by the couplings appearing already in ${\cal L}$ --- and it is finite. As such it does not introduce a new running coupling and 
does not affect renormalization. It is a physical process, a prediction of perturbation theory correcting the tree level scattering of 3 $\phi$ by 3 $\phi$ fields
by the exchange of a single $\phi$, analogous to the one-loop box correction to the 4-fermion scattering in QED.
Similarly, there is a Box diagram generating a finite $\l^4 \phi^8$ contribution (analogous to the light by light scattering in 
QED since it does not correct a tree level process) and so forth.
The entire perturbative, one-loop effective action is then
\be
{\cal L}_{\rm eff} = -\frac{1}{2} \phi (\Box + m^2) \phi - \frac{\lambda}{4!} \phi^4 +\sum_{n=3,\cdots} w_{2n} \frac{\l^{n}}{\L^{2n-4}} \phi^{2n}\, .
\ee 
The scale $\L$ is not an independent scale. It is an internal scale of mass dimension one, determined by $m^2$ and/or by dynamically generated scales
such as the vacuum expectation value (vev) of $\phi$ in a broken phase.
The coefficients of the infinite sum, the Wilson coefficients, are experimentally measurable quantum effects and are predictions of perturbation theory in a 
renormalizable quantum field theory. The corresponding operators are generally called Higher Dimensional Operators (HDO) because they have
classical dimensions higher than the relevant and (classicaly) marginal operators contained in ${\cal L}$. The $w_{2n}$ are numerical factors computable
in perturbation theory. 
Clearly, the only independent $\b$-functions are $\b_{m^2}$ and $\b_\l$, the running of the Wilson coefficients being determined by the running of $m^2$ and $\l$.
The resulting extension of the renormalizable theory is chracterized as an Effective Field Theory (EFT).
The nice thing about this simple scalar field theory is that there is only one independent HDO for each $n$. 
For the Standard Model the corresponding basis, the Warsaw basis \cite{Warsaw}, contains
2499 independent operators and the theory extended by this (or another, equivalent) basis is what we call the Standard Model EFT (SMEFT).

In the above we first renormalized the theory and then computed the effect of the HDO.
Let us now see what happens if we do things in the opposite order. Then we must quantize the classical Lagrangian
\be\label{LHDO}
{\cal L} = -\frac{1}{2} \phi_0 (\Box + m_0^2) \phi_0 - \frac{\l_0}{4!} \phi_0^4 +\sum_{n=3,\cdots} \frac{c_0^{(2n)}}{\L^{2n-4}} \phi_0^{2n}\, .
\ee
There are several new features in the above Lagrangian. It has the same structure at the relevant-marginal level but the Wilson coefficients
are now replaced by the new couplings $c_0^{(2n)}$ which are dimensionless (in $d=4$) due to the powers of the suppressing scale $\L$. 
Also, its HDO are pure polynomial terms even though we could have added also terms containing any
number of $\Box$ operators, consistently with the internal symmetry.
Regarding the role of such operators, especially of the ones of dimension higher than four, note that they typically
play a crucial role in the consistency of extensions of Einstein's theory of gravity, as well as in the consistency of the SMEFT. 
The reason is that both cases involve higher derivative operators which may induce non-physical degrees of freedom, that is ghosts, on their corresponding actions. 
In the current work we demonstrate a consistent way to deal with this problem from the point of view of Quantum Field Theory (QFT), 
using a simultaneous cancellation of the ghost's pole at tree and loop levels.

To begin, renormalization will return the same $\b$-functions $\b_{m^2}$ and $\b_\l$ to leading order in $\frac{1}{\L}$ with corrections
of order $\frac{1}{\L^{2n-4}}$ due to the fact that now the Wilson coefficients acquire a scale dependence dictated by independent
$\b$-functions $\b_{c^{(2n)}}$. We call such a non-renormalizable extension, a Generalized EFT (GEFT).
This is a point of view that has been taken before, especially in the context of the SMEFT \cite{Einhorn,Colangelo}, where
the Wilson coefficients get renormalized so that the associated counter-terms absorb the divergences
induced on SM processes by the HDO. As a consequence, one can define renormalization group equations 
for the Wilson coefficients which run independently. 
The one-loop Triangle diagram on the other hand is still finite and has the same expression as before
proportional to $\l^3$, with the effect of the HDO encoded indirectly in the modified running of $\l$.
At one-loop the modification goes to zero in the massless limit. 
Apparently this approach is 'more' than just perturbation theory. It is an extension of it, as it contains the
predictions of the renormalizable approach in the limit where the $c^{(2n)}\to w_{2n} \l^{n}$.
Note that $\L$ is still not an independent scale. In fact we could have inserted the HDO multiplied by dimensionfull couplings or suppress
them by a function of the regulating scale. 
In Dimensional Regularization (DR) we could consider for example
$\m=\L$ which amounts to identifying $\L$ with the scale regulating the loop diagrams.
In a cut-off regularization instead, $\L$ could be identified with the cut-off
used to regulate UV-divergent integrals. Or better, in a lattice regularization, $\L=1/a$ with $a$ the lattice spacing. Note that on the lattice
this is the natural context because the lattice action expanded near the naive continuum limit (i.e. for small $a$) 
yields a classical action analogous to \eq{LHDO} with coefficients that non-perturbatively depend on $a$. 
In lattice perturbation theory one then truncates the expansion in the lattice spacing at some order and performs the loop
expansion at some fixed order as usual, with the understanding that the truncation is only valid in domains of the phase diagram
where the lattice spacing is small. The important thing to keep in mind here is that by doing this one obtains a generalization
of perturbation theory with independently running Wilson coefficients at the price of having to deal with a 'non-renormalizable' effective action which 
generates an ever growing number of new counter-terms.
Of course none of this should matter if we could keep the entire series of HDO and be able to compute any number of loops.
In other words, the validity of the loop expansion comes from an assumption about the weakness of couplings and 
the appearance of an independent expansion in HDO is an artifact of truncation. Non-perturbatively the two expansions are related
and the decoupled double series may be valid only in certain regimes of the phase diagram. This may seem to be too restrictive
but it is perhaps one way to extend perturbation theory to obtain new pieces of the exact result.

A certain generalization of perturbation theory takes into account also the expectation value of operators $\langle  O \rangle$
\be\label{Opins}
\langle  O \rangle \sim \int D\phi\,  O\,  e^{i\int d^4x {\cal L}}\, ,
\ee
after the renormalization of ${\cal L}$. The operator $O$ (with an associated coupling $c$) develops an anomalous dimension when the arguments
of the fields it contains coincide, a fact slightly beyond basic, perturbative quantum field theory. 
Similarly, correlations of multiple operators at different points give divergences when some of those points coincide.
All these divergences contribute to a total anomalous dimension structure.
Finiteness of the insertions in the path integral demands then that the counter-tem of $c$
is determined by this anomalous dimension structure and vice versa. This allows us to use the kinetic term as the pivot Lagrangian
and treat any other operator as an insertion, including the mass and quartic interactions. In fact, for insertions of operators like $\phi^2$ and $\phi^4$
these operations are rather trivial as they just reproduce the known counter-terms of the mass and $\l$. More interesting things happen when
we insert operators of higher than four mass dimension, like $\phi^{2n}, n = 3,4,\cdots$ because then the anomalous dimension structure
imposes a scale dependence on the corresponding coupling $c^{(2n)}$. 
The non-trivial statement here is that if one takes into account all possible operator insertions then the phase diagram obtained should coincide
with the phase diagram obtained from the quantization of \eq{LHDO}.
So why we should use the operator insertions method to derive the phase diagram?
After all we claim that it gives the same  $\b$-functions as the direct renormalization of \eq{LHDO}, even in an EFT or a GEFT.
The motivation behind our choice to use \eq{Opins} has both a technical and a conceptual origin.
Regarding the technical part recall that following the standard renormalization procedure we end up with the $\b$-functions that depend on the couplings in a particular way.
However, in this procedure the relation between the anomalous dimension of an operator and the contribution of the associated coupling to the quantum-part of the $\b$-function 
is not transparent. In fact, given a $\b$-function, in the context of the direct renormalization of a classical Lagrangian containing perhaps HDO,
there is no way to determine the origin of its parts with respect to the insertion level of the corresponding operators.
This is true even for a simple model like that of \eq{L24} where the known result $\b_\l \sim \l^2$ does not indicate that $\phi^4$ has vanishing anomalous dimension at 1-loop.
As we show here, we can overcame this shortcoming using the framework of \eq{Opins} to get a complete view of the internal structure of the $\b$-functions.
Turning to the conceptual motivation, it is known that the running of the couplings with scale may drive the system into fixed points where a phase transition may occur.
The nature of the phase transition may be of the quantum or zero temperature type and it may be of second order with a continuum limit
or of first order with a nearly continuum limit.
Then an interesting question is how can we describe a QFT very close to such a phase transition.
To answer this recall that when all the couplings admit their fixed point value simultaneously, the theory becomes conformally invariant or
nearly conformally invariant, depending on the order of the phase transition. Then one could use instead of a QFT description, a Conformal Field Theory (CFT) description.
The departure from these fixed points (interacting or not) induces deformations of the corresponding CFT.
Thus, the quantum properties of the system near the phase transition can in principle be described by a deformed CFT. We plan to be more specific on this in the near future.
In a CFT now the natural language of deformations is that of operator insertions so if we wish to make a connection
between the two languages it is natural to use a similar algorithm in the QFT approach.
In short, the operator insertion method may be a useful tool to describe a field theory close to quantum phase transitions, may it be applied to
a QFT or a CFT language. Note finally that this is not necessarily just a conceptual exercise. Higher dimensional Yang-Mills theories for example possess
quantum phase transitions where they are also dimensionally reduced so physics is four-dimensional and exactly in a context as described above.

The equivalence of \eq{LHDO} and \eq{Opins}, the fact that $\L$ is an internal to the theory scale
together with the fact that in the standard renormalizable picture the HDO represent finite quantum effects, imply that 
in the extension \eq{Opins} the HDO may also be associated with quantum effects, 
thus contributing to the breaking of scale invariance a spontaneous amount.
This is to be contrasted with $m^2 \phi^2$ where $m^2$ is an explicit scale invariance breaking coupling that does not go away in the
classical limit. This is an important point for several reasons one of them being that in the massless limit and in a $\mathbb Z_2$ broken phase the breaking 
of the internal and scale symmetries have the same origin. 

In Section \ref{sobf} we describe the renormalization process and the derivation of the $\b$-functions in a GEFT using the operator insertion method.
In Section \ref{2-4-6} we derive the renormalization and the phase diagram of the $\phi^4$ theory extended by dimension-6 terms, using the algorithm of Section \ref{sobf}.
In this section we also tackle a problem associated with higher derivative operators that introduce ghosts in the spectrum.
In Section \ref{2-4-6-8} we discuss dimension-8 operators and in Section \ref{Concl} we state our Conclusions.
In Appendix \ref{Sc.In.} we show how scaleless integrals are treated in dimensional regularization.
To illustrate the method and notation used in the main part of the paper, 
in Appendix \ref{2-4} we review the renormalization of the $\phi^4$ theory using the operator insertion method.
This is the extended version of the much shorter letter \cite{FotisLetter} where many of the results derived here in detail can be found.

%---------------------------------------------------------------------------------------------------------------------------------------------------
\section{GEFT $\b$-functions and anomalous dimensions}\label{sobf}
%---------------------------------------------------------------------------------------------------------------------------------------------------

We are interested in the evaluation of $\b$-functions and anomalous dimensions
for the Lagrangian of \eq{LHDO} with $n=3,4$. This is just the massive $\phi^4$ theory extended by HDO of dimension-$(6,8)$.
The associated Wilson coefficients run independently of the marginal and relevant couplings, which determines the model as a GEFT.
An operator insertion of level-$w$ is an expectation value $\langle \cdots O^{(r_1)}\cdots O^{(r_w)}\rangle$, generated by the quantum theory,
perturbatively or non-perturbatively.

In the case of only one independent operator at every mass dimension, as it is the case for the scalar theory we are interested in here, the $\b$-function
of the coupling $c^{(l)}$ associated with the operator $O^{(l)}$ ($c^{(2)}=m^2$ and $c^{(4)}=\l$ etc.), 
as obtained from the renormalization of \eq{LHDO}, can be expressed as \cite{Manohar}
\be\label{totalbeta}
\b_{c^{(l)}} =  (d_{O^{(l)}}  - d) c^{(l)} + q_{l+2} \, \G_m^{l+2}\, m^2 + \G_{O^{(l)}} \, c^{(l)} +  \G_{l r s} \, c^{(r)} c^{(s)} + \cdots\, ,  %\G_m c^{(l+2)}
\ee
with $l,r,s=2,4,6,8,\cdots$ (sum over the indices $r$ and $s$ is implied),
\be
   q_{l+2} = 
   \begin{cases}
     0 \,\,{\rm when } \,\, l=2 
     \\
    \frac{ 2 }{l } \,\,\,{\rm otherwise }
   \end{cases}
 \ee
The dots represent higher than  level-2 insertions and 1-loop corrections. In case where there are several operators of a given dimension,
the couplings and the $\G$'s acquire extra indices. 
$\G_m^{l+2}$ and $\G_{O^{(l)}}$ are the anomalous dimensions of the operator $O^{(l)}$, induced by 
the relevant and marginal operators respectively. For the mass operator, when $l=2$, the terms 
$\G_m^{l+2}m^2$ and $ \G_{O^{(l)}} \, c^{(l)} $ coincide, so a coefficient $q_{l+2}$ in front of $\G_m^{l+2}$ is needed to avoid double counting. 
Moreover, $q_{l+2}$ protects $\G_m^{l+2}$ from numerical factors infiltrating from the classical dimension of $O^{(l)}$.
The quantity $\G_{l r s}$ in the non-linear term represents the anomalous dimension associated with 
the level-2 operator insertion $\langle \cdots O^{(r)} O^{(s)}\rangle$, $r,s \ge 2$. For $l\le 8$ we have only $r=s$, otherwise mixing
of HDO of different mass dimensions starts to appear.
Renormalization of \eq{LHDO} indeed gives this total structure but
it is not sensitive to the entire substructure implied by the $\G$-matrices. It is our intention here to reconstruct $\b_{c^{(l)}}$
to expose this substructure, by applying the operator insertion method.

Consider the operator $O^{(l)}_0$ and its associated coupling $c^{(l)}_0$.
Renormalization (with Dimensional Regularization) indicates that
\be\label{Phiclr}
{\phi}_0 = \sqrt{ Z_{\phi} } {\phi} = \sqrt{ (1+\d_{\phi}) } {\phi} \, , \,\,  c^{(l)}_0= Z_{c^{(l)}} \, c^{(l)} = (1+\d_{c^{(l)}}) c^{(l)}  \, ,
\ee 
and introduces the regulating scale $\m$ so that in $d$-dimensions $c^{(l)}_0 \to c^{(l)}_0 \m^{d -d_{O^{(l)} }}$ since (denoting the dimension of an operator by $[\cdots]$)
\be\label{dim.an.Oc}
[c^{(l)} \, O^{(l)}] =  d_{c^{(l)}} + d_{O^{(l)}} = d  \Rightarrow d_{c^{(l)}} = d - d_{O^{(l)}} \, , 
\ee
where $d_{O^{(l)}}$ and $d_{c^{(l)}}$ are the classical dimensions of $O^{(l)}$ and $c^{(l)}$ respectively. Now $c^{(l)}$ is dimensionless in $d$ dimensions.
An observation here is that \eq{Phiclr} is not the full story of renormalization since $O^{(l)}_0$ may include bare fields located at the 
same space-time point and this generates extra UV divergences. So Green's functions which 
include composite operators are divergent even in an already renormalized theory.
The simplest example that demonstrates this is $O^{(2)}_0(y)=\phi_0^2(y)$ (see \cite{Peskin}), as then $<O^{(2)}_0>$ is
\be
\langle 0 | T[ \phi_0^2(y)] | 0 \rangle \equiv \lim_{x\to y}  \langle 0 | T[ \phi_0(x)\phi_0(y)] | 0 \rangle =  \lim_{x\to y} \int\frac{d^4 k}{(2\pi)^4i} \frac{e^{i k(x-y)}}{k^2 - m^2} =  \int\frac{d^4 k}{(2\pi)^4 i} \frac{1}{k^2 - m^2} \, , \nonumber
\ee
which is the known (divergent) $A_0(m^2)$ Tadpole integral.
Thus, a bare operator, a product of $m$-fields ${\phi}_0$, should be renormalized as
\be\label{O0.r.}
O^{(l)}_0 = Z_{\phi}^{m/2} \, \overline {Z}_{O^{(l)}} O^{(l)} =  {Z}_{O^{(l)}} O^{(l)} \, ,
\ee
with 
\be\label{ZOdo1}
{Z}_{O^{(l)}} = Z_{\phi}^{m/2} \, \overline {Z}_{O^{(l)}} = 1 + \d_{O^{(l)}} \hskip .3cm {\rm or} \hskip .3cm ({Z}_{O^{(l)}})^{-1} = 1 - \d_{O^{(l)}} \, .
\ee
Let us now recall how the anomalous dimension $\G_{O^{(l)}}$ of an operator fits in the $\b$-function (see for example \cite{Neubert,Serone}).
To this effect, let us assume for a moment that level-1 insertions are the only ones present.
Acting on $c^{(l)}_0 \m^{d - d_{O^{(l)}} } = Z_{c^{(l)}} c^{(l)} \m^{d - d_{O^{(l)}} }$ with $\m \, d/d \m$ on both sides we have
\bea
\m \frac{d( c^{(l)}_0 \m^{d - d_{O^{(l)}} } ) }{d \m} &=& \m \frac{d( Z_{c^{(l)}} c^{(l)} \m^{d - d_{O^{(l)}} } ) }{d \m} \nonumber\\
0 &=& \m \frac{d Z_{c^{(l)}} }{d \m}  c^{(l)} \m^{d - d_{O^{(l)}} } + Z_{c^{(l)}} \left( \m \frac{d c^{(l)}  }{d \m} \right) \m^{d - d_{O^{(l)}}  }  + (d - d_{O^{(l)}} ) Z_{c^{(l)}} c^{(l)} \m^{d - d_{O^{(l)}} } \nonumber
\eea
while acting from the left with $ Z_{c^{(l)}}^{-1}$ and isolating the $  \m \frac{d c^{(l)}  }{d \m} \equiv \b_{c^{(l)}} $ term, we obtain
\be\label{vbc1}
\b_{c^{(l)}} = (  d_{O^{(l)}}  - d ) c^{(l)} - \left( Z_{c^{(l)}}^{-1} \, \m \frac{d  Z_{c^{(l)}}  }{d \m} \right) c^{(l)} \, .
\ee  
Since $c^{(l)}$ is contained in the insertion of $ c^{(l)}_0 O^{(l)}_0$, the extra UV divergences that $O^{(l)}_0$ carries are cancelled when
\be
Z_{c^{(l)}}^{-1} = Z_{O^{(l)}}\, .
\ee
Then \eq{vbc1} becomes 
\bea\label{vbc2}
\b_{c^{(l)}} &=&  (d_{O^{(l)}}  - d) c^{(l)} - \left( Z_{O^{(l)}} \, \m \frac{d(Z_{O^{(l)}})^{-1} }{d \m} \right) c^{(l)} \nonumber\\
&=&  (d_{O^{(l)}}  - d) c^{(l)} +  (Z_{O^{(l)}})^{-1} \,  \m \frac{d Z_{O^{(l)}} }{d \m} c^{(l)} \nonumber\\
&=&  (d_{O^{(l)}}  - d) c^{(l)} + \G_{O^{(l)}} c^{(l)}
\eea
where we defined
\be\label{An.d.m1}
\G_{O^{(l)}} = (Z_{O^{(l)}})^{-1} \,  \m \frac{d Z_{O^{(l)}} }{d \m} \, ,
\ee
the anomalous dimension of the operator $O^{(l)}$. In case where the only marginal operator is $O^{(4)}$, 
at 1-loop, the simple formula $\G_{O^{(l)}}= (d_{O^{(4)}}  - d) \frac{\d_{O^{(l)}}}{c^{(4)}}$ can be used.
The contribution of the mass insertion to $\b_{c^{(l)}}$ in \eq{totalbeta} is analogously defined by (see \eq{fnO2} below)
\be\label{Gm}
\G_m^{l+2} = \sum_k (Z_k^{l+2}  )^{-1} \m \frac{d Z_k^{l+2} }{d \m}\, .
\ee
Here one can use the 1-loop formula $\G_m^{l+2}= \sum_k (d_{O^{(l+2)}}  - d) \frac{\d^{(l+2)}_k}{ c^{(l+2)}}$, with $k= 2,4,6,8 \cdots$. 

This is of course just one part of \eq{totalbeta} because higher level insertions are also generated,
affecting the $\b$-function already at 1-loop. Such a case is the level-2 insertion.
Applying the OPE on the double insertion and repeating the steps above yields the contribution 
$\G_{l r s} \, c^{(r)} c^{(s)}\in \b_{c^{(l)}}$.
The level-2 anomalous dimension for every triplet of indices is then
\be
\G_{l r s} = \ve \d_{lrs} \, ,
\ee
where $\d_{lrs}=Z_{lrs}-1$ is the renormalization counter-term of the double insertion.

Now let us be more specific about the operator insertion method, according to which every operator can be considered as a deformation of the kinetic term. 
Let us assume for simplicity that $Z_\phi = 1$, which is true at 1-loop.
The basic path integral of a pure scalar theory with action $S[\phi_0]$ and inserted operators $O_0^{(l)}$ is given by 
\be
{\cal Z} [J,J^{(l)}] = \int {\cal D} \phi_0 \, e^{i S[\phi_0] + J \phi_0 + J^{(l)} O^{(l)}_0 } \, , \nonumber
\ee
with summation over $l$ implied. Then the $n$-point Green's function is
\be\label{ZJfn}
\langle  {\phi}(x_1)\cdots {\phi}(x_n) \rangle_0  = \frac{1}{{\cal Z} [J,J^{(l)}] i^n} \frac{\partial }{ \partial J(x_1)} \cdots \frac{\partial }{ \partial J(x_n)} {\cal Z} [J,J^{(l)}] \Big|_{J=J^{(l)} =0 }\, .
\ee
In order to insert $w$ times the $O_0^{(l)}$ operator we must act with $\partial^w/ i^w \partial J^{(l)}(y_1)\cdots J^{(l)}(y_w)  $ on the partition function.
$w$ defines the level of the insertion in the sense that $w=1$ is a single insertion, $w=2$ is a double insertion and so on.
A single insertion for example affects the $n$-point function as
\be\label{fnOy1}
\langle  {\phi}(x_1)\cdots {\phi}(x_n) O^{(l)}(y) \rangle_0  = \frac{1}{{\cal Z} [J,J^{(l)}] i^n} \frac{\partial }{ \partial J(x_1)} \cdots \frac{\partial }{ \partial J(x_n)} \frac{\partial }{ i \partial J^{(l)}(y)} {\cal Z} [J,J^{(l)}] \Big|_{J=J^{(l)} =0 } \, .
\ee
Since the functional derivatives wrt $J$ and $J^{(l)}$ are independent, each group of derivatives should be renormalized independently.
That is to say, any $n$-point Green's function can be affected by any operator $O_0^{(l)}$.

Now, the action depends on a Lagrangian which we split in two parts:
\be\label{S0L0}
S[\phi_0]= \int d^d x \Bigl [ {\cal L}_{\rm kin, 0} +  {\cal L}_{\rm int, 0} \Bigr ]\, ,
\ee
with ${\cal L}_{\rm kin, 0}  $ and $ {\cal L}_{\rm int, 0} $ the kinetic and interaction parts of the Lagrangian. 
The former determines the propagator while the latter contains the vertices.
The l.h.s of \eq{ZJfn} is given by
\be\label{G0.a1}
G^{(n)}_0(x_1,\cdots, x_n)  \equiv \langle  {\phi}(x_1)\cdots {\phi}(x_n) \rangle_0 = \langle 0 | T[ {\phi}_0(x_1)\cdots {\phi}_0(x_n) ] | 0 \rangle \, ,
\ee
where $T[  \cdots]$ denotes time-ordering.
Following \cite{Peskin}, the expectation value in \eq{G0.a1} is
\be\label{T.Or.}
\langle 0 | T[ {\phi}_0(x_1)\cdots {\phi}_0(x_n) ] | 0 \rangle = {\cal N}^{-1} \langle 0 | T[{\phi}_0(x_1)\cdots {\phi}_0(x_n) e^{i\int d^d x {\cal L}_{\rm int, 0} }  ] | 0 \rangle\, ,
\ee
with 
\be
{\cal N} = \langle 0 | T[e^{i\int d^d x {\cal L}_{\rm int, 0} }  ] | 0 \rangle \, , \nonumber
\ee
necessary for the cancellation of disconnected and bubble diagrams. ${\phi}$ represents the field as a quantum operator.

The algorithm is the following: start with a bare, free massless action so that any operator (including the mass term) added to the kinetic term can be viewed as an operator insertion.
Since the mass term is considered as part of the operator-insertion, massless propagators will run in the loops.
Then, ${\cal L}_{\rm int,0} =  \sum_l c_0^{(l)} O_0^{(l)}(y)$ which means that
\be\label{GnGnk.m.01}
\langle  {\phi} (x_1)\cdots {\phi}(x_n)  \rangle_0  =  
{\cal N}^{-1} \langle 0 | T[{\phi}_0(x_1)\cdots {\phi}_0(x_n) e^{i\int d^d y \sum_l c_0^{(l)} O_0^{(l)}(y)} ] | 0 \rangle  \, .  
\ee
Next, single out one operator with label $l$ from the sum in the exponent and apply the series expansion to it \cite{Serone}.
The rest of the operators remain summed over in the exponent, in a reduced sum over $l'$.
Then \eq{GnGnk.m.01} becomes
\bea\label{GnGnk.m1}
&& \langle  {\phi}(x_1)\cdots {\phi}(x_n)  \rangle_0 = 
{\cal N}^{-1} \sum_{w=0}^\infty \int d^d y_{1}\cdots d^dy_w \frac{(i c_0^{(l)} \m^{d - d_{O^{(l)}}})^w}{w!}\times  \nonumber\\
&&\langle 0 | T[{\phi}_0(x_1)\cdots {\phi}_0(x_n)
O_0^{(l)}(y_1)\cdots O_0^{(l)}(y_w) e^{i\int d^d y  \sum_{l'} c_0^{(l')} O_0^{(l')}(y)} ] | 0 \rangle  \nonumber\\
&=& \sum_{w=0}^\infty 
\int d^d y_{1}\cdots d^dy_w \frac{(i c_0^{(l)} \m^{d - d_{O^{(l)}}})^w}{w!}
\langle {\phi}(x_1)\cdots {\phi}(x_n) O^{(l)}(y_1)\cdots O^{(l)}(y_w) \rangle_0 \nonumber\\
&=& \sum_{w=0}^\infty 
\int d^d y_{1}\cdots d^dy_w \frac{(i c_0^{(l)} \m^{d - d_{O^{(l)}}})^w}{w!} G_0^{(n,w)}
\eea
The correlator $G^{(n,w)}$ on the r.h.s of \eq{GnGnk.m1} expresses the way that the $w$ times inserted operator $O^{(l)}$ affects the $n$-point Green's function, in the presence of
interactions specified by the sum $\sum_{l'}$.
For example, for a single insertion $w=1$ and $G^{(n,1)}$ gives the l.h.s of \eq{fnOy1}.

Now we are ready to evaluate the different parts of \eq{totalbeta}.
At level-1 of insertions, for $n$-external legs and operators $O^{(l)}$, there are two relevant cases:
\begin{itemize}
\item $n=l$\\
In this case the l.h.s of \eq{fnOy1} is
\be
\langle  {\phi}(x_1)\cdots {\phi}(x_n) O^{(l)}(y) \rangle_0 \equiv \langle  {\phi}(x_1)\cdots {\phi}(x_l) O^{(l)}(y) \rangle_0 \,  \nonumber
\ee
and its renormalization requires that
\be\label{flOl}
\langle  {\phi}(x_1)\cdots {\phi}(x_l) O^{(l)}(y) \rangle_0 = Z_{O^{(l)}} \langle  {\phi}(x_1)\cdots {\phi}(x_l) O^{(l)}(y) \rangle \, ,
\ee
from which the $\G_{O^{(l)}}$ part of \eq{totalbeta} can be extracted.
\item $n\ne 2$ and $l = 2$\\
The above shows the way that a mass operator affects $n$-external legs. Then the l.h.s of \eq{fnOy1} is given by
\be
\langle  {\phi}(x_1)\cdots {\phi}(x_n) O^{(2)}(y) \rangle_0 \, . \nonumber
\ee
After renormalization we have $n\ne2$ but all the possible insertions, so that
\be\label{fnO2}
\langle  {\phi}(x_1)\cdots {\phi}(x_n) O^{(2)}(y) \rangle_0 = \sum_k Z_k^{n+2} \langle  {\phi}(x_1)\cdots {\phi}(x_n) O^{(k)}(y) \rangle \, .
\ee
This will give us $\G_m^{n+2}$ of \eq{totalbeta} through \eq{Gm}.
\end{itemize}
The remaining contribution to $\b_{c^{(l)}}$ has its origin in double operator insertions and specifies the anomalous dimension $\G_{lrs}$.
Double insertions can be generated by acting one more time with $\partial/i \partial J^{(l)}$ on \eq{fnOy1}.
Then, for $n$-external legs and two $O^{(l)}$ operators, we obtain the following two cases:
\begin{itemize}
\item $n=l$\\
The relevant expectation value is
\be
\langle  {\phi}(x_1)\cdots {\phi}(x_n) O^{(l)}(y) O^{(l)}(z)  \rangle_0 \equiv \langle  {\phi}(x_1)\cdots {\phi}(x_l) O^{(l)}(y) O^{(l)}(z)  \rangle_0  \, , \nonumber
\ee
therefore renormalization indicates that
\be\label{flOlOl}
\langle  {\phi}(x_1)\cdots {\phi}(x_l) O^{(l)}(y) O^{(l)}(z)  \rangle_0 = Z_{lrs} \langle  {\phi}(x_1)\cdots {\phi}(x_l) O^{(r)}(y) O^{(s)}(z) \rangle \, ,
\ee
where we have set $ O_0^{(l)}(y) O_0^{(l)}(z) = Z_{lrs} O_0^{(r)}(y) O_0^{(s)}(z)$.
\item $n\ne k'$ with $k'$ the dimension of the inserted operators.\\
In this case double insertions are just
\be
\langle  {\phi}(x_1)\cdots {\phi}(x_n) O^{(k')}(y) O^{(k')}(z)  \rangle_0 \, , \nonumber
\ee
while renormalizing the above, now with $ O_0^{(k')}(y) O_0^{(k')}(z) = Z_{qrs} O_0^{(r)}(y) O_0^{(s)}(z)$, we get 
\be\label{fnOlOl}
\langle  {\phi}(x_1)\cdots {\phi}(x_n) O^{(k')}(y) O^{(k')}(z)  \rangle_0 = Z_{qrs} \langle  {\phi}(x_1)\cdots {\phi}(x_n) O^{(r)}(y) O^{(s)}(z) \rangle \, ,
\ee
where $Z_{qrs} \equiv Z_{q_{(n,k')}rs}$.
By the subscript $(n,k')$ on $q$ we stress the fact that the $\b$-function can in general be affected by the level-2
insertion of a dimension-$k'\ne n$ operator into the $n$-point function.
\end{itemize}
From these, the non-linear contribution $\G_{lrs} c^{(r)} c^{(s)}$ to $\b_{c^{(l)}}$ can be obtained.

There are a couple of subtle points here.
The first regards the possible contractions of the fields in the expectation value. 
When we consider double insertions there is no need to use ${\cal L}_{\rm int}$ to contract fields.
So couplings do not participate in the calculation.
The second refers to the absorption of divergences. In particular the two sides of \eq{fnOlOl}, when $l \le 8$, give the same loop integral but the l.h.s is divergent while the r.h.s is finite.
The equality of the two sides holds since $Z_{lrs}$ is divergent, which means that the diagram at the r.h.s should be considered as a finite vertex.
Finally, in the following calculations we meet only two kinds of divergent integrals.
These are the scalar integrals $A_0$ and $B_0$ evaluated in Appendix \ref{Sc.In.}.
When massless propagators participate in the loop, these integrals are scaleless.
When scaleless, the former is identically zero in DR, while the latter can be broken into a UV and an IR part.
Depending on the situation, we keep only its UV part (operator insertion) or we consider both its UV and IR parts
in which case $B_0$ is zero (renormalization of \eq{LHDO}) in DR.
There are two more 1-loop integrals, a Triangle ($C_0$) and a Box ($D_0$), already mentioned in the Introduction, which
are finite.

%---------------------------------------------------------------------------------------------------------------------------------------------------
\section{Inserting dimension-6 operators}\label{2-4-6}
%---------------------------------------------------------------------------------------------------------------------------------------------------

Consider the Lagrangian ${\cal L}={\cal L}^{(2)}+{\cal L}^{(4)} + {\cal L}^{(6)} $ 
which includes also the classically irrelevant HDO of dimension-6 \cite{phi6quantum}. We would like to determine its RG flows and symmetry breaking patterns.
There are three such operators (up to total derivatives), which respect Lorentz and $\mathbb{Z}_2$ symmetry:
\be
(O_1^{(6)}, O_2^{(6)}, O_3^{(6)}) = \left( \frac{\phi^2 \Box \phi^2}{ \L^2} , \frac{\phi \Box^2 \phi}{\L^2}, \frac{\phi^6}{\L^2} \right) 
\ee 
and they introduce the respective dimensionless couplings $c_1^{(6)}$, $c_2^{(6)}$ and $c_3^{(6)}$. 
Nevertheless, as we will see, only one of them is independent.
$\L$ is a cut-off scale and it is present to regulate the dimension of the inserted operators. 

With this in mind we add to ${\cal L}^{(4)} $ in \eq{L24}, the terms
\be\label{L6a}
{\cal L}^{(6)} = \frac{c_1^{(6)}}{ 4! \L^2} \phi^2 \Box \phi^2 + \frac{c_2^{(6)}}{ 2 \L^2} \phi \Box^2 \phi + \frac{c_3^{(6)}}{ 6! \L^2}  \phi^6 \, ,
\ee
so that the dimension-6 GEFT Lagrangian is
\bea\label{L6only1}
{\cal L} &=& -\frac{1}{2} \phi \Box \phi - \frac{1}{2} m^2\, \phi^2 - \frac{\l}{4!}\phi^4 \nonumber\\
&&+ \frac{c_1^{(6)}}{ 4! \L^2} \phi^2 \Box \phi^2 + \frac{c_2^{(6)}}{ 2 \L^2} \phi \Box^2 \phi + \frac{c_3^{(6)}}{ 6! \L^2}  \phi^6  \, 
\eea
and defines the $O^{(2)}$-$O^{(4)}$-$O^{(6)}$ system. Let us call the set of dimension-6 operators in \eq{L6a} the G-basis.
Notice that upon integration by parts and dropping total derivatives, the box operator commutes with $\phi$ and all positions of it
in the operator are equivalent.
There is an obstruction to compute directly RG flows in this basis since $O^{(6)}_1$ can result
in superluminal propagation when $c^{(6)}_1<0$ \cite{Nicolis}, in addition
$O_2^{(6)}$ contains an Ostrogradsky ghost (the O-ghost) \cite{Richard},
as it adds an extra, ghost-like pole to the propagator. The analysis of the phase diagram is expected to be 
simplified and more transparent in the basis where only $O_3^{(6)}$ exists. We call this the W-basis, by analogy 
to the Warsaw basis of dim-6 operators in the Standard Model \cite{Warsaw} which is ghost free.

Indeed, there is a freedom in \eq{L6only1} because we can perform a
field redefinition that allows to eliminate any HDO we want. 
To see how this works, take ${\cal L}_0$ (the bare version of \eq{L6only1}), introduce the field redefinition\footnote{Whether field redefinitions and renormalization commute is a fair question at this point and there are arguments that they do \cite{Anselmi}.}
\bea\label{f.red}
\phi_0 \rightarrow \phi_0 + \frac{x}{\L^2} \Box \phi_0 + \frac{y}{\L^2} \phi_0^3 
\eea
and substitute it back into ${\cal L}_0$. Keeping consistently terms up to ${\cal O}(1/\L^2)$, we obtain the transformed Lagrangian
\bea\label{L60}
{\cal L}_0 &=&  -\frac{1}{2} \left[ 1 + 2 x\frac{m_0^2}{\L^2} \right] \phi_0 \Box \phi_0 -\frac{1}{2} \phi_0\, m_0^2 \phi_0 - \left[\frac{\l_0}{4!} + \frac{y m_0^2}{\L^2} \right]\phi_0^4 
\nonumber\\
&+& \left[\frac{c_{1,0}^{(6)}-y-\frac{x}{6}\l_0}{4! \L^2} \right]\phi_0^2 \Box \phi_0^2 + \left[\frac{c_{2,0}^{(6)} - x}{2 \L^2}  \right]\phi_0 \Box^2 \phi_0 
+ \left[\frac{c_{3,0}^{(6)} - \frac{y \l_0}{6}}{6! \L^2}  \right] \phi_0^6 \, .
\eea
Perform also the extra redefinition 
\be\label{phired}
\phi_0 \to \frac{1}{\sqrt{1 + 2 x\frac{m_0^2}{\L^2}}}\, \phi_0
\ee
in order to bring the kinetic term into a canonical form. Now we can try to eliminate some operators in favour of others. 
The preferred elimination is that of the ones with higher derivatives.
This can be done by requiring the coefficients of those terms to vanish, which happens if
$x = c_{2,0}^{(6)} $ and $ y = c_{1,0}^{(6)} -( c_{2,0}^{(6)}/ 6) \l_0 $.
With these conditions we obtain the Lagrangian
\bea\label{L68noder}
{\cal L}_0 &=& -\frac{1}{2} \phi_0 \Box \phi_0 - \frac{m_0^2}{2} \left[ 1 - \frac{2 c_{2,0}^{(6)} m_0^2 }{\L^2}  
\right]\phi_0^2 + \left[ -\frac{\l_0}{24} + \frac{m_0^2 (-c_{1,0}^{(6)} + \frac{1}{3} \l_0 c_{2,0}^{(6)}) }{\L^2} \right] \phi_0^4 \nonumber\\
&+& \Biggl[ \frac{c_{3,0}^{(6)} - \frac{1}{6}c_{1,0}^{(6)}\l_0 + \frac{1}{36} c_{2,0}^{(6)} \l_0^2}{\L^2} \Biggr] \phi_0^6 \, ,
\eea
where indeed only $\phi^6$ remains as the only HDO.

Unfortunately this can not be the full story since in general, field redefinitions affect the measure of the path-integral \cite{phi6quantum}.
To clarify this aspect we renormalize ${\cal L}_0$ diagrammatically in the on-shell scheme, considering the 
HDO as part of the classical Lagrangian and we demonstrate how ghosts 
can be discarded from the spectrum. Having obtained a ghost-free theory with a reduced set of operators
(the W-basis), we can safely apply the operator insertion algorithm for the evaluation of the anomalous dimensions. 

%%%%%%%%%%%%%%%%%%%%%%%%%%%%%%%%%%%%%%%%%%%%%%%%%%%%%%%%%%%%%%%%%%%%%%%%%%%%%%
\subsection{Ghosts induced by higher derivative operators}\label{gairo}
%%%%%%%%%%%%%%%%%%%%%%%%%%%%%%%%%%%%%%%%%%%%%%%%%%%%%%%%%%%%%%%%%%%%%%%%%%%%%%

The main problem of \eq{L6only1} is the existence of the O-ghost, originating from $O^{(6)}_2$, 
which makes the model ill-defined. Consistency suggests that we must add a sector
whose purpose is to cancel the pole of the O-ghost. The sector that does the job is
\be\label{Reg}
{\cal L}_{Rg,0} =  - \frac{1}{2} \bar \chi_0\,\Box \chi_0 + \frac{m_{\chi,0}^2}{2} \bar \chi_0 \chi_0 - \frac{ \l_{\chi,0} }{4} \, \bar \chi_0 \chi_0 \, \phi_0^2 \, ,
\ee
featuring the Grassmann fields $\chi_0$ and $\bar \chi_0$ which we call the R-ghosts for reasons that will become clear. 
The total G-basis Lagrangian then becomes
\bea\label{L6xmx}
{\cal L}_0  &=& {\cal L}^{(4)}_{0} + {\cal L}^{(6)}_{0} + {\cal L}_{Rg,0} \nonumber\\
&=& -\frac{1}{2} \phi_0 \Box \phi_0 - \frac{m_0^2}{2}  \phi_0^2 -  \frac{\l_0}{4!} \phi_0^4 + 
\frac{ c_{1,0}^{(6)} }{4! \L^2} \phi_0^2 \Box \phi_0^2 + \frac{ c_{2,0}^{(6)} }{2 \L^2}  \phi_0 \Box^2 \phi_0  \nonumber\\
&&+ \frac{ c_{3,0}^{(6)} }{6! \L^2}  \phi_0^6 - \frac{1}{2} \bar \chi_0\,\Box \chi_0 + \frac{m_{\chi,0}^2}{2} \bar \chi_0 \chi_0 - \frac{ \l_{\chi,0} }{4} \, \bar \chi_0 \chi_0 \phi_0^2 \, .
\eea
The Feynman rules that this Lagrangian generates are:
\begin{itemize}
%------------------------------------- 
\item Scalar Propagator
%-------------------------------------
\begin{center}
\begin{tikzpicture}[scale=0.8]
\draw[dashed] (0,-0.19)--(2.5,-0.19) ;
%\node at (0,0.1) {$A$};
%\node at (2.5,0.1) {$B$};
%\draw[->] (0,-0.2)--(1.25,-0.2);
\node at (5.3,-0.3) {$= \hskip .1 cm \displaystyle
 \frac{i}{p^2 - m_0^2 + c_{2,0}^{(6)} \frac{p^4} {\L^2} } $};
\end{tikzpicture}
\end{center}
%------------------------------------- 
\item Ghost Propagator
%-------------------------------------
\begin{center}
\begin{tikzpicture}[scale=0.8]
\draw[] (0,-0.19)--(2.5,-0.19);
%\node at (0,0.1) {$A$};
%\node at (2.5,0.1) {$B$};
\draw [->,thick]  (1.25,-0.19)--(1.3,-0.19);
%\draw[->] (0,-0.2)--(1.25,-0.2);
\node at (4.5,-0.2) {$= \hskip .1 cm \displaystyle
 \frac{i}{p^2 + m_{\chi,0}^2 } $};
\end{tikzpicture}
\end{center}
%-------------------------------------
\item Four-point self interaction vertex
%-------------------------------------
\begin{center}
\begin{tikzpicture}[scale=0.7]
\draw [dashed] (0,0)--(1.5,1.4);
\draw [dashed] (0,0)--(1.5,-1.4);
\draw [dashed] (-1.5,1.4)--(0,0);
\draw [dashed] (-1.5,-1.4)--(0,0);
\draw [<-]  (0.7,0.3)--(1.3,0.9);
\node at (0.9,1.5) {$p_3$};
%\node at (2.1,1.7) {$N,C$};
\draw [<-]  (0.9,-0.4)--(1.5,-1);
\node at (1.1,-1.6) {$p_4$};
%\node at (2.3,-1.7) {$R,D$};
\draw [<-] (-0.9,0.3)--(-1.6,1);
\node at (-1,1.5) {$p_1$};
%\node at (-2.2,1.7) {$M,B$};
\draw [<-] (-0.8,-0.4)--(-1.5,-1);
\node at (-0.8,-1.5) {$p_2$};
%\node at (-2.1,-1.7) {$S,E$};
\node at (4.6,0) {$= \hskip .1 cm \displaystyle - i \l_0 - i c_{1,0}^{(6)} \frac{p^2} {\L^2}  \, .$};
\end{tikzpicture}
\end{center}
%-------------------------------------
\item Scalar-ghost vertex
%-------------------------------------
\begin{center}
\begin{tikzpicture}[scale=0.7]
\draw [] (0,0)--(1.5,1.4);
\draw [] (0,0)--(1.5,-1.4);
\draw [dashed] (-1.5,1.4)--(0,0);
\draw [dashed] (-1.5,-1.4)--(0,0);
\draw [<-]  (0.7,0.3)--(1.3,0.9);
\node at (0.9,1.5) {$p_3$};
%\node at (2.1,1.7) {$N,C$};
\draw [<-]  (0.9,-0.4)--(1.5,-1);
\node at (1.1,-1.6) {$p_4$};
%\node at (2.3,-1.7) {$R,D$};
\draw [<-] (-0.9,0.3)--(-1.6,1);
\node at (-1,1.5) {$p_1$};
%\node at (-2.2,1.7) {$M,B$};
\draw [<-] (-0.8,-0.4)--(-1.5,-1);
\node at (-0.8,-1.5) {$p_2$};
%\node at (-2.1,-1.7) {$S,E$};
\draw [->, very thick]  (0.7,0.7)--(0.8,0.8);
\draw [<-, very thick]  (0.7,-0.7)--(0.8,-0.8);
%\node at (-2.1,-1.7) {$S,E$};
\node at (4.2,0) {$= \hskip .1 cm \displaystyle - i \frac{ \l_{\chi,0} }{2} \, .$};
\end{tikzpicture}
\end{center}
%-------------------------------------
\item Six-point self interaction vertex
%-------------------------------------
\begin{center}
\begin{tikzpicture}[scale=0.7]
\draw [dashed] (0,0)--(1.5,1.4);
\draw [dashed] (0,0)--(1.5,-1.4);
\draw [dashed] (-1.5,1.4)--(0,0);
\draw [dashed] (-1.5,-1.4)--(0,0);
\draw [dashed] (0.02,-1.4)--(0.02,0);
\draw [dashed] (0.02,1.4)--(0.02,0);
\draw [<-]  (0.7,0.3)--(1.3,0.9);
\node at (0.9,1.5) {$p_5$};
%\node at (2.1,1.7) {$N,C$};
\draw [<-]  (0.9,-0.4)--(1.5,-1);
\node at (1.1,-1.6) {$p_6$};
%\node at (2.3,-1.7) {$R,D$};
\draw [<-] (-0.9,0.3)--(-1.6,1);
\node at (-1,1.5) {$p_1$};
%\node at (-2.2,1.7) {$M,B$};
\draw [<-] (-0.8,-0.4)--(-1.5,-1);
\node at (-0.9,-1.5) {$p_2$};
%\node at (-2.1,-1.7) {$S,E$};
%\node at (2.3,-1.7) {$R,D$};
\draw [<-] (-0.2,0.5)--(-0.2,1.2);
\node at (0,1.75) {$p_3$};
%\node at (-2.2,1.7) {$M,B$}
%\node at (2.3,-1.7) {$R,D$};
\draw [<-] (0.25,-0.5)--(0.25,-1.2);
\node at (0,-1.75) {$p_4$};
%\node at (-2.2,1.7) {$M,B$}
\node at (4.2,0) {$= \hskip .1 cm \displaystyle  i \frac{ c_{3,0}^{(6)} }{\L^2} \, ,$};
\end{tikzpicture}
\end{center}
\end{itemize}
We are about to renormalize \eq{L6xmx} using these rules. Clearly, this is in the spirit of \eq{LHDO} rather than in the 
spirit of the operator insertions. It is a more familiar process, which is however numb to the decomposition of
the $\b$-functions in terms of the various anomalous-dimension contributions that comprise it. At the end though,
we will obtain the same $\b$-functions from both methods.

The presence of the O-ghost is manifest through the scalar propagator which now has two poles. 
The extra pole can be exposed if we rewrite the scalar propagator as
\be\label{G2pm}
G^{(2)}_0(x_1,x_2) = \frac{i}{p^2 - m_0^2 + c_{2,0}^{(6)} \frac{p^4} {\L^2} } = \frac{\L^2}{ c_{2,0}^{(6)} }  \frac{i}{ ( p^2 - m^2_{+,0})(p^2 - m^2_{-,0} ) } 
\ee
and breaking the denominator as
\bea\label{G2pm2}
G^{(2)}_0(x_1,x_2) &=& \frac{1}{\sqrt{ 1 + 4 m^2_0 \frac{ c_{2,0}^{(6)} } {\L^2} }} \left( \frac{i}{ ( p^2 - m^2_{+,0}) } - \frac{i}{ (p^2 - m^2_{-,0} ) } \right) \nonumber\\
&=& \frac{i }{ b_0 ( p^2 - m^2_{+,0}) } - \frac{i }{ b_0 (p^2 - m^2_{-,0} ) }\, ,
\eea
where 
\be\label{bi0}
b_0 = \sqrt{ 1 + 4 m^2_0  \frac{ c_{2,0}^{(6)} } {\L^2} } \, .
\ee
This shows that the propagator describes two scalar fields with masses
\be\label{mplmin}
m_{\pm,0}^2 = \frac{\L^2}{2 c_{2,0}^{(6)} } \left( -1 \pm b_0  \right) \, ,
\ee
where $m_{+,0}^2$, $m_{-,0}^2$ correspond to the bare masses of $\phi$ and of the O-ghost respectively.

In order to arrive at a ghost-free and finite theory in this basis we should first deal with the extra pole and its possible connection to the R-ghosts.
Our expectation is that the unphysical pole of the O-ghost somehow cancels due to the presence of the R-ghosts and this should happen order by order in the loop expansion.
To fulfill this expectation we need the 1-loop part of the action, which includes  
the quantum corrections and the renormalized version of \eq{L6xmx}. Let us denote 
the 1-loop corrections to the 2-, 4- and 6-point functions of $\phi$ as ${\cal M}_{2,\phi}$, ${\cal B}_{4,\phi}$ and ${\cal B}_{6,\phi}$ respectively. 
The correction of $\chi$'s propagator is ${\cal M}_{2,\chi}$ while the $\phi$-$\chi$ vertex is corrected by ${\cal B}_{4,\chi}$. 
For the finite parts of the diagrams we use the notation $\left[ \, \,  \right]_{\rm f}$.
Relating bare and renormalized quantities in the usual way, \eq{L6xmx} gives in momentum space
\be\label{LRLC}
{\cal L}  =  {\cal L}_{R} + {\cal L}_{\rm ct}\, ,
\ee
where
\bea
{\cal L}_{R}  &=& \frac{p^2}{2}  \phi^2 - \frac{ m^2 }{2} \phi^2 - \frac{\l }{4!} \phi^4 - \frac{ c_1^{(6)} p^2 }{4! \L^2} \phi^4 + 
\frac{c_2^{(6)} p^4 }{2 \L^2}  \phi^2 + \frac{ c_3^{(6)} }{6! \L^2}  \phi^6  \nonumber\\
&&+ \frac{p^2}{2} \bar \chi  \chi + \frac{m_{\chi}^2 }{2}  \bar \chi \chi - \frac{ \l_\chi }{4} \, \bar \chi \chi \phi^2   \, ,\nonumber
\eea
featuring the renormalized field and couplings and 
\bea
{\cal L}_{\rm ct}  &=& \frac{ \d \phi \, p^2 }{2} \phi^2 - \frac{ \d m + m^2 \d \phi }{2} \phi^2 -  \frac{ \d \l + 2 \l \d \phi }{4!} \phi^4 - 
\frac{ ( \d c_1^{(6)} + 2 c_1^{(6)} \d \phi ) \, p^2 }{4! \L^2} \phi^4 + \frac{ ( \d c_2^{(6)} + c_2^{(6)} \d \phi ) \, p^4 }{2 \L^2}  \phi^2 \nonumber\\
&& + \frac{ \d c_3^{(6)} + 3 c_3^{(6)} \d \phi }{6! \L^2}  \phi^6 + \frac{ \d \chi \,  p^2 }{2} \bar \chi  \chi + \frac{ \d m_{\chi} 
+ m_{\chi}^2  \d \chi  }{2}  \bar \chi \chi - \frac{ \d \l_\chi + \l_\chi ( \d \chi + \d \phi) }{4} \, \bar \chi \chi \phi^2   \nonumber\\
&&+ \frac{ {\cal M}_{2, \phi} }{2} \phi^2 + \frac{ {\cal B}_{4,\phi} }{4!} \phi^4 + \frac{ {\cal B}_{6,\phi} }{6!} \phi^6 + 
\frac{ {\cal M}_{2, \chi} }{2} \bar \chi \chi + \frac{ {\cal B}_{4, \chi} }{4} \bar \chi \chi \phi^2 \nonumber
\eea
which contains the counter-terms and the 1-loop diagrams.
The next step is to use the above expressions to choose the renormalization conditions which simultaneously 
make vertices and propagators finite and leave $\phi$ with a single pole. 
Start with the 2-point Green's function. For the scalar field the renormalized, double pole, propagator is
\be
G^{(2)}(x_1,x_2) = \frac{i}{A p^4 + B p^2 + C } \nonumber
\ee
with 
\bea
A &=& \frac{c_{2}^{(6)} + \d c_{2}^{(6)} + c_{2}^{(6)} \d \phi } {\L^2}  \nonumber\\
B &=& 1 + \d \phi   \nonumber\\
C &=& {\cal M}_{2, \phi} - m^2 - \d m - m^2 \d \phi   \nonumber
\eea
and the R-ghost propagator is
\be
G^{(2)}_{{\rm R}g}(x_1,x_2) = \frac{i}{p^2 + m_\chi^2 + \d m_\chi + \d \chi (p^2 + m_\chi^2) + {\cal M}_{2, \chi} } \, .
\ee
Following steps analogous to \eq{G2pm}, we break $G^{(2)}$ in two parts to obtain the renormalized version of \eq{G2pm2} and we get
\be\label{G2re}
G^{(2)}(x_1,x_2) = \frac{i }{ A [ M_{+} - M_{-} ] ( p^2 - M_{+}) } - \frac{i }{ A [ M_{+} - M_{-} ] (p^2 - M_{-} ) }
\ee
with 
\be
M_{\pm} = \frac{B}{2 A} \left[  -1 \pm \sqrt{1 - 4 \frac{A C}{B^2}}  \right] \, .
\ee
The pole-cancelation condition that we want to impose amounts to forcing $G^{(2)}_{{\rm Rg}}$ to cancel the wrong-sign term in \eq{G2re}.
In other words we need
\be
A [ M_{+} - M_{-} ] (p^2 - M_{-} ) = p^2 + m_\chi^2 + \d m_\chi + \d \chi (p^2 + m_\chi^2) + {\cal M}_{2, \chi} \nonumber\, .
\ee
This contains two pole-cancellation conditions. The first is
\bea
p^2 &=& A [ M_{+} - M_{-} ]  p^2 \nonumber\\
&=& b_0 \, p^2 + \hbar \left[ b_0 \frac{2 m^2 ( \d c_2^{(6)} + 2 c_2^{(6)} \d \phi ) +  c_2^{(6)}  
( -2 {\cal M}_{2, \phi} + 2 \d m + \frac{\L^2 }{c_2^{(6)} } \d \phi  ) }{c_2^{(6)} ( 4 m^2 + \frac{\L^2}{c_2^{(6)} } )  } \right] p^2 \, ,\nonumber
\eea
which, comparing the two sides order by order in $\hbar$, gives
\bea
b_0 &=& 1 \nonumber\\
2 m^2 ( \d c_2^{(6)} + 2 c_2^{(6)} \d \phi ) +  c_2^{(6)}  ( -2 {\cal M}_{2, \phi} + 2 \d m + \frac{\L^2 }{c_2^{(6)} } \d \phi  ) &=& 0 \, . \nonumber
\eea
Checking back to \eq{bi0} it is easy to see that the first constraint is fulfilled when $c_2^{(6)} = 0$ or when $m^2 = 0$. 
The former is rejected since it gets us to a basis where there are no O-ghosts from the beginning and the 
pole cancelation is meaningless. Therefore, we have to choose the $m^2=0$ solution. Substituting this in the second constraint we have
\be\label{syn1}
2 \frac{c_2^{(6)}}{\L^2} \left(- {\cal M}_{2, \phi} +  \d m\right) + \d \phi = 0 \, .
\ee
The second pole-cancellation condition is
\be
- A [ M_{+} - M_{-} ]  M_{-} = m_\chi^2 + \d m_\chi + \d \chi (p^2 + m_\chi^2) + {\cal M}_{2, \chi} \, , \nonumber
\ee
which, keeping terms up to ${\cal O}(\hbar)$ and using the constraints from the first condition, becomes 
\be
\frac{\L^2 }{c_2^{(6)} } + \hbar \left[ 3 ( \d m - {\cal M}_{2, \phi} ) -  \frac{\L^2 }{c_2^{(6)} } \left( \frac{\d c_2^{(6)} }{c_2^{(6)}} \right) +  
\frac{\L^2 }{c_2^{(6)} } \d \phi  \right]  = m_\chi^2 + \hbar \left( \d m_\chi + \d \chi ( p^2 + m_\chi^2 ) + {\cal M}_{2, \chi}  \right) \, . \nonumber
\ee
At ${\cal O}(\hbar^0)$ the constraint 
\be
m_\chi^2 =  \frac{\L^2 }{c_2^{(6)}}\, 
\ee
is obtained, a relation that connects the O- and R-ghosts and defines the mass of the latter in terms of the coupling of the former.
It also implies the useful identity 
\be
\frac{\d m_\chi }{ m_\chi^2} = -  \frac{\d c_2^{(6)} }{ c_2^{(6)}}\, .
\ee
At ${\cal O}(\hbar)$ level we obtain
\bea\label{syn2}
3 (  \d m - {\cal M}_{2, \phi} ) - m_\chi^2  \frac{\d c_2^{(6)} }{c_2^{(6)}}  +  m_\chi^2 \d \phi &=& \d m_\chi + \d \chi ( p^2 + m_\chi^2 ) + {\cal M}_{2, \chi}  \Rightarrow \nonumber\\
3 (  \d m - {\cal M}_{2, \phi} )  +  m_\chi^2 \d \phi &=& \d \chi ( p^2 + m_\chi^2 ) + {\cal M}_{2, \chi} \, .
\eea
Notice that even though we are forced to set the renormalized mass to zero, there is a mass counterterm $\d m$ left over, whose purpose is to absorb $ {\cal M}_{2, \phi}$.
\eq{syn1} resembles the way that the single-pole propagator of a pure scalar theory gets renormalized after resummation. 
This behaviour is inherited by the G-basis propagator when
\be\label{dmdfi}
\d m = {\cal M}_{2, \phi} \, \, \, {\rm and} \, \, \, \d \phi = 0 \, .
\ee
Then \eq{syn2} fixes the counter-term $\d \chi$:
\be\label{dxi}
\d \chi = - \frac{ {\cal M}_{2, \chi} }{p^2 + m_\chi^2} \, .
\ee
It is now easy to check that using all the constraints, the renormalized $\phi$-propagator takes the form
\be\label{spsp}
G^{(2)}(x_1,x_2) = \frac{i}{p^2} - \frac{i}{p^2 + m_\chi^2 } \, ,
\ee
with the second term the same as the R-ghost propagator, with the opposite sign.
Bringing the $\phi$ and $\chi$ propagators in this relative form is a necessary condition but its sufficiency is not yet obvious. 

The only ingredient left at the 2-point function level is the computation of $ {\cal M}_{2, \phi} $ and $ {\cal M}_{2, \chi}$.
Starting with the former, we consider contributions from two vertices.
The first comes from the 4-point vertex given by 
%-------------------------------------
\vskip .5cm
\begin{center}
\begin{tikzpicture}[scale=0.7]
\draw [dashed] (0,0)--(1.8,0);
\draw [dashed] (0,0.9) circle [radius=0.9];
\draw [dashed] (-1.8,0)--(0,0);
\draw [->]  (-1.9,0.3)--(-1.2,0.3);
\node at (-1.7,-0.4) {$q$};
\node at (1.7,-0.4) {$q$};
\node at (0,1.4) {$k$};
\draw [<-]  (1.9,0.3)--(1.2,0.3);
%\node at (0.9,0) {$ > $};
%\node at (-0.9,0) {$ > $};
\node at (4,0) {$=  \hskip .1 cm i {\cal M}^1_{2,\phi}$};
\end{tikzpicture}
\end{center}
%-------------------------------------
whose evaluation gives
\bea\label{M1fx}
i {\cal M}^1_{2,\phi} &=& - i  \left(  \l +  c_{1}^{(6)} \frac{q^2} {\L^2}  \right) S_{{\cal M}^1} \int \frac{d^4 k }{(2 \pi)^4} \left[ \frac{i }{ k^2} - \frac{i }{ k^2 + m_{\chi}^2 } \right] \Rightarrow \nonumber\\
{\cal M}^1_{2,\phi} &=& \frac{1}{2} \left(  \l +  c_{1}^{(6)} \frac{q^2} {\L^2}  \right) \int \frac{d^4 k }{(2 \pi)^4 i } \left[ \frac{1 }{ k^2} - \frac{1 }{ k^2 + m_{\chi}^2 } \right] \Rightarrow \nonumber\\
{\cal M}^1_{2,\phi} &=&  \frac{1}{2} \left(  \l +  c_{1}^{(6)} \frac{q^2} {\L^2}  \right) \left[ A_0 (0) - A_0(- m_{\chi}^2 ) \right] \, ,
\eea
with the symmetry factor $S_{{\cal M}^1} = \frac{1}{2}$ and \eq{spsp} used for the scalar propagator.
The second two-point contribution comes from the $\phi$-$\chi$ vertex:
%-------------------------------------
\vskip .5cm
\begin{center}
\begin{tikzpicture}[scale=0.7]
\draw [dashed] (0,0)--(1.8,0);
\draw [] (0,0.9) circle [radius=0.9];
\draw [dashed] (-1.8,0)--(0,0);
\draw [->]  (-1.9,0.3)--(-1.2,0.3);
\node at (-1.7,-0.4) {$q$};
\node at (1.7,-0.4) {$q$};
\node at (0,1.3) {$k$};
\draw [<-]  (1.9,0.3)--(1.2,0.3);
%\node at (0.9,0) {$ > $};
\draw [<-,very thick]  (-0.1,1.8)--(0,1.8);
%\node at (-0.9,0) {$ > $};
\node at (4,0) {$=  \hskip .1 cm i {\cal M}^2_{2,\phi}$};
\end{tikzpicture}
\end{center}
%------------------------------------- 
It evaluates to
\bea\label{M2fx}
i {\cal M}^2_{2,\phi} &=& (-1) (- i) \frac{ \l_{\chi}}{2} S_{{\cal M}^2} \int \frac{d^4 k }{(2 \pi)^4} \frac{i}{ k^2 + m_{\chi}^2 }\Rightarrow \nonumber\\
{\cal M}^2_{2,\phi} &=& - \frac{ \l_{\chi}}{2} \int \frac{d^4 k }{(2 \pi)^4 i } \frac{1}{ k^2 + m_{\chi}^2 } \equiv  - \frac{ \l_{\chi}}{2}  \, A_0( - m_{\chi}^2 ) \, , 
\eea
with symmetry factor $S_{{\cal M}^2} =1$. Notice the extra minus sign in the above due to the anti-commuting ghost which runs in the loop.
Adding \eq{M1fx} and \eq{M2fx} we get the complete 1-loop correction to the scalar two-point function
\be\label{Mfx}
{\cal M}_{2,\phi} =   \frac{1}{2} \left(  \l +  c_{1}^{(6)} \frac{q^2} {\L^2}  \right) \left[ A_0 (0) - A_0(- m_{\chi}^2 ) \right]  - \frac{ \l_{\chi}}{2}  \, A_0( - m_{\chi}^2 ) 
\ee
or, in DR,
\be\label{Mfx2}
{\cal M}_{2,\phi} =   \frac{ \l m_{\chi}^2}{16 \pi^2 \ve} + \frac{ c_{1}^{(6)} }{ c_{2}^{(6)} } \frac{q^2} {16 \pi^2 \ve}  
+ \frac{ \l_{\chi} m_{\chi}^2 }{{16 \pi^2 \ve }}  + \left[ {\cal M}_{2,\phi}  \right]_{\rm f} \, .
\ee
We complete the 2-point function analysis with the correction to R-ghost's propagator. 
In this case only one digram contributes, coming from the $\phi$-$\chi$ vertex:
%-------------------------------------
\vskip .5cm
\begin{center}
\begin{tikzpicture}[scale=0.7]
\draw [] (0,0)--(1.8,0);
\draw [dashed] (0,0.9) circle [radius=0.9];
\draw [] (-1.8,0)--(0,0);
\draw [->]  (-1.9,0.3)--(-1.2,0.3);
\node at (-1.7,-0.4) {$q$};
\node at (1.7,-0.4) {$q$};
\node at (0,1.4) {$k$};
\draw [<-]  (1.9,0.3)--(1.2,0.3);
%\node at (0.9,0) {$ > $};
\draw [->, very thick]  (-0.9,0)--(-0.8,0);
\draw [->, very thick]  (0.9,0)--(1,0);
%\node at (-0.9,0) {$ > $};
\node at (4,0) {$=  \hskip .1 cm i {\cal M}_{2,\chi}$};
\end{tikzpicture}
\end{center}
%-------------------------------------
Its explicit form is given by
\bea\label{Mx1}
i {\cal M}_{2,\chi} &=& (- i) \frac{ \l_{\chi}}{2} S_{{\cal M}_\chi} \int \frac{d^4 k }{(2 \pi)^4 }  \left[ \frac{i }{ k^2} - \frac{i }{ k^2 + m_{\chi}^2 } \right]  \Rightarrow \nonumber\\
{\cal M}_{2,\chi} &=& \frac{ \l_{\chi}}{2}  \int \frac{d^4 k }{(2 \pi)^4 i}  \left[ \frac{1 }{ k^2} - \frac{1 }{ k^2 + m_{\chi}^2 } \right] \Rightarrow \nonumber\\
{\cal M}_{2,\chi} &=&  \frac{ \l_{\chi}}{2}  \left[ A_0 (0) - A_0(- m_{\chi}^2 ) \right]  \, ,
\eea
with symmetry factor $S_{{\cal M}_\chi} =1$. In DR this is
\be\label{Mx2}
{\cal M}_{2,\chi}  = \frac{ \l_{\chi} m_{\chi}^2 }{{16 \pi^2 \ve }} + \left[ {\cal M}_{2,\chi} \right]_{\rm f} \, .
\ee
Using these results, \eq{dmdfi} and \eq{dxi} become
\bea\label{dmfixi}
\d m &=&  \frac{ \l m_{\chi}^2}{16 \pi^2 \ve} + \frac{ c_{1}^{(6)} }{ c_{2}^{(6)} } \frac{q^2} {16 \pi^2 \ve}  
+ \frac{ \l_{\chi} m_{\chi}^2 }{{16 \pi^2 \ve }}  + \left[ {\cal M}_{2,\phi}  \right]_{\rm f}   \nonumber\\
\d \chi &=& \frac{-1}{p^2 + m_{\chi}^2} \left[  \frac{ \l_{\chi} m_{\chi}^2 }{{16 \pi^2 \ve }} + \left[ {\cal M}_{2,\chi} \right]_{\rm f}  \right]   \nonumber\\
\d \phi &=&  0 \, .
\eea
Up to this point we have managed to secure a scalar-field whose extra propagator component is of the opposite sign with respect to the propagator of the R-ghost 
and to renormalize these propagators.
To render the theory finite, renormalization conditions associated with the 4- and 6-point functions are also needed.

Starting with the $\phi^4$ vertex, ${\cal L}_{\rm ct}$ indicates that 
\bea\label{dlB4}
\d \l + \d c_1^{(6)} \frac{p^2}{\L^2} + 2  ( \l + c_1^{(6)} \frac{p^2}{\L^2} ) \d \phi  &=& {\cal B}_{4,\phi} \Rightarrow \nonumber\\
\d \l + \d c_1^{(6)} \frac{p^2}{\L^2} &=& {\cal B}_{4,\phi} \, ,
\eea
where we have used \eq{dmdfi}.
The diagrammatic set of ${\cal B}_{4,\phi}$ consists of three 1-loop box diagrams. 
The first contribution is a pure scalar box with pinched external legs, i.e. a Candy. It has three channels given by \eq{3channels}. 
The $s$-channel, denoted by ${\cal B}^{1,s}_{4,\phi}$
%-------------------------------------
\vskip .5cm
\begin{center}
\begin{tikzpicture}[scale=0.7]
\draw [dashed] (0.9,0)--(2.5,1.5);
\draw [dashed] (0.9,0)--(2.5,-1.5);
\draw [dashed] (-2.5,1.5)--(-0.9,0);
\draw [dashed] (-2.5,-1.5)--(-0.9,0);
\node at (2.7,1.1) {$p_3$};
\node at (2.7,-1.1) {$p_4$};
\node at (-2.7,1.2) {$p_1$};
\node at (-2.7,-1) {$p_2$};
\draw [<- ] (-1.65,0.35)--(-2.3,0.95);
\draw [<- ] (-1.65,-0.35)--(-2.3,-0.95);
\draw [<-] (1.65,0.35)--(2.3,0.95);
\draw [<-] (1.65,-0.35)--(2.3,-0.95);
\draw [dashed] (0,0) circle [radius=0.9];
\node at (0,1.4) {$k+q$};
\node at (0,-1.4) {$k$};
\node at (5,0) {$=\, \,i {\cal B}^{1,s}_{4,\phi} $  };
\end{tikzpicture}
\end{center}
%-------------------------------------
is evaluated as 
\bea\label{B1fxs}
i {\cal B}^{1,s}_{4,\phi} &=& (-i) (-i) \left(  \l + c_{1}^{(6)} \frac{q^2} {\L^2}  \right)^2  S_{{\cal B}^1_{\phi}}  
\int \frac{d^4 k }{(2 \pi)^4} \left[ \frac{i }{ k^2} - \frac{i }{ k^2 + m_{\chi}^2 } \right] \left[ \frac{i }{ (k+q)^2} - \frac{i }{ (k+q)^2 + m_{\chi}^2 } \right] \Rightarrow \nonumber\\
{\cal B}^{1,s}_{4,\phi} &=& \frac{1}{2}  \left(  \l + c_{1}^{(6)} \frac{q^2} {\L^2}  \right)^2 \int \frac{d^4 k }{(2 \pi)^4 i} 
\left[ \frac{1 }{ k^2} - \frac{1 }{ k^2 + m_{\chi}^2 } \right] \left[ \frac{1 }{ (k+q)^2} - \frac{1 }{ (k+q)^2 + m_{\chi}^2 } \right]   \Rightarrow \nonumber\\
{\cal B}^{1,s}_{4,\phi} &=&  \frac{1}{2}  \left(  \l + c_{1}^{(6)} \frac{q^2} {\L^2}  \right)^2  
\left[ B_0(q^2, 0, 0 ) - B_0( q^2 , 0, - m_{\chi}^2 ) - B_0(q^2 , -m_{\chi}^2 , 0 ) + B_0(q^2, - m_{\chi}^2, - m_{\chi}^2 )   \right] \, ,\nonumber
\eea
where the symmetry factor is $S_{{\cal B}^1_{\phi}} = \frac{1}{2}$ and $p_1 + p_2 + p_3 + p_4 = 0$.
Including the other two channels, we obtain
\be\label{B1fx}
{\cal B}^1_{4,\phi} = {\cal B}^{1,s}_{4,\phi} + {\cal B}^{1,t}_{4,\phi} + {\cal B}^{1,u}_{4,\phi} \equiv \sum_{q^2 = s,t,u}  \frac{1}{2}  \left(  \l + c_{1}^{(6)} \frac{q^2} {\L^2}  \right)^2 F(q^2, m_{\chi}^2) \, ,
\ee
with 
\be
F(q^2, m_{\chi}^2) = B_0(q^2, 0, 0 ) - B_0( q^2 , 0, - m_{\chi}^2 ) - B_0(q^2 , -m_{\chi}^2 , 0 ) + B_0(q^2, - m_{\chi}^2, - m_{\chi}^2 ) \, . \nonumber
\ee 
The next contribution to the four-point vertex comes from the $\phi$-$\chi$ interaction. The diagram is another Candy whose $s$-channel is
%-------------------------------------
\vskip .5cm
\begin{center}
\begin{tikzpicture}[scale=0.7]
\draw [dashed] (0.9,0)--(2.5,1.5);
\draw [dashed] (0.9,0)--(2.5,-1.5);
\draw [dashed] (-2.5,1.5)--(-0.9,0);
\draw [dashed] (-2.5,-1.5)--(-0.9,0);
\node at (2.7,1.1) {$p_3$};
\node at (2.7,-1.1) {$p_4$};
\node at (-2.7,1.2) {$p_1$};
\node at (-2.7,-1) {$p_2$};
\draw [<- ] (-1.65,0.35)--(-2.3,0.95);
\draw [<- ] (-1.65,-0.35)--(-2.3,-0.95);
\draw [<-] (1.65,0.35)--(2.3,0.95);
\draw [<-] (1.65,-0.35)--(2.3,-0.95);
%\node at (0.9,0) {$ > $};
\draw [->, very thick]  (0,0.9)--(0.1,0.9);
\draw [<-, very thick]  (-0.1,-0.9)--(0,-0.9);
%\node at (-0.9,0) {$ > $};
\draw [] (0,0) circle [radius=0.9];
\node at (0,1.4) {$k+q$};
\node at (0,-1.4) {$k$};
\node at (5,0) {$=\, \,i {\cal B}^{2,s}_{4,\phi} $  };
\end{tikzpicture}
\end{center}
%-------------------------------------
evaluating to
\bea\label{B2fxs}
i {\cal B}^{2,s}_{4,\phi} &=& (-1) (-i ) (-i) \frac{ \l_{\chi}^2}{4} S_{{\cal B}^2_{\phi}}  \int \frac{d^4 k }{(2 \pi)^4} \frac{i}{ k^2 + m_{\chi}^2 } \frac{i}{ (k + q)^2 + m_{\chi}^2 } \Rightarrow \nonumber\\
{\cal B}^{2,s}_{4,\phi} &=& - \frac{ \l_{\chi}^2}{4} \int \frac{d^4 k }{(2 \pi)^4 i}  \frac{1 }{ [k^2 + m_{\chi}^2 ] [(k+q)^2 + m_{\chi}^2 ] } \equiv - \frac{ \l_{\chi}^2}{4} B_0(q^2, - m_{\chi}^2 ,- m_{\chi}^2 ) \, , 
\eea
with symmetry factor $S_{{\cal B}^2_{\phi}} = 1$ and momentum conservation $p_1 + p_2 + p_3 + p_4 = 0$.
The complete contribution of this diagram is
\be\label{B2fx}
{\cal B}^2_{4,\phi} = - \frac{ \l_{\chi}^2}{4} \sum_{q^2 = s,t,u} B_0(q^2, - m_{\chi}^2 ,- m_{\chi}^2 )  \, .
\ee
The third, and last, diagram which corrects the $\phi^4$-vertex comes from the six-point function.
The $s$-channel diagram is 
%-------------------------------------
\vskip .5cm
\begin{center}
\begin{tikzpicture}[scale=0.7]
\draw [dashed] (0,0)--(1.8,0);
\draw [dashed] (0,0.9) circle [radius=0.9];
\draw [dashed] (-1.8,0)--(0,0);
\draw [->]  (-1.9,0.3)--(-1.2,0.3);
\node at (-1.7,-0.4) {$p_1$};
\node at (1.7,-0.4) {$p_3$};
\node at (0,1.4) {$k$};
\draw [->]  (1.9,0.3)--(1.2,0.3);
%\node at (0.9,0) {$ > $};
\draw [dashed] (-1,-1)--(0,0);
\draw [dashed] (1,-1)--(0,0);
%\node at (-0.9,0) {$ > $};
\draw [->]  (-0.7,-1)--(-0.2,-0.5);
\draw [<-]  (0.2,-0.5)--(0.7,-1);
%\node at (-0.9,0) {$ > $};
\node at (-0.8,-1.4) {$p_2$};
\node at (0.8,-1.4) {$p_4$};
\node at (4,0) {$=  \hskip .1 cm i {\cal B}^{3,s}_{4,\phi} $};
\end{tikzpicture}
\end{center}
%-------------------------------------
This contribution is the same as that of \eq{M1fx} except from an extra vertex factor.
We can then directly write
\be\label{B3fxs}
{\cal B}^{3,s}_{4,\phi} = - \frac{ c_{3}^{(6)} }{ 2 \L^2} \left[ A_0 (0) - A_0(- m_{\chi}^2 ) \right] \, . 
\ee
This diagram gives an $A_0$-integral which is channel-independent, so its total contribution is just three times ${\cal B}^{3,s}_{4,\phi}$:
\be\label{B3fx}
{\cal B}^3_{4,\phi} = - \frac{ 3 c_{3}^{(6)} }{ 2 \L^2 }  \left[ A_0 (0) - A_0(- m_{\chi}^2 ) \right] \, .
\ee
Adding \eq{B1fx}, \eq{B2fx} and \eq{B3fx} we end up with the full 1-loop correction to the pure scalar 4-point vertex
\bea\label{Bfx}
{\cal B}_{4,\phi} &=& \sum_{q^2 = s,t,u}  \frac{1}{2}  \left(  \l + \frac{q^2} {m_{\chi}^2} \frac{c_{1}^{(6)}} {c_{2}^{(6)}}  \right)^2 
F(q^2, m_{\chi}^2) \, - \frac{ \l_{\chi}^2}{4} \sum_{q^2 = s,t,u} B_0(q^2, - m_{\chi}^2 ,- m_{\chi}^2 ) \nonumber\\
&-& \frac{ 3 c_{3}^{(6)} }{ 2 \L^2  }  \left[ A_0 (0) - A_0(- m_{\chi}^2 ) \right] \, .
\eea
Decompressing and calculating the integrals in DR, we obtain
\bea\label{Bfx2}
{\cal B}_{4,\phi} =  - \frac{3 \l^2}{16 \pi^2 \ve} 
- \frac{3 \l_\chi^2}{32 \pi^2 \ve} - \frac{ 3 c_{3}^{(6)} }{\L^2} \frac{ m_{\chi}^2 }{16\pi^2 \ve }  + \left[ {\cal B}_{4,\phi} \right]_{\rm f} \, ,
\eea
where we have used the identity $s+t+u=0$ and kept terms up to ${\cal O}(1/\L^2)$.

For the $\phi$-$\chi$ vertex, according to ${\cal L}_{\rm ct}$, we get the condition
\bea\label{dlxiBxi}
\d \l_\chi + \l_\chi ( \d \chi + \d \phi ) = {\cal B}_{4,\chi} \Rightarrow \nonumber\\
\d \l_\chi + \l_\chi \d \chi  = {\cal B}_{4,\chi} \, ,
\eea
using \eq{dmdfi}.
In this case the 1-loop correction comes only from one box diagram. It is again a Candy and its $s$-channel is  
%-------------------------------------
\vskip .5cm
\begin{center}
\begin{tikzpicture}[scale=0.7]
\draw [] (0.9,0)--(2.5,1.5);
\draw [] (0.9,0)--(2.5,-1.5);
\draw [dashed] (-2.5,1.5)--(-0.9,0);
\draw [dashed] (-2.5,-1.5)--(-0.9,0);
\node at (2.7,1.1) {$p_3$};
\node at (2.7,-1.1) {$p_4$};
\node at (-2.7,1.2) {$p_1$};
\node at (-2.7,-1) {$p_2$};
\draw [<- ] (-1.65,0.35)--(-2.3,0.95);
\draw [<- ] (-1.65,-0.35)--(-2.3,-0.95);
\draw [<-] (1.65,0.35)--(2.3,0.95);
\draw [<-] (1.65,-0.35)--(2.3,-0.95);
%\node at (0.9,0) {$ > $};
\draw [->, very thick]  (1.7,0.75)--(1.8,0.85);
\draw [<-, very thick]  (1.7,-0.75)--(1.8,-0.85);
%\node at (-0.9,0) {$ > $};
\draw [dashed] (0,0) circle [radius=0.9];
\node at (0,1.4) {$k+q$};
\node at (0,-1.4) {$k$};
\node at (5,0) {$=\, \,i {\cal B}^s_{4,\chi} $  };
\end{tikzpicture}
\end{center}
%------------------------------------- 
It is of the form \eq{B1fx} except from the vertex factor. Thus,
\be\label{Bxs}
{\cal B}^s_{4,\chi} =  \frac{\l_{\chi}}{4}  \left(  \l + c_{1}^{(6)} \frac{q^2} {\L^2}  \right)  F(q^2, m_{\chi}^2) \, .
\ee
Even though this diagram is a Candy with scalar fields in the loop, it has only two channels, so its total contribution is
\be\label{Bxi}
{\cal B}_{4,\chi} = \sum_{q^2 = s,t}  \frac{\l_{\chi}}{4}  \left(  \l + c_{1}^{(6)} \frac{q^2} {\L^2}  \right)  F(q^2, m_{\chi}^2) \, .
\ee
Computing in DR the integrals, we get
\be\label{Bchi2}
{\cal B}_{4,\chi} = - \frac{ \l_{\chi} \,\l }{16\pi^2 \ve} %- \l_{\chi} c_{1}^{(6)} \frac{m_{\chi}^2} {\L^2} 
+ \left[ {\cal B}_{4,\chi} \right]_{\rm f}  \, ,
\ee
where we have used that $s+t= 0$.

The renormalization condition that ensures the finiteness of the 6-point vertex is
\bea\label{dc3Bi6}
\d c_{3}^{(6)} + 3 c_{3}^{(6)} \d \phi = - \L^2 {\cal B}_{6,\phi} \Rightarrow \nonumber\\
\d c_{3}^{(6)}  = - \L^2 {\cal B}_{6,\phi} \, .
\eea
The related diagrammatic set includes ${\cal K}^1_{\phi^6} $, ${\cal K}^2_{\phi^6} $ and ${\cal B}_{\phi^6} $. 
The first and the second come from the $\phi^4$ and $\phi$-$\chi$ vertices respectively, while the third is a combination of the $\phi^4$ and $\phi^6$ vertices.
${\cal K}^1_{\phi^6} $ is a triangle diagram with only one channel. It is
%-------------------------------------
\vskip .5cm
\begin{center}
\begin{tikzpicture}[scale=0.7]
\draw [dashed] (-1,0) -- (1,1);
\draw [dashed] (-1,0) -- (1,-1);
\draw [dashed] (1,1) -- (1,-1);
%\draw [<-, very thick ] (-1.65,0.45)--(-1.85,0.65);
%\draw [->, very thick ] (-1.65,-0.45)--(-1.85,-0.65);
%%%%%%
\draw [dashed] (1,1) -- (2.5,1);%p3
\draw [dashed] (1,-1) -- (2.5,-1);%p4
%%%%%%
\draw [dashed] (1,1) -- (1,2.5);
\draw [dashed] (1,-1) -- (1,-2.5);
%%%%%%
\draw [dashed] (-2.5,-1.2)--(-1,0);
\draw [dashed] (-2.5,1.2)--(-1,0);
%%%%%%
%\draw [->, very thick] (1,-0.1) -- (1,0.1);
%\draw [<-, very thick] (1.7,1) -- (1.5,1);
\draw [<-] (-1.4,0.9)--(-2.1,1.4);
\node at (-1.2,1.5) {$p_1$};
%\node at (-2.2,1.7) {$M,B$};
\draw [<-] (-1.4,-0.9)--(-2.1,-1.4);
\node at (-1,-1.5) {$p_2$};
%\draw [<-, very thick] (1.5,-1) -- (1.7,-1);
\draw [<-] (1.3,1.3)--(2.1,1.3);
\node at (2.1,1.8) {$p_5$};
%%%%%%
\draw [<-] (1.3,-1.3)--(2.1,-1.3);
\node at (2.1,-1.8) {$p_6$};
%%%%%%
\draw [<-] (0.7,1.3)--(0.7,2.1);
\node at (0.5,2.3) {$p_3$};
%%%%%%
\draw [<-] (0.7,-1.3)--(0.7,-2.1);
\node at (0.5,-2.3) {$p_4$};
%%%%%%
\node at (0,-1) {$k$};
\node at (0,1) {$k'$};
\node at (1.5,0) {$k''$};
%%%%%%
\node at (4,0) {$=  \hskip .1 cm i {\cal K}^1_{\phi^6}$};
\end{tikzpicture}
\end{center}
%-------------------------------------
where $k' = k + p_1 + p_2 $, $k'' = k' +p_3 + p_5 $. In addition, momentum conservation demands that $p_1 + p_2 + p_3 + p_4 + p_5 + p_6 =0$.
Now the propagators in the loop come from \eq{spsp} so the calculation of this diagram involves
eight parts. It is useful to see how ${\cal K}^1_{\phi^6}$ scales as a function of the loop momentum. Each one of the contributing parts contributes as
${\cal K}^1_{\phi^6} \sim \int d^4 k  \frac{1}{k^6}$
which is just a finite correction, proportional to a $C_0$-integral. In other words, the above 1-loop Triangle is finite and we can write it as
\be\label{K1f6}
{\cal K}^1_{\phi^6} \equiv \left[ {\cal K}^1_{\phi^6} \right]_{\rm f} \, .
\ee
This is the diagram mentioned in the Introduction and it is in fact the only correction associated with a dimension-6 operator in the usual EFT, at 1-loop.
The second diagram in the GEFT is another Triangle diagram, originating from the interaction of the scalar field with the R-ghost. It is a contribution with only one channel, given by 
%-------------------------------------
\vskip .5cm
\begin{center}
\begin{tikzpicture}[scale=0.7]
\draw [] (-1,0) -- (1,1);
\draw [] (-1,0) -- (1,-1);
\draw [] (1,1) -- (1,-1);
%\draw [<-, very thick ] (-1.65,0.45)--(-1.85,0.65);
%\draw [->, very thick ] (-1.65,-0.45)--(-1.85,-0.65);
%%%%%%
\draw [dashed] (1,1) -- (2.5,1);%p3
\draw [dashed] (1,-1) -- (2.5,-1);%p4
%%%%%%
\draw [dashed] (1,1) -- (1,2.5);
\draw [dashed] (1,-1) -- (1,-2.5);
%%%%%%
\draw [dashed] (-2.5,-1.2)--(-1,0);
\draw [dashed] (-2.5,1.2)--(-1,0);
%%%%%%
%\draw [->, very thick] (1,-0.1) -- (1,0.1);
%\draw [<-, very thick] (1.7,1) -- (1.5,1);
\draw [<-] (-1.4,0.9)--(-2.1,1.4);
\node at (-1.2,1.5) {$p_1$};
%\node at (-2.2,1.7) {$M,B$};
\draw [<-] (-1.4,-0.9)--(-2.1,-1.4);
\node at (-1,-1.5) {$p_2$};
%\draw [<-, very thick] (1.5,-1) -- (1.7,-1);
\draw [<-] (1.3,1.3)--(2.1,1.3);
\node at (2.1,1.8) {$p_5$};
%%%%%%
\draw [<-] (1.3,-1.3)--(2.1,-1.3);
\node at (2.1,-1.8) {$p_6$};
%%%%%%
\draw [<-] (0.7,1.3)--(0.7,2.1);
\node at (0.5,2.3) {$p_3$};
%%%%%%
\draw [<-] (0.7,-1.3)--(0.7,-2.1);
\node at (0.5,-2.3) {$p_4$};
%%%%%%
%\node at (0.9,0) {$ > $};
\draw [<-, very thick]  (1,-0.1)--(1,0);
\draw [->, very thick]  (0,0.5)--(0.1,0.55);
\draw [<-, very thick]  (0,-0.5)--(0.1,-0.55);
%%%%%%
\node at (0,-1) {$k$};
\node at (0,1) {$k'$};
\node at (1.7,0) {$k''$};
%%%%%%
\node at (4,0) {$=  \hskip .1 cm i {\cal K}^2_{\phi^6}$};
\end{tikzpicture}
\end{center}
%-------------------------------------
evaluating to
\bea\label{K2f6}
i {\cal K}^2_{\phi^6} &=& (-1) (-i)^3 \frac{\l_{\chi}^3}{8} \, S_{{\cal K}^2_{\phi^6}} \int \frac{d^4 k }{(2 \pi)^4 }  \frac{i^3}{[ k^2 + m_{\chi}^2 ] [ (k')^2 + m_{\chi}^2 ] [ (k'')^2 + m_{\chi}^2 ] } \Rightarrow \nonumber\\ 
{\cal K}^2_{\phi^6} &=& - \frac{\l_{\chi}^3}{8} \int \frac{d^4 k }{(2 \pi)^4 i}  \frac{1}{[ k^2 + m_{\chi}^2 ] [ (k')^2 + m_{\chi}^2 ] [ (k'')^2 + m_{\chi}^2 ] } \equiv   - \frac{\l_{\chi}^3}{8} \, C_0(P_1,P_2, - m_{\chi}^2, - m_{\chi}^2, - m_{\chi}^2 ) \, ,  \nonumber
\eea 
with symmetry factor $S_{{\cal K}^2_{\phi^6}} = 1 $ and momentum conservation $p_1 + p_2 + p_3 + p_4 + p_5 + p_6 =0 $.\\
It also gives a finite $C_0$-integral which has no effect on the $\b$-function of the corresponding coupling. So,
\be\label{K2f6}
{\cal K}^2_{\phi^6} = \left[ {\cal K}^2_{\phi^6} \right]_{\rm f} \, .
\ee
This calculation ends with one more 1-loop diagram which corrects the $\phi^6$-vertex and has five channels,
labelled by $s_\s$ with $\s = 1,2,3,4,5$. The diagram itself is a $2 \to 4$ Candy whose first channel is
%-------------------------------------
\vskip .5cm
\begin{center}
\begin{tikzpicture}[scale=0.7]
\draw [dashed] (0.9,0)--(2.5,1.5);
\draw [dashed] (0.9,0)--(2.5,-1.5);
\draw [dashed] (-2.5,1.5)--(-0.9,0);
\draw [dashed] (-2.5,-1.5)--(-0.9,0);
%\draw [<-, very thick ] (-1.65,0.45)--(-1.85,0.65);
%\draw [<-, very thick ] (-1.65,-0.45)--(-1.85,-0.65);
\draw [dashed] (0.9,0)--(2.8,0.8);
\draw [dashed] (0.9,0)--(2.8,-0.8);
%\draw [<-, very thick ] (1.65,0.45)--(1.85,0.65);
%\draw [<-, very thick ] (1.65,-0.45)--(1.85,-0.65);
\node at (2.7,1.8) {$p_3$};
\node at (2.7,-1.8) {$p_4$};
\node at (-2.5,1.1) {$p_1$};
\node at (-2.5,-0.9) {$p_2$};
\node at (3.2,1) {$p_5$};
\node at (3.2,-1) {$p_6$};
%\draw [<-, very thick] (1.7,1) -- (1.5,1);
\draw [<-] (-1.4,0.8)--(-2.1,1.4);
%\node at (-1.2,1.5) {$p_1$};
\draw [<-] (-1.4,-0.8)--(-2.1,-1.4);
%\node at (-1,-1.5) {$p_2$};
\draw [<-] (1.4,0.9)--(2.1,1.5);
%\node at (-1.2,1.5) {$p_1$};
\draw [<-] (1.4,-0.9)--(2.1,-1.5);
%\draw [<-, very thick] (1.5,-1) -- (1.7,-1);
%\node at (-1.2,1.5) {$p_1$};
\draw [<-] (2,0.15)--(2.6,0.35);
%\draw [<-, very thick] (1.5,-1) -- (1.7,-1);
%\node at (-1.2,1.5) {$p_1$};
\draw [<-] (2,-0.15)--(2.6,-0.35);
%\draw [<-, very thick] (1.5,-1) -- (1.7,-1);
\draw [dashed] (0,0) circle [radius=0.9];
\node at (0,1.4) {$k+q$};
\node at (0,-1.4) {$k$};
\node at (5,0) {$=\, \,i {\cal B}^{s_1}_{\phi^6} $  };
\end{tikzpicture}
\end{center}
%-------------------------------------
It is equal to ${\cal B}^{1,s}_{4,\phi}$ barring the vertex factor. Therefore, it is equal to
\be
{\cal B}^{s_1}_{\phi^6} = - \frac{1}{2} \left(  \l + \frac{q^2} {m_{\chi}^2} \frac{c_{1}^{(6)}} {c_{2}^{(6)}}  \right) \frac{c_{3}^{(6)}}{\L^2} F(q^2, m_{\chi}^2)\, .
\ee
Adding all channels we get
\be\label{BFi6}
{\cal B}_{\phi^6} =  - \sum_{q^2 = s_1, \cdot \cdot \cdot, s_5}  \frac{1}{2}   \left(  \l + \frac{q^2} {m_{\chi}^2} \frac{c_{1}^{(6)}} {c_{2}^{(6)}}  \right) \frac{ c_{3}^{(6)} }{\L^2} F(q^2, m_{\chi}^2) \, .
\ee
Adding \eq{K1f6}, \eq{K2f6} and \eq{BFi6} in DR, we have 
\bea\label{Bfi6}
\L^2 {\cal B}_{6,\phi} &=& \frac{5 \l c_{3}^{(6)}  }{16\pi^2 \ve } + \L^2 \left[ {\cal K}^1_{\phi^6} \right]_{\rm f} + \L^2 \left[ {\cal K}^2_{\phi^6} \right]_{\rm f} + \L^2 \left[ {\cal B}_{\phi^6} \right]_{\rm f} \Rightarrow \nonumber\\ 
\L^2 {\cal B}_{6,\phi} &=&  \frac{5 \l c_{3}^{(6)}  }{16\pi^2 \ve } + \L^2 \left[ {\cal B}_{6,\phi} \right]_{\rm f} \, ,
\eea
where $ s_1 + s_2 + s_3 + s_4 + s_5 = 0$ and
\be
\left[ {\cal B}_{6,\phi} \right]_{\rm f} = \left[ {\cal K}^1_{\phi^6} \right]_{\rm f} + \left[ {\cal K}^2_{\phi^6} \right]_{\rm f} + \left[ {\cal B}_{\phi^6} \right]_{\rm f} \, . \nonumber
\ee
An important comment is in order. We calculated the 1-loop diagrams using the function
$F(q^2,m_\chi^2)$ that contains the scaleless version of the $B_0$-integral. To extract the counter-terms we 
use the fact that in DR this integral is zero. This seems to be in contrast with the operator insertion 
method where they also contributed to the renormalization but we took them to be non-zero. 
The reason is that here the loop integrals see momenta from IR to UV and this is why they give zero in DR.
In contrast the renormalized correlation function of a composite 
operator has extra UV divergences that are possible to extract by separating the UV from the IR part 
of the scaleless $B_0$-integrals. 
The $A_0$ integral on the other hand does not have this flexibility because its UV and IR limits are non-separable,
therefore it vanishes identically when scaleless.

Having completed the computation of the 1-loop corrections we can now determine the counter-terms of the vertices.
Substituting the four-point vertex in \eq{dlB4} we have
\be
\d \l = \frac{ 3 \l^2}{16 \pi^2 \ve} + \frac{ 3 c_{3}^{(6)} }{16 \pi^2 \ve}  \frac{ m_{\chi}^2 }{\L^2} + 
\left[ {\cal B}_{4,\phi} \right]_{\rm f}    \, \, \, {\rm and} \, \, \,  \frac{p^2}{\L^2} \d c_{1}^{(6)} =  
- \frac{6 \l^2}{16 \pi^2 \ve} - \frac{3 \l_\chi^2}{32 \pi^2 \ve} - \frac{ 6 c_{3}^{(6)} }{\L^2} \frac{ m_{\chi}^2 }{16\pi^2 \ve } 
\ee
where we have added $\pm 3 ( \l^2 + c_{3}^{(6)}  \frac{ m_{\chi}^2 }{\L^2}  ) /(16 \pi^2 \ve ) $, 
taking advantage of the arbitrariness of $\d c_{1}^{(6)}$.
The $\phi$-$\chi$ vertex, substituted in \eq{dlxiBxi}, gives the condition
\be
\d \l_\chi = - \frac{ \l_{\chi} \,\l }{16\pi^2 \ve} 
+ \left[ {\cal B}_{4,\chi} \right]_{\rm f} + \frac{\l_\chi}{p^2 + m_{\chi}^2} \left(  \frac{ \l_{\chi} m_{\chi}^2 }{{16 \pi^2 \ve }} + \left[ {\cal M}_{2,\chi} \right]_{\rm f}  \right) \, ,
\ee
and the 6-point vertex substituted in \eq{dc3Bi6} gives
\be
\d c_{3}^{(6)} = - \frac{ 5\l  c_{3}^{(6)} }{16 \pi^2 \ve} - \L^2 \left[ {\cal B}_{6,\phi} \right]_{\rm f} \, .
\ee
Notice here that the above renormalization conditions should be enough to render the theory finite. 
This means that $ \d c_{2}^{(6)} $, or equivalently $ \d m_\chi $, remains undetermined. Nevertheless it has a significant role which we will uncover soon.

Using \eq{dmdfi}, \eq{dxi}, \eq{dlB4}, \eq{dlxiBxi} and \eq{dc3Bi6} in \eq{LRLC}, the renormalized Lagrangian becomes
\bea\label{L4-6fi}
{\cal L} &=& \frac{1}{2} \left( p^2 + m_{\chi}^2  \right) \left( \frac{p^2}{m_{\chi}^2} \phi^2 + \bar \chi \, \chi  \right) 
- \frac{1}{4!} \left( \l + c_{1}^{(6)} \frac{p^2}{\L^2}  \right)  \phi^4 + \frac{c_{3}^{(6)} }{6! \L^2} \phi^6 \nonumber\\
&&+ \frac{ \d c_{2}^{(6)} }{2 c_{2}^{(6)}} \left( \frac{p^4}{m_{\chi}^2} \phi^2 - m_{\chi}^2 \, \bar \chi \, \chi  \right) \, . 
\eea
An observation from the above analysis is that the $\phi$-$\chi$ vertex did not contribute to the pole-cancelation conditions of the scalar propagator. 
Recall that the ghost-like pole did not disappear through cancellation between diagrams, rather only the kinetic part of the R-ghost was used. 
In other words even if $\chi$ was decoupled, we would have ended up with the above, single-pole, Lagrangian. 
This is why in ${\cal L}$, without loss of generality, we have set $\l_\chi = 0$. 
However, \eq{L4-6fi} still needs refinement since the pole cancellation and renormalization left $\d c_{2}^{(6)}$ undetermined in the Lagrangian. 
In fact, in the last term in \eq{L4-6fi} it appears as a Lagrange multiplier so we can integrate it out. By doing so we obtain
\be
\bar \chi \, \chi = \frac{p^4}{m_{\chi}^4} \phi^2 \, ,\nonumber
\ee
a condition that relates the R-ghost to $\phi$. Substituting back into \eq{L4-6fi} and defining the composite field 
\be
\Phi^2 = \frac{p^2}{m_{\chi}^2} \left( \frac{p^2}{m_{\chi}^2} + 1 \right)  \phi^2 \nonumber
\ee
the Lagrangian, in position-space, becomes
\be\label{LRF}
{\cal L}_\Phi = - \frac{1}{2} \Phi \left( \Box - m_{\chi}^2  \right) \Phi - \frac{\l'}{4!} \Phi^4 +  \frac{c_{3}^{(6)'} }{6! \L^2} \Phi^6
\ee
with 
\be
\l' = \frac{\l - c_{1}^{(6)} \frac{\Box}{\L^2} }{ \left[ \frac{\Box}{m_{\chi}^2} \left( \frac{\Box}{m_{\chi}^2} - 1 \right) \right]^2 } \, \, \, 
{\rm and} \, \, \, c_{3}^{(6)'} =  \frac{ c_{3}^{(6)} }{ \left[ \frac{\Box}{m_{\chi}^2} \left( \frac{\Box}{m_{\chi}^2} - 1 \right) \right]^3 }  \, . \nonumber
\ee
Finally, the counter-terms determine the 1-loop contributions to the $\b$-functions (with $\l_\chi=0$)
\bea\label{untbf}
\b^q_{m^2} &=&  \frac{ \l m_{\chi}^2}{16 \pi^2} + \frac{ c_{1}^{(6)} }{ c_{2}^{(6)} } \frac{q^2} {16 \pi^2 } \nonumber\\
\b^q_{\l} &=& \frac{ 3 \l^2}{16 \pi^2} + \frac{ 3 c_{3}^{(6)} }{16 \pi^2 }  \frac{ m_{\chi}^2 }{\L^2}\nonumber\\
\b^q_{ c_{3}^{(6)} } &=& - \frac{ 5\l  c_{3}^{(6)} }{16 \pi^2} \, .
\eea
A few additional comments are in order.
The comparison between ${\cal L}$ and ${\cal L}_{\Phi}$ shows that the original local theory 
with all the dimension-6 operators and R-ghosts present is equivalent to a non-local and ghost-free theory including only the polynomial dimension-6 operator.
Moving from ${\cal L}$ to ${\cal L}_{\Phi}$ the pole is shifted and from the viewpoint of the non-local field $\Phi$, is located 
at $- m_\chi^2$. 
On the other hand, even though the propagator of the original local field $\phi$ had a pole at $m^2=0$, there is 
a non-trivial counter-term $\d m$. This is reminiscent of the Coleman-Weinberg model \cite{Coleman}.
Note also that the original basis of \eq{L6xmx} becomes consistent
only after renormalization, having imposed both tree and loop level conditions. This can be considered as a hint of the quantum nature of the HDO.
Note finally that the equivalence of the two Lagrangians renders the operators $O_1^{(6)}$ and $O_{2}^{(6)}$ redundant since
\eq{LRF} contains only one dimension-6 operator, $O_{3}^{(6)}$. 

The only nuisance is the non-local nature of \eq{LRF} which makes the computation of RG flows difficult.
In the next section we will show that ${\cal L}_\Phi$ is equivalent to a local and ghost-free theory, connected to \eq{L6xmx} via the redefinition of \eq{f.red}.

%%%%%%%%%%%%%%%%%%%%%%%%%%%%%%%%%%%%%%%%%%%%%%%%%%%%%%%%%%%%%%%%%%%%%%%%%%%%%%
\subsection{The pure polynomial W-basis}\label{onlyphi}
%%%%%%%%%%%%%%%%%%%%%%%%%%%%%%%%%%%%%%%%%%%%%%%%%%%%%%%%%%%%%%%%%%%%%%%%%%%%%%

In the previous section we showed how the non-physical degrees of freedom present in the G-basis are cancelled
by the R-ghost Lagrangian, which was added though by hand. How can this structure be better motivated?
Suppose that we start with the local, ghost-free Lagrangian
\be\label{L6phi6}
\wt {\cal L}_0 = -\frac{1}{2} \phi_0 \Box \phi_0 - \frac{1}{2} \wt m_0^2 \phi_0^2
- \frac{\wt \l_0 }{4!} \phi_0^4 + \frac{\wt c_{3,0}^{(6)} }{6! \L^2} \phi_0^6 \, ,
\ee
that includes only one, non-derivative dimension-6 operator. This "tilded" basis is just our previous W-basis.
We have seen that the field redefinition of \eq{f.red} transforms us back to G-basis but without the R-ghosts.
Not to miss those one must be careful since after the field transformation a non-trivial Jacobean develops.
This can be made clear if we consider the path-integral in the absence of sources
\be
{\cal Z} [0] = \int {\cal D} \phi_0 \, e^{i \wt S[\phi_0]}\, , \hskip 1cm  \wt S[\phi_0] = \int d^4 x \, \wt {\cal L}_0 \, . \nonumber
\ee
Now since \eq{L6phi6} is ghost-free then the field redefinition\footnote
{In the presence of sources the term $J_0 \, \phi_0$ should also be properly transformed, otherwise the $S$-matrix can not be kept unchanged.
Also, one could reduce the operator basis by using the equations of motion and dropping operators that are proportional to them. 
This would give the basis of reduced $\b$-functions \cite{Einhorn}.
This can also reduce the derivative order of the HDO avoiding the O-ghost but it is not clear if it leads to an equivalent theory \cite{phi6quantum}.}
\be
\phi_0 \rightarrow \phi'_0 = \phi_0 + \frac{x}{\L^2} \Box \phi_0 + \frac{y}{\L^2} \phi_0^3 \nonumber
\ee
should move us to another ghost-free basis without changing anything, if performed consistently. This means that if in
\be\label{Z0phi1}
{\cal Z} [0] = \int {\cal D} \phi_0 \left| \frac{d  \phi'_0}{d  \phi_0}\right| e^{i \wt S[ \phi_0 + \frac{x}{\L^2} \Box \phi_0 + \frac{y}{\L^2} \phi_0^3]} 
\ee
the Jacobean of the transformation is non-trivial, it must be accounted for.
Now the new degrees of freedom $\phi_0'$ and $(\Box\phi_0)'$ are a function of two independent functions, the field $\phi_0$ and $\Box \phi_0$.
Clearly since everything (both the Lagrangian and the field redefinition) is valid up to total derivatives there is an arbitrariness involved in the choice of the independent field variables.
To choose a gauge in the equivalence class of the $\partial^{2n} \phi$ variables we demand that in the transformed basis a pole cancellation
mechanism should be at work. 
This amounts to a "diagonal" or "physical" gauge where $\phi_0'$ is a function only of $\phi_0$ and $(\Box \phi_0)'$ is a function only of $\Box \phi_0$.
In the diagonal gauge the Jacobean is a diagonal matrix given by 
\be
\frac{d  \phi'_0}{d  \phi_0} = \begin{bmatrix}
   1+ \frac{x}{\L^2} \Box + \frac{3 y}{\L^2} \phi_0^2 & 
  0 \\
                  0   &1+ \frac{x}{\L^2} \Box + \frac{3 y}{\L^2} \phi_0^2  \\
\end{bmatrix}  \nonumber
\ee
with determinant
\be
D(\phi_0) = \det \left[ \frac{d  \phi'_0}{d  \phi_0} \right] = \left( 1+ \frac{x}{\L^2} \Box + \frac{3 y}{\L^2} \phi_0^2 \right)^2 \equiv 1+ \frac{2 x}{\L^2} \Box + \frac{6 y}{\L^2} \phi_0^2\, ,
\ee
keeping terms to ${\cal O}(1/\L^2)$.
Now recall that an operator depending on commuting variables, like $D(\phi_0)$, can be rewritten as a path integral of the form 
\be
D(\phi_0) =  \int {\cal D} \bar \chi'_0 {\cal D} \chi'_0\, e^{- i \int d^4 x \bar \chi'_0 D(\phi_0)  \chi'_0} \, ,\nonumber
\ee
with $\bar \chi'_0, \chi'_0$ two real Grassmann variables.
Then the transformed path integral of \eq{Z0phi1} includes also the anti-commuting, "reparametrization" ghost-fields. It becomes
\be
{\cal Z} [0] = \int {\cal D} \phi_0 {\cal D} \bar \chi'_0 {\cal D} \chi'_0 \, e^{i \wt S[ \phi_0 
+ \frac{x}{\L^2} \Box \phi_0 + \frac{y}{\L^2} \phi_0^3] - i \int d^4 x \bar \chi'_0 D(\phi_0)  \chi'_0 } \, 
\ee
and the correct version of \eq{L6phi6} is
\be\label{L6}
\wt {\cal L}_0  = \wt {\cal L}_{{\rm red},0} + {\cal L}_{{\rm R}g,0}
\ee
with
\bea
\wt {\cal L}_{{\rm red},0} &=& -\frac{1}{2} \left[ 1 + 2 x\frac{ \wt m_0^2 }{\L^2} \right] \phi_0 \Box \phi_0  - 
\frac{1}{2} \wt m_0^2 \, \phi_0^2 - \left[\frac{\wt \l_0}{4!} + \frac{y \, \wt m_0^2 }{\L^2} \right]\phi_0^4 
+ \left[\frac{ - y - \frac{x}{6} \wt \l_0}{\L^2} \right]\phi_0^2 \Box \phi_0^2 \nonumber\\
&&+ \left[\frac{- x}{\L^2}  \right]\phi_0 \Box^2 \phi_0
+ \left[ \frac{ \wt c_{3,0}^{(6)} }{6! \L^2}   -  \frac{y \wt \l_0 }{6 \L^2}  \right] \phi_0^6 \, , \nonumber\\
 {\cal L}_{{\rm R}g,0} &=& - \frac{2 x}{\L^2} \, \bar \chi'_0 \Box \chi'_0 - \bar \chi'_0 \chi'_0 - \frac{6 y}{\L^2} \, \bar \chi'_0 \chi'_0 \phi_0^2 \, .
\eea
The kinetic term of $\phi_0$ and $\chi'_0$ can be made canonical using \eq{phired} and the redefinition 
$ (\bar \chi'_0, \chi'_0) \to \sqrt{ \frac{\L^2}{2 x} } ( \bar \chi_0, \chi_0 )$. Then 
\bea\label{L6c}
\wt {\cal L}_0 &=& -\frac{1}{2} \phi_0 \Box \phi_0 - \frac{ \wt m_0^2 }{2} \left[ 1 - 2 x \frac{ \wt m_0^2 }{\L^2}  \right]  
\phi_0^2 + \left[ - \frac{\wt \l_0}{4!} - \frac{y \wt m_0^2 }{\L^2} + \frac{x \wt \l_0\wt m_0^2 }{6\L^2} \right]\phi_0^4 \nonumber\\
&&+ \left[\frac{ - y - \frac{x}{6} \wt \l_0}{\L^2} \right]\phi_0^2 \Box \phi_0^2 
+ \left[\frac{ - x}{\L^2}  \right]\phi_0 \Box^2 \phi_0
+ \left[ \frac{ \wt c_{3,0}^{(6)} }{6! \L^2}  -  \frac{y \wt \l_0 }{6 \L^2} \right] \phi_0^6 \nonumber\\
&&- \bar \chi_0 \, \Box \chi_0 - \frac{\L^2}{2 x} \bar \chi_0 \chi_0 - \frac{6 y}{2 x} \left[  1 - 2 x \frac{\wt m_0^2 }{\L^2} \right] \, \bar \chi_0 \chi_0 \phi_0^2  \nonumber\\
&=& -\frac{1}{2} \phi_0 \Box \phi_0 - \frac{m_0^2}{2}  \phi_0^2 -  \frac{\l_0}{4!} \phi_0^4 + 
\frac{ c_{1,0}^{(6)} }{4! \L^2} \phi_0^2 \Box \phi_0^2 + \frac{ c_{2,0}^{(6)} }{2 \L^2}  \phi_0 \Box^2 \phi_0  \nonumber\\
&&+ \frac{ c_{3,0}^{(6)} }{6! \L^2}  \phi_0^6 - \bar \chi_0\,\Box \chi_0 + \frac{\L^2}{c_{2,0}^{(6)}} \bar \chi_0 \chi_0 - \frac{ \l_{\chi,0} }{2} \, \bar \chi_0 \chi_0 \phi_0^2 \, ,
\eea
where in the last expression we made the identifications
\bea\label{allcou}
m_0^2 &=& \wt m_0^2 \left[ 1 - 2 x \frac{ \wt m_0^2 }{\L^2}  \right] \nonumber\\
\l_0 &=& \wt \l_0 + 4! \frac{ \wt m_0^2 }{\L^2} \left( y  - \frac{x  }{6} \wt \l_0  \right)  \nonumber\\ 
c_{1,0}^{(6)} &=& 4! \left( - y - \frac{x}{6} \wt \l_0  \right) \nonumber\\
c_{2,0}^{(6)} &=& 2 \left( - x  \right) \nonumber\\
c_{3,0}^{(6)} &=&  \wt c_{3,0}^{(6)} - 6! \frac{ y }{6} \wt \l_0  \nonumber\\ 
\l_{\chi,0} &=&  \frac{6 y}{x} \left[  1 - 2 x \frac{\wt m_0^2 }{\L^2} \right] \, .
\eea
As we expected, we started from a ghost-free basis and the redefinition gave us back exactly
\eq{L6xmx} (up to a harmless normalization of $\chi$), justifying the original insertion of ${\cal L}_{{\rm R}g}$. 
To complete the connection between the G and W-bases we need to relate the corresponding couplings. 
Using the renormalized form of \eq{allcou} to move from the tilded to the un-tilded Lagrangian, we got 
\be
x = - \frac{1}{2} c_{2}^{(6)} \,\, {\rm and} \,\, y = - \frac{1}{4!} (c_{1}^{(6)} - 2 \wt \l c_{2}^{(6)}) \, .
\ee
In addition, recall that the R-ghost decouples if $\l_\chi = 0$ which can be achieved when $y=0$ or ${\wt m}^2=-m_\chi^2$.
The former happens if $c^{(6)}_1=2{\wt \l}c^{(6)}_2$ and the latter ensures that $m^2=0$ and
we need both in order to get an equivalent to \eq{LRF} theory.
With these constraints, we have $\l= - {\wt\l}$ and $c_3^{(6)}= {\wt c}_3^{(6)}$.

We are now ready to use directly the W-basis given by the Lagrangian \eq{L6phi6} for the evaluation of the phase diagram and of the RG flows.
For this, we return to the algorithm of Sect. \ref{sobf}.
Like we considered the mass and $\phi^4$ term deformations of the kinetic term, we will do the same for the $\phi^6$ operator.
We can fully exploit the detailed example of Sect. \ref{2-4} where the operator insertion method was applied and the notation was set up.
Our Lagrangian now has the form 
\be
\wt {\cal L} = \wt {\cal L}_4 + \wt c^{(6)}_3 \wt O^{(6)}_3  \nonumber
\ee
with $ \wt {\cal L}_4$ the tilded version of \eq{L4fi0}. $\wt c^{(2)} \wt O^{(2)} $ and $\wt c^{(4)} \wt O^{(4)} $ 
were defined in Sect. \ref{2-4} and the new entry here is $\wt O^{(6)}_3 = \phi^6/ (6! \L^2 )$ with coupling $\wt c^{(6)}_3$.
The classical dimension of the coupling is
\be\label{dc63cl}
d_{\wt c_3^{(6)}} = 2(4-d) = 2\ve\, ,
\ee
while the classical dimension of its associated operator, using $d_\phi=\frac{d-2}{2}$, is
\be\label{do63cl}
d_{\wt O_3^{(6)}} = 3 d - 8 = 4 - 3\ve \, .
\ee
Since the extra deformation is just another polynomial term there is no new contribution to $\d_\phi$ which remains zero at 1-loop.

Here we admit three $\b$-functions, since $l=2,4,6$, so let us start with the determination of $\b_{\wt c^{(2)}}$. Its general form is given by \eq{bc2a}.
Only the non-linear part is affected since, apart from $\wt Z_{222}$ and $\wt Z_{244}$, there is now one more contribution, $\wt Z_{266}$. In particular, this is given by \eq{fnOlOl} with $n = 2$, $k'=r=s=6$ and $q_{(2,6)}=2$ whose effect starts at 3-loop level. Hence $\wt Z_{266}$ and $\wt \G_{266}$ are trivial.
Then $\b_{\wt c^{(2)}}$ is the same as in \eq{bc2b} and reads
\be\label{wtbc2b}
\b_{\wt c^{(2)}} = -2 \wt m^2 + \frac{\wt \l \wt m^2}{16 \pi^2} \, .
\ee
Next we look at the $\b$-function of $ \wt c^{(4)} $ whose general form is
\be\label{wbc4a}
\b_{\wt c^{(4)}} =  -\ve \wt c^{(4)}  + \frac{ \wt m^2}{2} \wt \G_m^{6} + \wt \G_{\wt O^{(4)}} \, \wt c^{(4)} +  
\wt \G_{4 2 2} \, \wt c^{(2)} \wt c^{(2)} + \wt \G_{4 4 4} \, \wt c^{(4)} \wt c^{(4)} + \wt \G_{4 6 6} \, \wt c^{(6)}_3 \wt c^{(6)}_3 \, .
\ee
In this case there are two extra contributions to $\b_{\wt c^{(4)}}$ with respect to Appendix \ref{2-4}. 
The first refers to $\wt \G^6_m$ which is non-zero here and the second to $\wt \G_{466}$.

The former is calculated using \eq{fnO2} for $n=4$ and $k=2,4,6$
\bea\label{Z6246}
\langle \phi(x_1) \phi(x_2) \phi(x_3) \phi(x_4) \wt O^{(2)}(y) \rangle_0 &=& \wt Z_2^6 \langle \phi(x_1) \cdots 
\phi(x_4) \wt O^{(2)}(y) \rangle + \wt Z_4^6 \langle \phi(x_1) \cdots \phi(x_4) \wt O^{(4)}(y) \rangle \nonumber\\
&+& \wt Z_6^6 \langle \phi(x_1) \cdots \phi(x_4) \wt O^{(6)}_3 (y) \rangle
\eea
where now the bare part of the expression involves $\wt {\cal L}_{\rm int,0} = - \wt c_0^{(4)} \wt O_0^{(4)} + \wt c^{(6)}_{3,0} \wt O^{(6)}_{3,0} $.
Expanding the exponential in the expectation value to order ${\cal O}(\wt c^{(4)}_0, \wt c^{(6)}_{3,0})$, the l.h.s becomes
\be
\langle \phi(x_1) \phi(x_2) \phi(x_3) \phi(x_4) \frac{\phi^2(y)}{2} \frac{-i \wt \l \, \phi^4(z)}{4!} \rangle_0 + 
\langle \phi(x_1) \phi(x_2) \phi(x_3) \phi(x_4) \frac{\phi^2(y)}{2} \frac{i \wt c^{(6)}_3 \, \phi^6(z)}{6!\L^2} \rangle_0 \nonumber
\ee
The first term above contributes only disconnected and bubble diagrams while the second term gives
\be\label{O3d23}
%-------------------------------------
\begin{tikzpicture} [scale=0.9]
\draw [dashed] (6,0)--(7.6,0);
\draw [dashed] (7.6,0.5) circle [radius=0.5];
\draw [dashed] (7.6,0)--(9.1,0);
\node at (7.6,0.04) {$\times $};
\node at (7.6,0.04) {$\times $};
\node at (7.6,1) {$\times $};
\node at (7.6,1) {$\times $};
%\draw [thick] [fill=black] (3.7,0) circle [radius=0.1];
\draw [dashed] (7.6,0)--(7,-1);
\draw [dashed] (7.6,0)--(8.2,-1);
\end{tikzpicture}
%------------------------------------- 
\ee
corresponding to a scaleless $B_0$-integral. Its explicit form is given by \eq{M2.f}, where remember, we must now extract its UV divergent part. 
Apart from this, it has three channels given in \eq{3channels}.
From the r.h.s of \eq{Z6246} only the second term gives a counter-term at 1-loop which can absorb the remnant divergence.
This fixes $\wt Z_4^6 = 3 \wt c^{(6)}_3 / (16 \pi^2 \L^2 \ve) $ and from the definition of \eq{Gm} we get
\be\label{G6m}
\wt \G^6_m = - 6 \frac{\wt c^{(6)}_3 }{16 \pi^2\L^2} \, .
\ee
Next we consider the non-linear term, $\wt \G_{466}$.
For the evaluation of this contribution we use \eq{fnOlOl} for $n=4$, $k'=r=s=6$ and $q_{(4,6)}=4$ getting back
only greater than 1-loop scaleless Tadpoles. 
Collecting all the ingredients, the picture of $\b_{\wt c^{(4)}}$ is complete:
\be\label{wtbc4b}
\b_{\wt c^{(4)}} = -\ve \wt \l + \frac{3 \wt \l^2}{16 \pi^2} - \frac{3 \wt c_3^{(6)} \wt m^2}{16 \pi^2\L^2}\, .
\ee
As a last step we consider \eq{totalbeta} for $l=6$ which gives the general form 
\be\label{bc6a}
\b_{\wt c^{(6)}_3} =  -2 \ve \wt c^{(6)}_3  + \frac{ \wt m^2}{3} \wt \G_m^{8} + \wt \G_{\wt O^{(6)}_3} \,\wt c^{(6)}_3 +  
\wt \G_{6 2 2} \, \wt c^{(2)} \wt c^{(2)} + \wt \G_{6 4 4} \, \wt c^{(4)} \wt c^{(4)} + \wt \G_{6 6 6} \, \wt c^{(6)}_3 \wt c^{(6)}_3 \, .
\ee
Since the Lagrangian does not include dim-8 operators, the second term of \eq{bc6a} is absent.
We start with the linear part using \eq{flOl} for $n=l=6$ and the corresponding expectation value is
\be
\langle \phi(x_1) \cdots \phi(x_6) \wt O^{(6)}_3(y) \rangle_0 = \wt Z_{ \wt O^{(6)}_3 } \langle \phi(x_1) \cdots \phi(x_6) \wt O^{(6)}_3(y) \rangle  \, , 
\ee
or using \eq{ZOdo1},
\be\label{GZ66}
\langle \phi(x_1) \cdots \phi(x_6) \wt O^{(6)}_3(y) \rangle_0 - \wt \d_{ \wt O^{(6)}_3 } \langle \phi(x_1) \cdots  \phi(x_6) \wt O^{(6)}_3(y) \rangle =  \langle \phi(x_1) \cdots  \phi(x_6) \wt O^{(6)}_3(y) \rangle  \, .
\ee
The l.h.s of this expression should be evaluated with $\wt {\cal L}_{\rm int,0} = - \wt c_0^{(2)} \wt O_0^{(2)} - \wt c_0^{(4)} \wt O_0^{(4)} $.
Doing so, we have
\be
{\cal N}^{-1}  \langle 0 | T[ \phi_0(x_1) \cdot\cdot\cdot \phi_0(x_6) \frac{\phi_0^6(y)}{6! \L^2} 
e^{- i\int d^d z \frac{\wt m_0^2}{2} \phi_0^2(z) + \frac{\wt \l_0}{4!} \phi_0^4(z)} ] | 0 \rangle - \wt \d_{ \wt O^{(6)}_3 } 
\langle \phi(x_1) \cdots  \phi(x_6) \frac{\phi^6(y)}{6! \L^2} \rangle  \nonumber
\ee
with disconnected and 2-loop diagrams to ${\cal O}(\wt m^2, \wt \l^0)$.
On the other hand, to ${\cal O}((\wt m^2)^0, \wt \l)$ it gives
\be
%-------------------------------------
\begin{tikzpicture} [scale=0.9]
\draw [dashed] (6.4,0)--(7.4,0);
\draw [dashed] (6.4,1)--(7.4,1);
\draw [dashed] (7.6,0.5) circle [radius=0.5];
\draw [dashed] (7.9,0)--(9,0);
\draw [dashed] (7.9,1)--(9,1);
\node at (7.6,1) {$\times $};
\node at (7.6,1) {$\times $};
\node at (7.6,0) {$\times $};
\node at (7.6,0) {$\times $};
\draw [dashed] (6.9,2)--(7.6,1);
\draw [dashed] (7.6,1)--(8.3,2);
%\draw [thick] [fill=black] (3.7,0) circle [radius=0.1];
\draw [dashed] (11,1)--(11.7,0);
\draw [dashed] (11.7,0)--(12.4,1);
\draw [dashed] (11,0)--(12.5,0);
\node at (10,0) {$- $};
\draw [dashed] (11,-1)--(11.7,0);
\draw [dashed] (11.7,0)--(12.4,-1);
%\node at (10.5,0) {$ \d^{22}_O \cdot$};
\node at (11.7,0.01) {$\times $};
\node at (11.7,0.01) {$\times $};
\end{tikzpicture}
%------------------------------------- 
\nonumber
\ee
where apart from the 6 external legs, which imply 5 different channels and the coupling multiplying the diagram, it gives the same contribution as in \eq{O3d23}.
Keeping only the UV part of the scaleless integral involved, finiteness forces 
\be\label{dO66}
\wt \d_{ \wt O^{(6)}_3 }  = - \frac{5 \wt \l}{16 \pi^2} \frac{1}{\ve} \, ,
\ee
while using \eq{An.d.m1} or its 1-loop version below, we get
\be\label{GaO66}
\wt \G_{ \wt O^{(6)}_3 }  = \frac{5 \wt \l}{16 \pi^2}  \, .
\ee
The only thing left is the evaluation of the non-linear part in \eq{bc6a}.
The first two of the three non-linear terms, $\wt \G_{622}$ and $\wt \G_{644}$, are determined from \eq{fnOlOl} with $n=6$, $k'=r=s=2$, $q_{(6,2)}=6$ and $n=6$, $k'=r=s=4$, $q_{(6,4)}=6$ respectively.
Both expectation values admit disconnected (finite or 1-loop) diagrams. As a consequence, $\wt \G_{622}$ and $\wt \G_{644}$ vanish.
The third non-linear term, $\wt \G_{666}$, is given by \eq{flOlOl} for $l=r=s=6$
and has the same fate with the previous cases since it gives a 2-loop diagram as the lowest order contribution.
Putting all the pieces together we get
\be\label{wtbc6b}
\b_{\wt c^{(6)}_3} = - 2 \ve \wt c^{(6)}_3 + \frac{5 \wt \l \, \wt c^{(6)}_3}{16 \pi^2} \, .
\ee
Using \eq{wtbc2b}, \eq{wtbc4b} and \eq{wtbc6b} we summarize. Then the quantum parts of the $\b$-functions are 
\be\label{bqlmc3}
\vec{\b}^q_{\wt c^{(l)}} = \begin{bmatrix}
   \b^q_{\wt c^{(2)}}      \\
   \b^q_{\wt c^{(4)}}    \\
   \b^q_{\wt c^{(6)}_3}  \\ 
\end{bmatrix}   = \begin{bmatrix}
    \frac{\wt \l \wt m^2}{16 \pi^2}      \\
 - \frac{3 \wt c_3^{(6)} \wt m^2}{16 \pi^2\L^2} +  \frac{3 \wt \l^2}{16 \pi^2}  \\
 \frac{5  \wt \l \, \wt c_3^{(6)} }{16 \pi^2}    \\
\end{bmatrix} 
\ee
while the full 1-loop $\b$-functions are
\be\label{blmc3}
\vec{\b}_{\wt c^{(l)}} = \begin{bmatrix}
   \b_{\wt c^{(2)}}      \\
   \b_{\wt c^{(4)}}    \\
   \b_{\wt c^{(6)}_3}  \\ 
\end{bmatrix}   = \begin{bmatrix}
    -2 \wt m^2 + \frac{\wt \l \wt m^2}{16 \pi^2}      \\
 -\ve \wt \l + \frac{3 \wt \l^2}{16 \pi^2}  - \frac{3 \wt c_3^{(6)} \wt m^2}{16 \pi^2\L^2}  \\
 -2\ve \wt c_3^{(6)} + \frac{5  \wt \l \, \wt c_3^{(6)} }{16 \pi^2}    \\
\end{bmatrix} \, .
\ee
Check that using the relations below \eq{allcou} which connect the tilded and untilded couplings, we have arrived at the same $\b$-functions as in \eq{untbf}.

One of our original goals was to search for a broken phase in the absence of an explicit mass term. 
Let us then take $\wt m^2 \to 0$ in \eq{bqlmc3} and \eq{blmc3} so that we are left with only two operator insertions, $\wt O^{(4)}$ and $\wt O^{(6)}_{3}$.
This defines the $O^{(4)}$-$O^{(6)}$ system. 
Now we can finally evaluate the phase diagram of the theory in $d = 3,4\, \, {\rm and}\, \,  5$ dimensions.
Regarding the breaking of the $\mathbb Z_2$ internal symmetry, its fate is governed by the renormalized scalar potential
\be\label{wtV1l2}
\wt V^{(6)} = \frac{\wt \l }{4!}\phi^4 - \frac{\wt c_3^{(6)} }{6! \L^2} \phi^6 \, .
\ee
According to the signs of $\wt \l$ and $\wt c_3^{(6)}$, there are the following cases:
%%%%%%%%%%%%%%%%%%%%%%%%%%%%
\begin{itemize}
\item $\wt \l > 0$ and $\wt c_3^{(6)} > 0$\\
$\phi^6$ is dominant so the overall sign of the potential is minus. As a consequence, it becomes unstable and since the relative sign of the couplings is negative there are two maxima.
\item $\wt \l > 0$ and $\wt c_3^{(6)} < 0$\\
The overall sign of the potential and the relative sign of the couplings is positive. Therefore, the potential is stable but the theory does not exhibits broken phase.

\item $\wt \l < 0$ and $\wt c_3^{(6)} > 0$\\
This case resembles the previous one with the difference that now the overall sign of the potential is negative making the system unstable.

\item $\wt \l < 0$ and $\wt c_3^{(6)} < 0$\\
The potential is stable since the dominant term, $\phi^6$, has a positive sign. The relative sign of the couplings is negative and the theory has a broken phase with two minima. 
\end{itemize}
%%%%%%%%%%%%%%%%%%%%%%%%%%%%
In the last case the minimization of the potential inherits to the scalar field a non-trivial vev, given by
\be\label{vevf6}
\wt v = \pm \sqrt{20} \sqrt{\frac{ | \wt \l | }{ | \wt c_3^{(6)} | }} \L\, .
\ee
Observe that the vev depends explicitly on the cutoff, $\L$, which in general should be considered a function of the renormalization scale. 
Performing now the expansion $\phi \to \phi \pm \wt v $ in \eq{wtV1l2} and looking at the extrema of the potential, we obtain the physical mass
\be\label{mphylc6}
\wt m_\phi^2 = \frac{ | \wt \l | }{3} \wt v^2\, ,
\ee
which also depends on $\L$. 

\begin{figure}[!htbp]
\centering
\includegraphics[width=8cm]{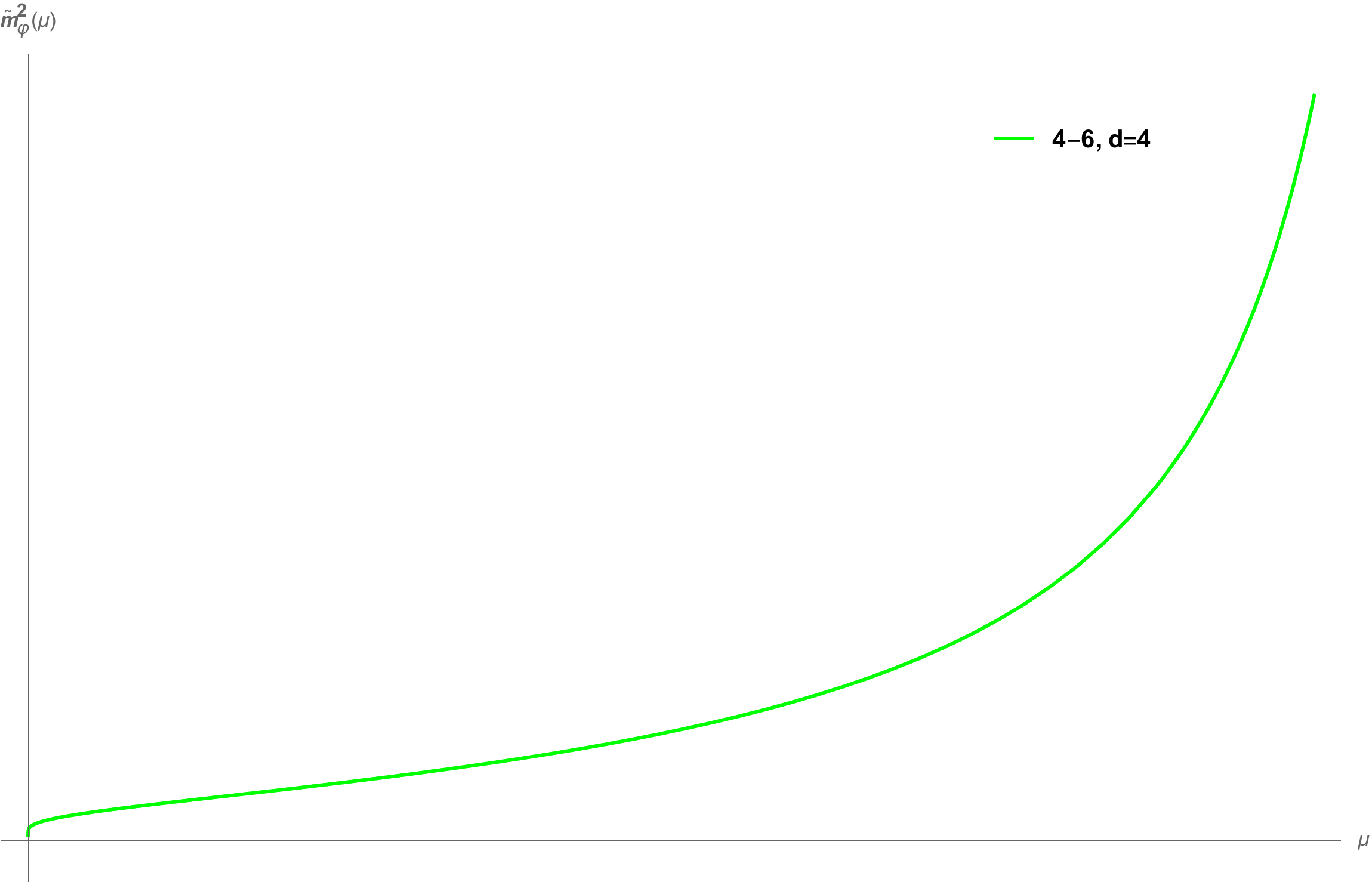}
\caption{\small The running of the physical mass, after SSB, as a function of the renormalization scale in four dimensions. \eq{mfi46} shows that the running of $\wt m_\phi$ is dominated by $\wt \l$ which diverges when $\m$ reaches the Landau pole. This is a representation of the naturalness problem.
\label{mulc6d4}}
\end{figure}

We proceed by constructing the phase diagram.
Following the same steps as for the $O^{(2)}$-$O^{(4)}$ system, the main ingredients are the RG equations of the couplings 
$\wt \l, \wt c_3^{(6)} $, the possible fixed points of the RG flow and the determination of the nature of the associated operators.
Regarding the first of these ingredients, we need to solve the RG equations $\m \, d {\wt \l} / d \m = \b_{\wt \l} $ and $\m \, d { \wt c_3^{(6)} } / d \m = \b_{ \wt c_3^{(6)} }$.
The solution for the RG evolution of $\wt \l$ in $d=4$ is
\be\label{wtlmue0}
\wt \l(\m) = \frac{\wt \l}{ 1 - \frac{3 \wt \l^2}{16 \pi^2}  \ln \frac{\m}{\m_{\rm R}} } \, ,
\ee
while in $d \ne 4$ it is
\be\label{wtlmu}
\wt \l(\m) = \frac{\wt \l}{\frac{3\wt \l}{16 \pi^2\ve} + \frac{\m^\ve}{\m_{\rm R}^\ve} \Bigl( 1 - \frac{3\wt \l}{16 \pi^2\ve} \Bigr ) }\, .
\ee
The solution for $\wt c_3^{(6)}$ in $d = 4- \ve$ is
\be\label{c6mue}
\wt c_3^{(6)}(\m) = \wt c_3^{(6)} \left(  \frac{\m}{\m_{\rm R}}  \right)^{-2 \ve + \frac{5 \wt \l}{16 \pi^2}}\, ,
\ee
with $\m_{\rm R}$ the usual arbitrary renormalization scale. Recalling that the RG equation of the mass in the $\phi^4$ theory is given by
\be\label{mmu46}
m^2(\m) = m^2 \Bigl(  \frac{\m}{m}  \Bigr )^{-2 + \frac{\l}{16 \pi^2}} 
\ee
and comparing it with \eq{c6mue} one observes that $\wt c_3^{(6)}$ acts like a mass term in $d=3$ and as an inverse mass in $d=5$.
The above solutions allow us to combine \eq{mphylc6} with \eq{wtlmue0} and \eq{c6mue}, 
and compute the running of $\wt m_\phi^2$ as a function of $\m$. To be more specific, in $d = 4$ we get
\be\label{mfi46}
\wt m_\phi^2(\m) = \frac{ | \wt \l(\m) | }{3} \wt v^2\, , 
\ee
whose running is depicted on \fig{mulc6d4} and shows that it sees the Landau pole in the UV.
%%%%%%%%%%%%%%%%%%%%%%%%%%%%%%%%%%%%%%%
\FloatBarrier
%%%%%%%%%%%%%%%%%%%%%%%%%%%%%%%%%%%%%%%
The next step is to check for possible fixed points in the RG flow, which is done by demanding that the $\b$-functions vanish for specific values of the couplings.
In four dimensions $\b_{\wt \l}$ vanishes only when 
\be\label{lambu1}
\wt \l = \wt \l_\bu = 0\, 
\ee
and on that point $ \b_{ \wt c_3^{(6)} } $ vanishes for any value of $ \wt c_3^{(6)} $.
A special case is when both couplings are zero since then, the flow reaches a trivial fixed point, $(\wt \l_\bu,\wt c_{3\bu}^{(6)})$.
Regarding the $d=4$ case, the picture therefore is that of a WF line formed by all possible values of $\wt c_3^{(6)}$, when \eq{lambu1} is true.
This line ends on the G fixed-point where both couplings vanish simultaneously.
On the other hand, in $d \ne 4$, $\b_{\wt \l}$ has both $\bu$ and $\star$ points corresponding to
\be\label{wtlbulst.}
\wt \l_\bu =0  \hskip .3cm {\rm and} \hskip .3cm  \wt \l_\star = \frac{16 \pi^2}{3} \ve\, ,
\ee
while $\b_{ \wt c_3^{(6)} }$ vanishes only for vanishing $ \wt c_3^{(6)}$. This is true in any dimension, except four, and indicates that $ \wt c_3^{(6)}$ has only one fixed point given by
\be
\wt c_{3\star}^{(6)} = \wt c_{3\bu}^{(6)} = 0 \, .\nonumber
\ee
The last ingredient we need for the description of the phase diagram is the behaviour 
of the operators $\wt O^{(4)}$ and $\wt O^{(6)}_{3}$ with respect to the fixed points, as we move from the UV to the IR.
For this we apply the rules reviewed in Appendix \ref{2-4}.
In what follows we present these results for $-1 \le \ve \le 1$.
\be\label{wtlcases}
   \frac{\partial \beta_{\wt \l}}{\partial \wt \l} = -\ve + \frac{6}{16 \pi^2} \wt \l =
   \begin{cases}
     \bullet : -\ve & \begin{cases}
                            d=4:0\,\, (\text{IR from \eq{wtlmue0}}) \\
                            d=3:-1 < 0 \rightarrow {\rm UV} \\
                            d=5: +1 >0 \rightarrow {\rm IR}
                            \end{cases}
     \\
     \star : +\ve & \begin{cases}
                           d=4: - \\
                           d=3: +1 >0 \rightarrow {\rm IR}\\
                           d=5: -1 < 0 \rightarrow {\rm UV} 
                          \end{cases}
   \end{cases}
 \ee
%%%%%%%%%%%%%%%%%%%%%%%%%%%%%%%%%%%%%%%
\be\label{wtc63cases}
\frac{\partial \beta_{\wt c_3^{(6)}}}{\partial \wt c_3^{(6)}} = -2 \ve + \frac{5 \wt \l}{16 \pi^2} 
\begin{cases}
     \bullet : -2 \ve & \begin{cases}
                            d=4:0\,\, (\text{IR from \eq{c6mue} for $\ve = 0$}) \\
                            d=3:-2 < 0 \rightarrow {\rm UV} \\
                            d=5: +2 >0 \rightarrow {\rm IR}
                            \end{cases}
     \\
     \star : -\frac{\ve}{3} & \begin{cases}
                           d=4: - \\
                           d=3: -\frac{1}{3} < 0 \rightarrow {\rm UV}\\
                           d=5: \frac{1}{3} > 0 \rightarrow {\rm IR} 
                          \end{cases}
\end{cases}
\ee
where \eq{wtc63cases} shows that $\wt O^{(6)}_{3}$ sees both fixed points as UV and as IR in $d=3$ and $d=5$ respectively.
The way to determine the nature of the two operators is through their scaling dimensions as reviewed in Appendix \ref{2-4}.
From \eq{sc.d.Ol} and \eq{an.d.1} we get
\be
\Delta_{\wt O^{(4)}} = 4 -2\ve + \frac{6 \wt \l}{16 \pi^2}
\ee
for $\wt O^{(4)}$, while for $\wt O^{(6)}_{3}$, keeping in mind \eq{do63cl}, 
\be
\Delta_{\wt O^{(6)}_{3}} = 4 - 3\ve + \frac{5 \wt \l}{16 \pi^2}\, .
\ee
Applying these expressions to the two fixed points we obtain the 1-loop exponent that determines the nature of an operator:
%%%%%%%%%%%%   TABLE   %%%%%%%%%%
\hskip 2cm
\begin{center}
\begin{tabular}{|c|c|c|c|}
\hline 
$$ & $\Delta_O-d$ & $\bullet$ & $\star$ \\
\hline \hline   
$\wt O^{(4)}$  & $-\ve + 2 \frac{3\wt \l}{16 \pi^2}$ &$-\ve $ & $\ve$\\ \hline
$\wt O^{(6)}_{3}$  & $- 2\ve + \frac{5\wt \l}{16 \pi^2}$ &$- 2\ve$ & $- \frac{\ve}{3}$\\ \hline
\end{tabular}
%\center{TABLE 4.}
\end{center}
%%%%%%%%%%%%%%%%%%%%%%%%%%%%
and this fixes the following characterizations:
%%%%%%%%%%%%   TABLE   %%%%%%%%%%
\hskip 2cm
\begin{center}
\begin{tabular}{|c|c|c|c|}
\hline 
$$ & $d=4$ & $d=3$ & $d=5$ \\
\hline \hline   
%%%%
$\wt O^{(4)}$  & $ \bullet: {\rm irrel} \hskip .5cm  \vline \hskip .5cm \star: {-}$ &$ \bullet: {\rm rel} \hskip .5cm  \vline \hskip .5cm \star: {\rm irrel}$ & $ \bullet: {\rm irrel} \hskip .5cm  \vline \hskip .5cm \star: {\rm rel}$\\ \hline
$\wt O^{(6)}_{3}$  & $ \bullet: {\rm irrel} \hskip .5cm  \vline \hskip .5cm \star: - $ & $\bullet: {\rm rel} \hskip .5cm  \vline \hskip .5cm \star: {\rm rel} $ & $ \bullet: {\rm irrel} \hskip .5cm  \vline \hskip .5cm \star: {\rm irrel} $\\ \hline
\end{tabular}
%\center{TABLE 5. Classifications of the operators at the fixed points according to Table 4.}
%\center{TABLE 3. Classification of operators at the fixed points according to Table 2.}
\end{center}
%%%%%%%%%%%%%%%%%%%%%%%%%%%%
The construction of the phase diagram follows the rules given in the last part of Appendix \ref{2-4}.

\begin{figure}[!htbp]
\centering
\includegraphics[width=8cm]{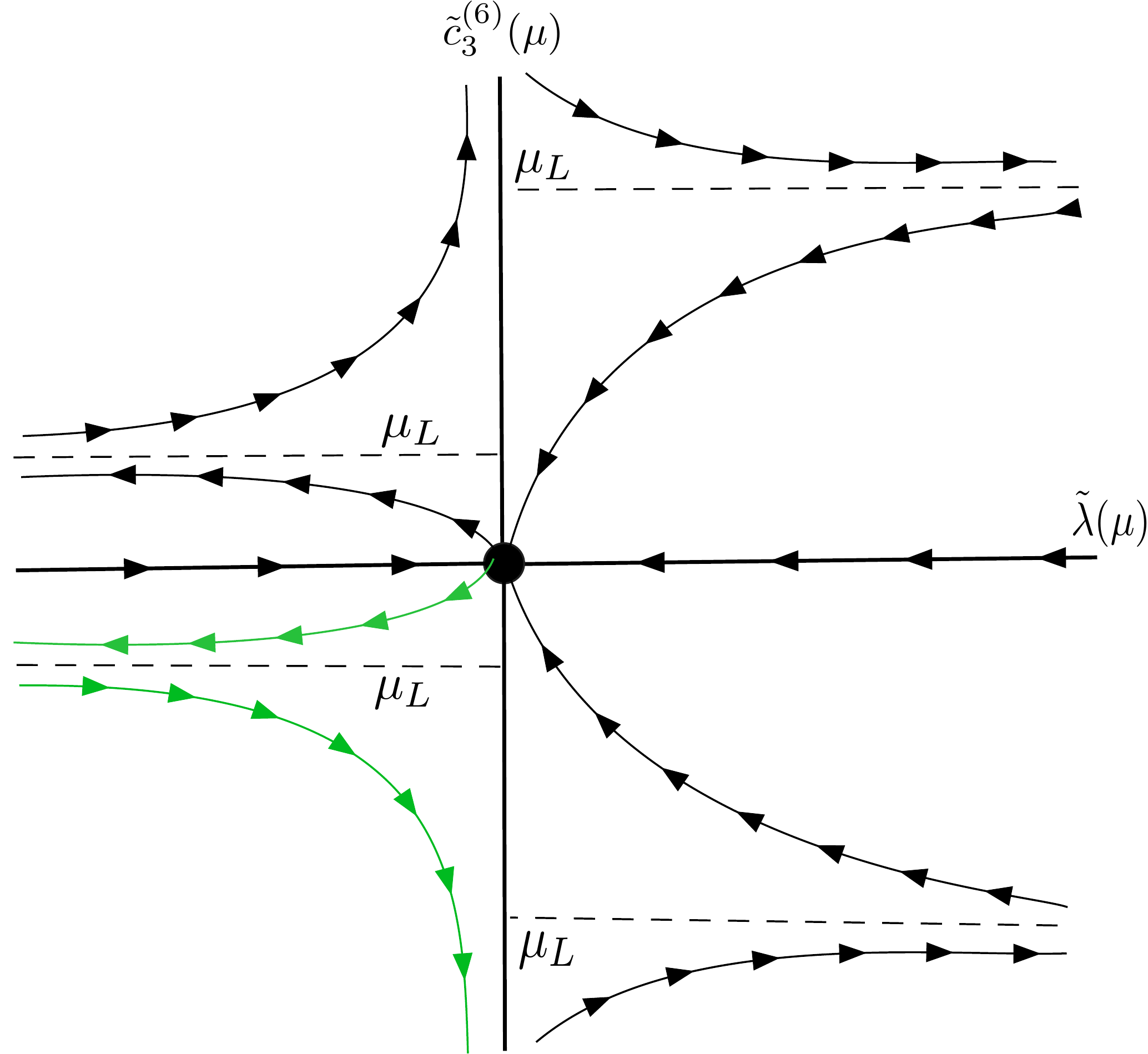}
\caption{\small RG flows for ${\wt {\cal L}}$, in $d=4$. The ${\wt \l}, {\wt c_3^{(6)}} < 0$ (green) flow corresponds to the broken phase. 
\label{d4c63lpo}}
\end{figure}

\begin{figure}[!htbp]
\begin{minipage}{.5\textwidth}
\includegraphics[width=8cm]{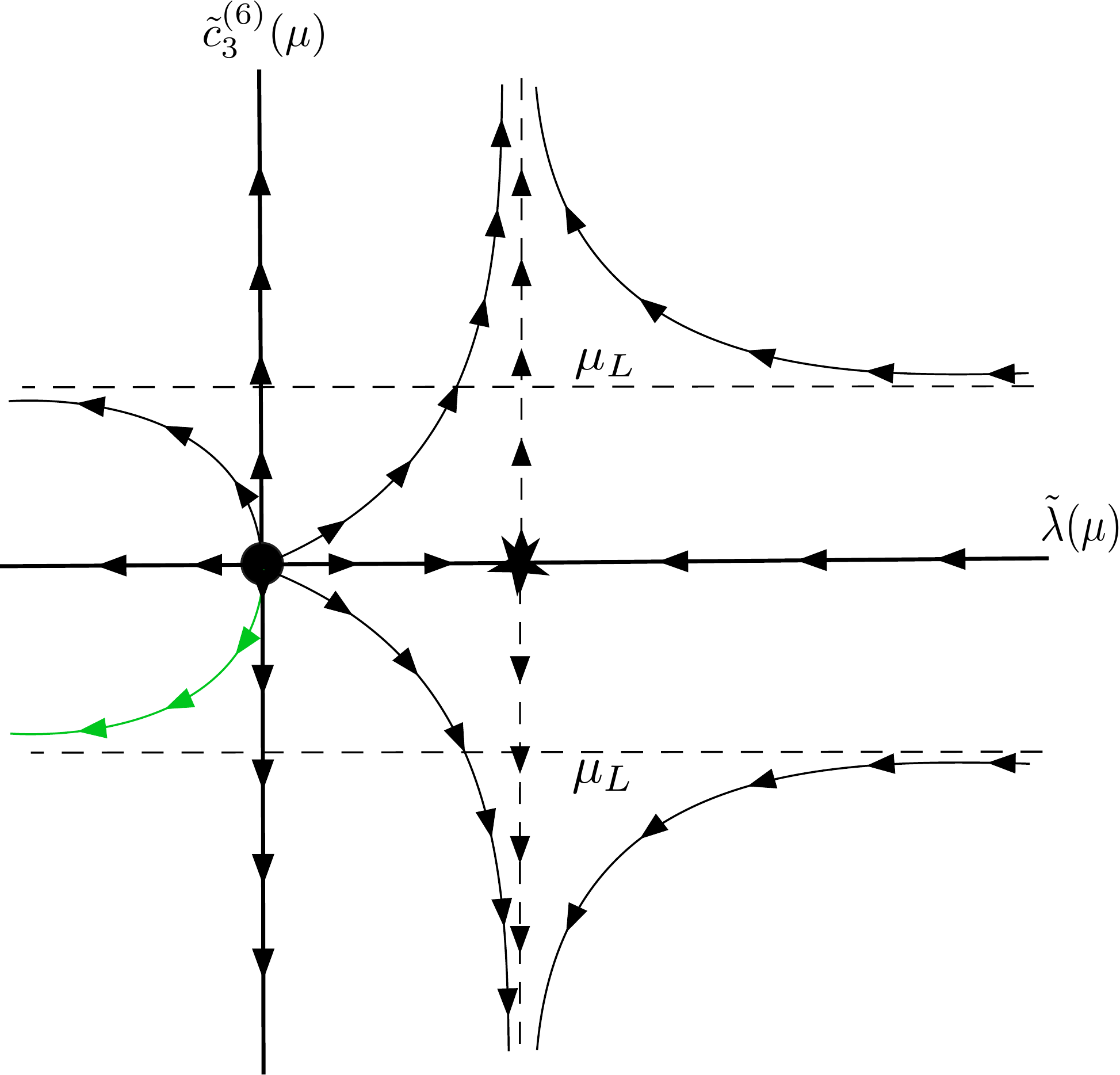}
\end{minipage}
\begin{minipage}{.5\textwidth}
\includegraphics[width=8cm]{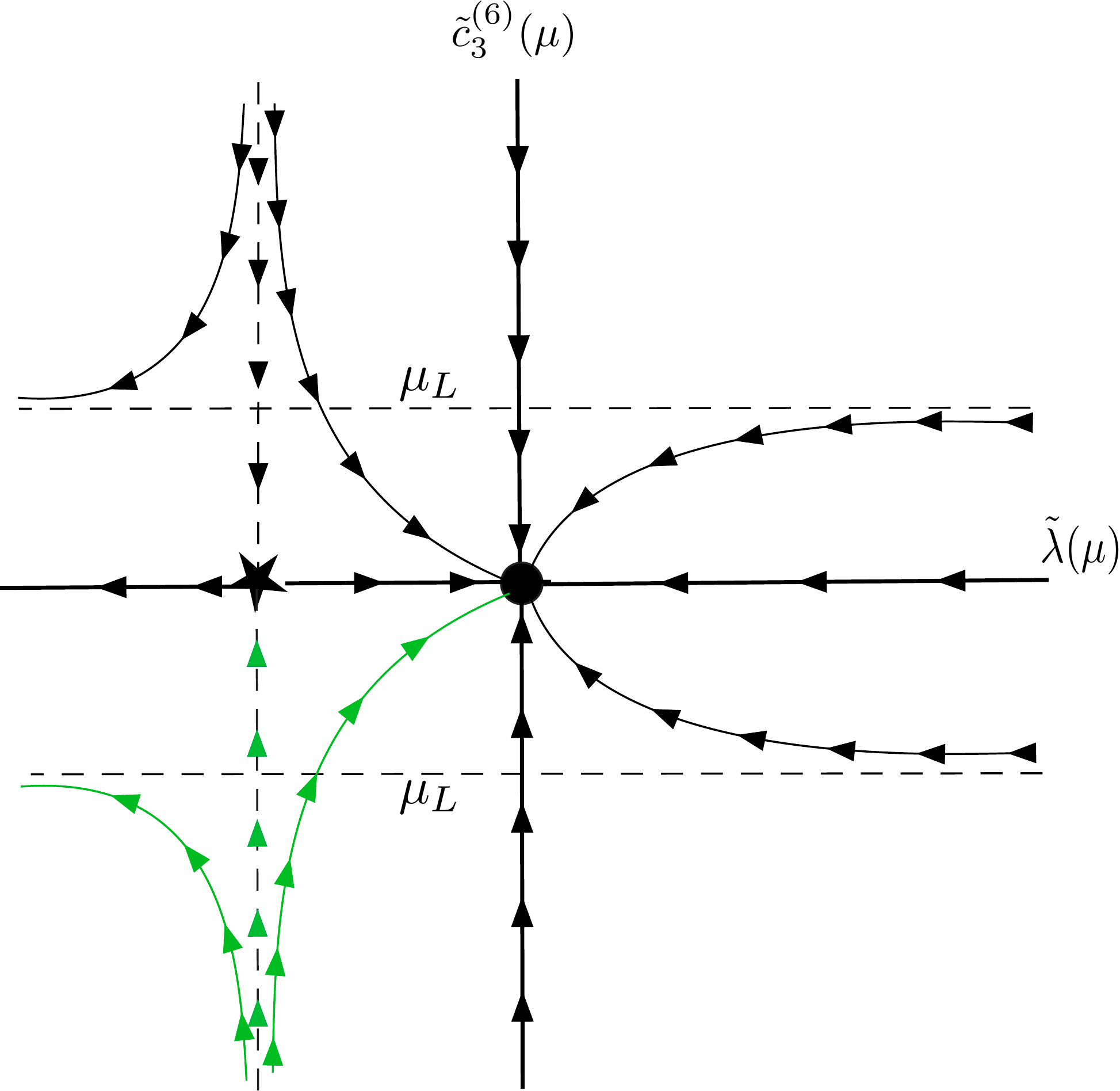}
\end{minipage}
\caption{\small Left: RG flows for ${\wt {\cal L}}$, in $d=3$. The ${\wt \l}, {\wt c_3^{(6)}} < 0$ (green) flow corresponds to the broken phase. 
Here, $\star$ denotes the WF fixed point where scale invariance is restored while the system is still interacting.
Right: RG flows for ${\wt {\cal L}}$, in $d=5$.  
\label{PDc35}}
\end{figure}

Let us start with $d=4$ where the RG flows are determined by \eq{wtlmue0} and \eq{c6mue} for $\ve = 0$. 
There are four possible combinations for the sign of the couplings, two for $\wt \l > 0$ and two for $\wt \l < 0$. 
In each case the theory has a Landau pole when the scale reaches $\m_{\rm L} = \exp[16 \pi^2/ 3\wt \l] \m_{\rm R}$. 
Notice that the broken phase here mimics the one of ${\cal L}^{(4)}$ for $\l <0$ in \fig{Ph.D.e0}. In fact, 
when $\wt \l < 0$ the absolute value of the coupling $\wt c_3^{(6)}(\m)$ increases towards the IR which indeed resembles the behaviour of a mass operator. 
These features are depicted in \fig{d4c63lpo}. 
%
%%%%%%%%%%%%%%%%%%%%%%%%%%%%%%%%%%%%%%%
%%%%%%%%%%%%%%%%%%%%%%%%%%%%%%%%%%%%%%%
%
\FloatBarrier
Next we consider the $d=3$ RG flows, given by \eq{wtlmu} and \eq{c6mue} for $\ve = 1$. 
The RG equation of $\wt \l(\m)$ is exactly the same as that of $\l(\m)$ in Appendix \ref{2-4}. Moreover, notice that for 
$\wt \l < \frac{32 \pi^2}{5}$, the coupling $\wt c_3^{(6)}(\m)$ has the same behaviour as the mass, \eq{mmu46}. 
Therefore the phase diagram on the left of Fig. \ref{PDc35} is qualitatively similar to the left part of \fig{PDepos}.
The last case regards the $d=5$ RG flows given by \eq{wtlmu} and \eq{c6mue} but now for $\ve = -1$. 
The flows are depicted on the right of Fig. \ref{PDc35}.
In this case the WF point is negative and we can take
$\wt \l$ as negative we want, however we restrict ourselves here to $- \frac{32 \pi^2}{5} < \wt \l $, to keep the exponent 
in the RG equation of $\wt c_3^{(6)} (\m)$ positive. Similarly to the $d=3$ case, we see two Landau 
branches, for $\wt \l < 0$, $\wt \l < \wt \l_\star$ and $\wt \l > 0$.  The flow of the broken phase here starts off G in the IR where
both couplings vanish and tends to a constant $\wt \l$ and a diverging $ \wt c_3^{(6)} $ in the UV.
%
%%%%%%%%%%%%%%%%%%%%%%%%%%%%%%%%%%%%%%%
%%%%%%%%%%%%%%%%%%%%%%%%%%%%%%%%%%%%%%%
%
\FloatBarrier

%------------------------------------------------------------------------------------------------------
\subsection{Filling a couple of remaining holes}\label{2-4-6_finalcomments}
%------------------------------------------------------------------------------------------------------

We have repeatedly argued that $\L$ is not an external to the system scale but we have not given any quantitative argument to that effect.
In DR we argued in particular that at a general point of the phase diagram $\L$ is a function of $\m$. At special points it may be 
easily related to specific scales like a Landau pole or a vev (which is a function of $\m$) but in general there is an arbitrariness in fixing this function.
Let us analyze a bit one simple case here, the case $\L=\m$ and demonstrate that this choice does not affect qualitatively the phase diagrams.
To see this, recall that for the construction of the phase diagrams in $d$-dimensions we used \eq{wtlmu} and \eq{c6mue} 
and focus on $\wt c_3^{(6)} (\m)$. In \eq{c6mue}, the number in the first term of the exponent is determined by the dimensionality of $\wt c_3^{(6)} (\m)$
and $\ve$ determines the nature of the running. Suppose now that the dimension-6 operator insertion is
\be
\wt c_3^{(6)} \wt O_3^{(6)} = \wt c_3^{(6)} \frac{\phi^6}{\m^2} \, . \nonumber
\ee
Dimensional analysis shows that $d_{\wt c_3^{(6)}}$ is still given by \eq{dc63cl}. Therefore in this case the running of 
$\wt c_3^{(6)}$ follows \eq{c6mue}, leading to the same phase diagrams.

There is another possibility, one where the inserted coupling $\wt c_3^{(6)}$ is dimensionfull, with $[\wt c_3^{(6)}] = -2$.
In such scenario the $\b$-function is given by
\be
\b_{ \hat c_3^{(6)} }   = ( 2 -2 \ve) \hat c_3^{(6)} + \b_{ \hat c_3^{(6)} }^q  \nonumber
\ee
since in $d$-dimensions $\wt c_3^{(6)} = \hat c_3^{(6)} \m^{d - d_{\wt O^{(6)}_3} } \equiv \hat c_3^{(6)} \m^{ -2(1-\ve) } $, with $\hat c_3^{(6)}$ dimensionless.
In any case, the above leads to the RG equation
\bea
\hat c_3^{(6)}(\m) &=& \hat c_3^{(6)} \left(  \frac{\m}{\m_{\rm R}}  \right)^{2 -2 \ve + \b_{ \hat c_3^{(6)} }^q / \hat c_3^{(6)} } \Rightarrow  \nonumber\\ 
\frac{ \hat c_3^{(6)}(\m) }{\m^2} &=& \frac{ \hat c_3^{(6)}}{\m_{\rm R}^2} \left(  \frac{\m}{\m_{\rm R}}  \right)^{ -2 \ve + \b_{ \hat c_3^{(6)} }^q / \hat c_3^{(6)} } \, , \nonumber
\eea
indicating that the RG equation for the dimensionfull coupling reads
\be
\wt c_3^{(6)}(\m) = \wt c_3^{(6)} \left(  \frac{\m}{\m_{\rm R}}  \right)^{-2 \ve + \b_{ \hat c_3^{(6)} }^q / \hat c_3^{(6)} } \nonumber
\ee
which is equivalent to \eq{c6mue}.

Finally, we have said nothing about unitarity in GEFT.
It would be interesting to establish the unitarity bounds that the theory respects in the presence of the HDO. 
This is quite important since unitarity violation imposes on the theory an energy scale above which it cannot be trusted.
Focusing on our case, recall that the $O^{(4)}$-$O^{(6)}$ system contains only one HDO, $O^{(6)}_{3}$. 
This operator was introduced along with $\L$ which regulates its strength.
Therefore, connecting this scale with unitarity bounds indicates how the internal scale $\m$ is related to the violation of unitarity.
Moreover, as we have seen, both the vev and the physical mass of the scalar field depend on $\L$. 
This means that through unitarity we can obtain bounds for the scale where SSB takes place and see how the bound affects $\wt m_\phi$.
To make these arguments more concrete, we follow Appendix A of \cite{Luty}. 
There, a general formula for placing bounds in tree-level scattering amplitudes was derived using unitarity. 
In our case we want to constrain the six-point vertex which in potential-form is given by
\be
\wt V \sim \frac{\wt c_3^{(6)}}{6!\L^2} \phi^6 \nonumber
\ee
and resembles Eq. (A.12) of \cite{Luty} for $n_1 = 6$, $n_2 = \cdot \cdot \cdot = n_r = 0$ and 
$n = n_1 +n_2 \cdot \cdot \cdot n_r =6$ with $\l_6 = \frac{\wt c_3^{(6)}}{\L^2}$.\
Treating the 6-point vetrex as a $3 \to 3$ scattering process 
corresponds to the case $k_1 = \frac{n_1}{2} = 3$ hence Eq. (A.18) of \cite{Luty} can be used directly leading to the optimal energy bound
\be\label{unE3}
E_{3} \le \frac{16\sqrt{6} \pi^{3/2}}{\sqrt{| \wt c_3^{(6)}}|} \L  \approx 218 \frac{\L}{\sqrt{| \wt c_3^{(6)}}|}\, .
\ee 
Therefore, the W-basis remains unitary as long as the energy scale respects this bound.

%---------------------------------------------------------------------------------------------------------------------------------------------------
\section{Dimension-8 operators}\label{2-4-6-8}
%---------------------------------------------------------------------------------------------------------------------------------------------------

Consider now the most general (up to total derivatives) dimension-8 extension of \eq{L6only1}
\be\label{d8L}
{\cal L}^{(8)} = \frac{c_{1}^{(8)}}{\L^4} \phi^3 \Box \phi^3 + \frac{c_{2}^{(8)}}{\L^4} \phi^2 \Box^2 \phi^2 + \frac{c_{3}^{(8)}}{\L^4} \phi \Box^3 \phi + \frac{c_{4}^{(8)}}{\L^4} \phi^8 \, .
\ee
Dropping tildes from now on, we immediately attempt to rotate onto a pure polynomial basis. The most general field redefinition up to total derivatives is 
\be
\phi \rightarrow \phi + \frac{x}{\L^2} \Box \phi + \frac{y}{\L^2} \phi^3 + \frac{z}{\L^4} \Box^2 \phi + \frac{u}{\L^4} \phi \Box \phi^2 + \frac{w}{\L^4} \phi^5 \, .
\ee
Then we follow steps similar to those that lead us to \eq{L68noder}. We will not go through the analysis of operator elimination. Things should work out 
in an analogous, alas more complicated way. The final result is coefficients $x,y,z,u,w$ that can be chosen\footnote{Even though we have explicitly calculated $x,y,z,u,\, {\rm and} \, w$ eliminating the derivative operators of the redefined Lagrangian, we do not adduce their values here since they are not illuminating. However, in \cite{FotisLetter} a simple example is presented where these coefficients have a transparent form. A relevant calculation is given in Passarino's paper in \cite{phi6quantum}.} so that
after the transformation the Lagrangian becomes
\be\label{L2468}
{\cal L} = - \frac{1}{2} \phi \Box \phi - \frac{1}{2} m^2 \phi^2 - \frac{ \l }{4!}\phi^4  + \frac{ c_{3}^{(6)}}{6! \L^2} \phi^6 + \frac{ c^{(8)}_{4}}{8! \L^4} \phi^8 \, .
\ee
The potential in this Lagrangian is
\be\label{V8L}
V^{(8)} = \frac{1}{2} m^2 \phi^2 + \frac{ \l}{4!}\phi^4 - \frac{ c_3^{(6)}}{6! \L^2} \phi^6 - \frac{ c^{(8)}_4}{8! \L^4} \phi^8  \, .
\ee
This is now our dimension-8 W-basis.
Regarding insertions, the dimension-8 extension has the three operator insertions of \eq{L6phi6} plus $O_{4}^{(8)}$ with its associated coupling $c_{4}^{(8)}$.
The extra operator, as a function of the field, is $O_{4}^{(8)} = \phi^8/( 8!\L^4 )$ and its classical dimension is
\be\label{do84cl}
d_{O_4^{(8)}} = 4 d - 12 = 4(1 - \ve) \, ,
\ee
while the classical dimension of the corresponding coupling is
\be\label{dc84cl}
d_{c_4^{(8)}} = 3(4-d) = 3\ve\, .
\ee
In this case we have $l=2,4,6,8$ in \eq{totalbeta} while the interaction Lagrangian will always include three terms.
At 1-loop $\phi^8$ does not affect the one-point function and $\d_\phi$ stays zero.
We have demonstrated in great detail how the insertion algorithm works in the previous sections, so here we will be more concise.

Starting with $\b_{c^{(2)}}$, we have the form of \eq{wtbc2b} plus one more non-linear contribution, $\G_{288}$.
This is given by \eq{fnOlOl} for $n=2$, $k'=r=s=8$ and $q_{(2,8)}=2$ which gives higher than 1-loop diagrams at lowest order.
This means that $Z_{288}$ and $\G_{288}$ are trivial. Then, the $\b$-function of $c^{(2)}$ remains as in the dimension-6 model:
\be\label{bc2c}
\b_{c^{(2)}} = -2 m^2 + \frac{ \l m^2}{16 \pi^2} \, .
\ee
The same goes for $\b_{c^{(4)}}$ as \eq{wbc4a} is enhanced by the non-linear term $\G_{488} c^{(8)}_4 c^{(8)}_4$. \eq{fnOlOl} with $n=4$ and $k'=r=s=8$ along with $q_{(4,8)}= 4$ generates higher than 1-loop diagrams hence,
\be\label{bc4c}
\b_{ c^{(4)}} = -\ve \l - \frac{3 c_3^{(6)} m^2}{16 \pi^2\L^2} + \frac{3 \l^2}{16 \pi^2} \, .
\ee
The first non-trivial contribution here comes from the effect of $O^{(8)}_4$ on $O^{(6)}_3$ due to the mass operator.
\eq{totalbeta} for $l=6$ reads
\be
\b_{ c^{(6)}_3} =  -2 \ve c^{(6)}_3  + \frac{ m^2}{3}  \G_m^{8} + \G_{ O^{(6)}_3} \, c^{(6)}_3 + 
\G_{6 2 2} \, c^{(2)}  c^{(2)} + \G_{6 4 4} \,  c^{(4)}  c^{(4)} + \G_{6 6 6} \, c^{(6)}_3 c^{(6)}_3 + \G_{6 8 8} \, c^{(8)}_4 c^{(8)}_4 
\ee
where now $\G_m^{8} $ is non-zero and $\G_{6 8 8}$ is added. The former is evaluated using \eq{fnO2} for $n=6$ and $k=2,4,6,8$
\bea\label{Z64Z68}
\langle \phi(x_1) \cdots \phi(x_6) O^{(2)}(y) \rangle_0 &=& Z_2^8 \langle \phi(x_1) \cdots \phi(x_6) O^{(2)}(y) \rangle + Z_4^8 \langle \phi(x_1) \cdots \phi(x_6) O^{(4)}(y) \rangle \nonumber\\
&+& Z_6^8 \langle \phi(x_1) \cdots \phi(x_6) O^{(6)}_3 (y) \rangle + Z_8^8 \langle \phi(x_1) \cdots \phi(x_6) O^{(8)}_4 (y) \rangle \, . \nonumber\\
\eea
Now the bare part of the above expression involves $ {\cal L}_{\rm int,0} = - c_0^{(4)} O_0^{(4)} + c^{(6)}_{3,0} O^{(6)}_{3,0} + c^{(8)}_{4,0} O^{(8)}_{4,0}  $.
Expanding the exponential in the expectation value to order ${\cal O}( c^{(4)}_0, c^{(6)}_{3,0},c^{(8)}_{4,0})$, the l.h.s becomes
\bea
\langle \phi(x_1) \cdots \phi(x_6) \frac{\phi^2(y)}{2} \frac{-i  \l \, \phi^4(z)}{4!} \rangle_0 &+& \langle \phi(x_1) \cdots \phi(x_6) \frac{\phi^2(y)}{2} \frac{i  c^{(6)}_3 \, \phi^6(z)}{6!\L^2} \rangle_0 \nonumber\\
&+& \langle \phi(x_1) \cdots \phi(x_6) \frac{\phi^2(y)}{2} \frac{i  c^{(8)}_4 \, \phi^8(z)}{8!\L^4} \rangle_0
\eea
The first and second terms give only disconnected 1-loop diagrams while the third gives
\be\label{M268}
%-------------------------------------
\begin{tikzpicture} [scale=0.9]
\draw [dashed] (6,0)--(7.6,0);
\draw [dashed] (7.6,0.5) circle [radius=0.5];
\draw [dashed] (7.6,0)--(9.1,0);
\node at (7.6,0.04) {$\times $};
\node at (7.6,0.04) {$\times $};
\node at (7.6,1) {$\times $};
\node at (7.6,1) {$\times $};
%\draw [thick] [fill=black] (3.7,0) circle [radius=0.1];
\draw [dashed] (7.6,0)--(7,-1);
\draw [dashed] (7.6,0)--(8.2,-1);
\draw [dashed] (7.6,0)--(6,-0.7);
\draw [dashed] (7.6,0)--(9.2,-0.7);
\end{tikzpicture}
%-------------------------------------
\nonumber
\ee
This is once more a scaleless $B_0$-integrals, given by \eq{M2.f} and we keep only its UV part. There are five independent channels.
From the r.h.s of \eq{Z64Z68} only the third term gives, at 1-loop, a counter-term which absorbs the remnant divergence.
Notice that on one side there is a $1/\L^4$ contribution while on the other only $1/\L^2$ which fixes 
$ Z_6^8 = 5 c^{(8)}_4 / (16 \pi^2 \L^2 \ve) $. From the definition of \eq{Gm} we then get
\be\label{G8m}
 \G^8_m = - 15 \frac{ c^{(8)}_4 }{16 \pi^2\L^2} \, .
\ee
This expression generates the opposite sign term in the $\b$-function compared to the diagrammatic calculation. 
The mismatch is an artifact of the operator insertion method, for level-1 insertions are defined up to an overall sign.
For the determination of $ \G_{688}$ we need \eq{fnOlOl} with $n=6$, $k'=r=s=8$ and $q_{(6,8)}= 6$ that involves higher loop diagrams
so it does not contribute to $\b_{ c^{(6)}_3}$. Then,
\be\label{bc6c}
\b_{ c^{(6)}_3} = - 2 \ve c^{(6)}_3 - \frac{5 \, c_4^{(8)}}{16 \pi^2} \frac{ m^2 }{\L^2} + \frac{5 \l \, c^{(6)}_3}{16 \pi^2}  \, .
\ee
The last $\b$-function is of the form \eq{totalbeta} for $l=8$:
\be\label{bc8a}
\b_{c^{(8)}_4} =  -3 \ve c^{(8)}_4  + \frac{ m^2}{4} \G_m^{10} + \G_{ O^{(8)}_4} \, c^{(8)}_4 + 
\G_{8 22} \, c^{(2)} c^{(2)} + \G_{8 4 4} \, c^{(4)} c^{(4)} +  \G_{8 6 6} \, c^{(6)}_3 c^{(6)}_3 + \G_{8 8 8} \, c^{(8)}_4 c^{(8)}_4\, .
\ee
There is no dim-10 operator in the action so the second term does not contribute. 
However, the third term is non-trivial and the same is true only for the non-linear term $\G_{866}$, at 1-loop.
The former is evaluated from \eq{flOl} for $n=l=8$ and the associated expectation value is
\be
\langle \phi(x_1) \cdots \phi(x_8) O^{(8)}_4(y) \rangle_0 = Z_{ O^{(8)}_4 } \langle \phi(x_1) \cdots \phi(x_8) O^{(8)}_4(y) \rangle  \, . 
\ee
From \eq{ZOdo1} this is
\be\label{ZO88}
\langle \phi(x_1) \cdots \phi(x_8) O^{(8)}_4(y) \rangle_0 - \d_{ O^{(8)}_4 } \langle \phi(x_1) \cdots \phi(x_8) O^{(8)}_4(y) \rangle =  \langle \phi(x_1) \cdots \phi(x_8) O^{(8)}_4(y) \rangle  \, ,
\ee
where the l.h.s involves $ {\cal L}_{\rm int,0} = - c_0^{(2)} O_0^{(2)} -  c_0^{(4)} O_0^{(4)} + c^{(6)}_{3,0} O^{(6)}_{3,0}$.
Then we get
\be
{\cal N}^{-1}  \langle 0 | T[ \phi_0(x_1) \cdot\cdot\cdot \phi_0(x_8) \frac{\phi_0^8(y)}{8! \L^4} e^{- i\int d^d z \frac{ m_0^2}{2} \phi_0^2(z) + \frac{ \l_0}{4!} \phi_0^4(z) - c^{(6)}_{3,0} \frac{ \phi^6_0}{6!\L^2}  } ] | 0 \rangle - \d_{ O^{(8)}_4 } \langle \phi(x_1) \cdots \phi(x_8) O^{(8)}_4(y) \rangle \,  \nonumber
\ee
and the only non-trivial contribution comes at order ${\cal O}( ( m^2)^0, \l, (c^{(6)}_3)^0)$:
\be
%-------------------------------------
\begin{tikzpicture} [scale=0.9]
\draw [dashed] (6.4,0)--(7.4,0);
\draw [dashed] (6.4,1)--(7.4,1);
\draw [dashed] (7.6,0.5) circle [radius=0.5];
\draw [dashed] (7.9,0)--(9,0);
\draw [dashed] (7.9,1)--(9,1);
\node at (7.6,1) {$\times $};
\node at (7.6,1) {$\times $};
\node at (7.6,0) {$\times $};
\node at (7.6,0) {$\times $};
\draw [dashed] (6.9,2)--(7.6,1);
\draw [dashed] (7.6,1)--(8.3,2);
%\draw [thick] [fill=black] (3.7,0) circle [radius=0.1];
\draw [dashed] (6.2,2)--(7.5,1);
\draw [dashed] (7.6,1)--(9.2,2);
%\draw [thick] [fill=black] (3.7,0) circle [radius=0.1];
\draw [dashed] (11,1)--(11.7,0);
\draw [dashed] (11.7,0)--(12.4,1);
\draw [dashed] (11,0)--(12.5,0);
\node at (10,0) {$- $};
\draw [dashed] (11,-1)--(11.7,0);
\draw [dashed] (11.7,0)--(12.4,-1);
\draw [dashed] (11.7,0)--(11.7,1);
\draw [dashed] (11.7,0)--(11.7,-1);
%\node at (10.5,0) {$ \d^{22}_O \cdot$};
\node at (11.7,0.01) {$\times $};
\node at (11.7,0.01) {$\times $};
\end{tikzpicture}
%------------------------------------- 
\nonumber
\ee
The first diagram has seven channels and belongs to the scaleless integrals of \eq{M2.f}. 
The second is just a vertex multiplied with a counter-term. Keeping only the UV part, finiteness requires
\be\label{dO88}
\d_{ O^{(8)}_4 }  = - \frac{7 \l}{16 \pi^2} \frac{1}{\ve} \, 
\ee
and \eq{An.d.m1} (or its 1-loop version below) indicates that
\be\label{GaO88}
\G_{ O^{(8)}_4 }  = \frac{7 \l}{16 \pi^2}  \, .
\ee
The last step is the evaluation of $\G_{866}$. We take \eq{fnOlOl} for $n=8$ and $k'=r=s=6$, with $q_{(8,6)}= 8$, so that
\be\label{Z866}
\langle \phi(x_1) \cdots \phi(x_8) O^{(6)}_3 (y) O^{(6)}_3 (z) \rangle_0 = Z_{866} \langle \phi(x_1)  \cdots \phi(x_8) O^{(6)}_3 (y) O^{(6)}_3 (z) \rangle  \, .
\ee
The l.h.s gives the divergent diagram
\be
%-------------------------------------
\begin{tikzpicture} [scale=0.9]
\draw [dashed] (6.4,0)--(7.4,0);
\draw [dashed] (6.4,1)--(7.4,1);
\draw [dashed] (7.6,0.5) circle [radius=0.5];
\draw [dashed] (7.9,0)--(9,0);
\draw [dashed] (7.9,1)--(9,1);
\node at (7.6,1) {$\times $};
\node at (7.6,1) {$\times $};
\node at (7.6,0) {$\times $};
\node at (7.6,0) {$\times $};
\draw [dashed] (6.9,2)--(7.6,1);
\draw [dashed] (7.6,1)--(8.3,2);
%\draw [thick] [fill=black] (3.7,0) circle [radius=0.1];
\draw [dashed] (6.9,-1)--(7.6,0);
\draw [dashed] (7.6,0)--(8.3,-1);
%\draw [thick] [fill=black] (3.7,0) circle [radius=0.1];
\end{tikzpicture}
%------------------------------------- 
\nonumber
\ee
another scaleless $B_0$-integral with seven channels. 
The r.h.s of \eq{Z866} gives the finite version of the above diagram.
Since $Z_{866}$ is divergent, according to the discussion below \eq{fnOlOl}, the latter diagram should be thought as a finite vertex.
Then the remnant divergence is absorbed by $\d_{866} = 7 / (16 \pi^2 \ve) $ which determines
\be
\G_{866} = \frac{7}{ 16 \pi^2} \, .
\ee 
Assembling all parts of $\b_{ c^{(8)}_4} $, we arrive at the final expression
\be\label{bc8b}
\b_{ c^{(8)}_4} = - 3\ve c^{(8)}_4 + \frac{7 \l \, c^{(8)}_4 }{16 \pi^2} + \frac{7 (c^{(6)}_3 )^2 }{16 \pi^2} \, .
\ee
Let us collect for easier reference all $\b$-functions from \eq{bc2c}, \eq{bc4c}, \eq{bc6c} and \eq{bc8b} into 
\be\label{blmc4}
\vec{\b}_{ c^{(l)}}   = \begin{bmatrix}
  -2 m^2 + \frac{ \l m^2}{16 \pi^2}      \\
 -\ve  \l + \frac{3  \l^2}{16 \pi^2}  - \frac{3 c_3^{(6)} m^2}{16 \pi^2\L^2}  \\
 -2\ve c_3^{(6)} + \frac{5 \l  c_3^{(6)} }{16 \pi^2} - \frac{5 \, c_4^{(8)}}{16 \pi^2} \frac{ m^2 }{\L^2}   \\
 -3\ve  c_4^{(8)} +  \frac{7 ( c_3^{(6)})^2 }{16 \pi^2}  + \frac{7 \l  c_4^{(8)} }{16 \pi^2} \\
\end{bmatrix} 
\ee
The phase diagram is a four-dimensional space and its full analysis is straightforward but quite complicated.
As in the previous section we are particularly interested in the massless limit which reduces the dimensionality of
the phase diagram by one unit. A further limit that can be considered is the one of vanishing quartic coupling.
The resulting two-dimensional phase diagram could be just the projection on the $\l=0$ plane in the three-dimensional
phase diagram or it could be the phase diagram of a system that for some reason, generates only irrelevant interactions.
The potentially interesting fact is that even in this drastic limit there is a non-trivial scalar potential
\be\label{wtV1lc6c8}
V^{(8)} = - \frac{ c_3^{(6)}}{6! \L^2} \phi^6 - \frac{ c^{(8)}_4}{8! \L^4} \phi^8  \, .
\ee
The $\mathbb Z_2$ broken phase is the one where $ c_3^{(6)} > 0$ and $ c_4^{(8)} < 0$
since the dominant term, $\phi^8$, has a positive sign. 
Then the scalar field acquires a non-trivial vev at the minimum 
\be\label{vevf8}
v = \pm \sqrt{42} \sqrt{\frac{ c_3^{(6)}}{ | c_4^{(8)}| }} \L \, .
\ee
Expanding the field around its vev $\phi \to \phi \pm v $ we obtain the mass of the scalar field:
\be\label{mphyc6c8}
 m_\phi^2 = \frac{7}{10} \frac{( c_3^{(6)})^2 }{ | c_4^{(8)}| } v^2\, ,
\ee
which shows that it depends on $\L$ through the vev, as expected.

RG flows in this context have not been examined before to our knowledge, so let us proceed further.
In the double limit the $\b$-functions simplify to
\be\label{bfc6c8}
\vec{\b}_{ c^{(l)}}  = \begin{bmatrix}
 -2\ve c_3^{(6)}  \\
 -3\ve c_4^{(8)} + \frac{7 ( c_3^{(6)})^2 }{16 \pi^2}  \\
\end{bmatrix} 
\ee
and according to \eq{an.d.1} the anomalous dimensions of $O^{(6)}_3$ and $O^{(8)}_4$ are
\be\label{G33}
\g_{O^{(6)}_3} = 0\, , \hskip 1cm \g_{O^{(8)}_4} = 0 \, ,
\ee
which means that the scaling dimensions of both operators coincide with the classical ones.
The RG flows are determined by the solution to
\be\label{bc63bc84}
\m \frac{d c_3^{(6)} (\m) }{d \m} =  \b_{ c_3^{(6)} } = -2 \ve  c_3^{(6)}   \hskip .3cm {\rm and}  \hskip .3cm 
\m \frac{d  c_4^{(8)} (\m) }{d \m} =  \b_{ c_4^{(8)}} = -3 \ve c_4^{(8)} + \frac{7 ( c_3^{(6)})^2 }{16 \pi^2} \, .
\ee
Here we will meet behaviour of the couplings that can be quite exotic, depending on $d$.

\begin{figure}[!htbp]
\centering
\includegraphics[width=7cm]{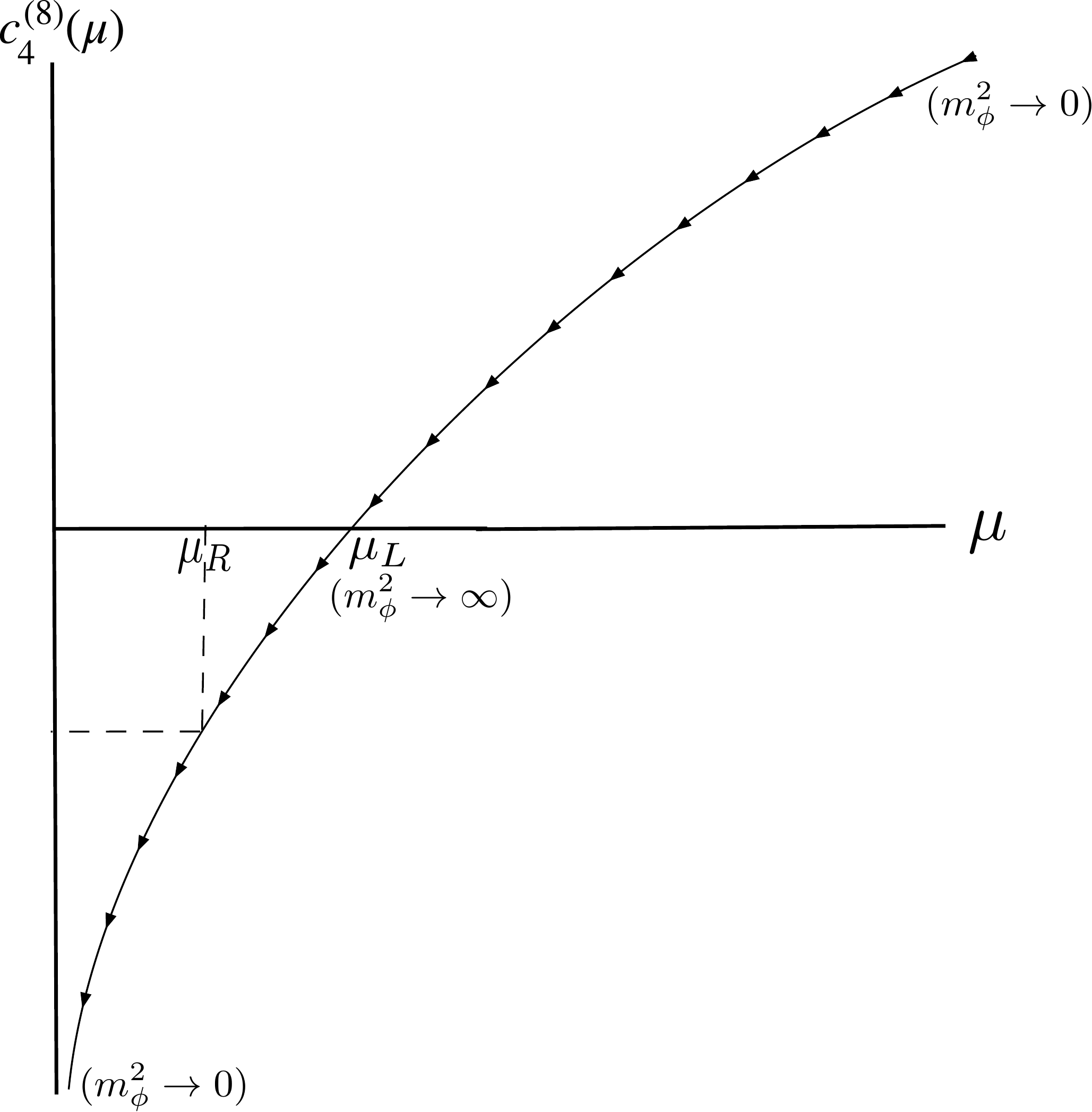}
\caption{\small The running of $c_4^{(8)}(\m)$ as a function of the renormalization scale $\m$ 
when the physical coupling, $ c_4^{(8)}$, is positive and the potential is stable. 
As the scale decreases, the coupling becomes less positive and at some small scale $\m_L$ it vanishes. 
After that scale, $ c_4^{(8)}$ flips sign and as $\m \to 0$ it increases up to minus infinity.  \label{c84mu1}}
\end{figure}

\begin{figure}[!htbp]
\centering
\includegraphics[width=9cm]{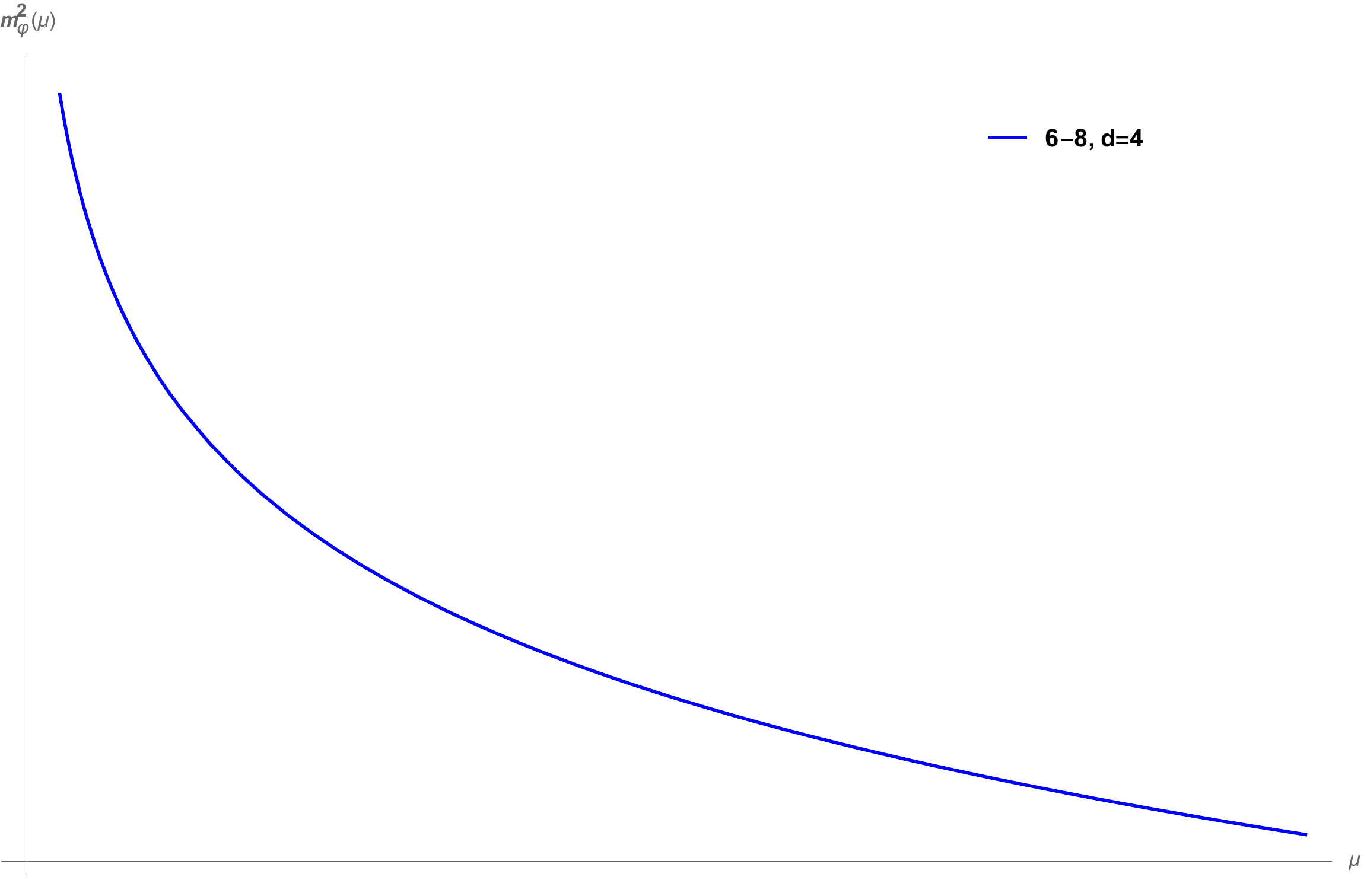}
\caption{\small The running of the physical mass in $d = 4$, after SSB, as a function of the renormalization scale.
\label{muc6c8d4}}
\FloatBarrier
\end{figure}

\begin{figure}[!htbp]
\centering
\includegraphics[width=8cm]{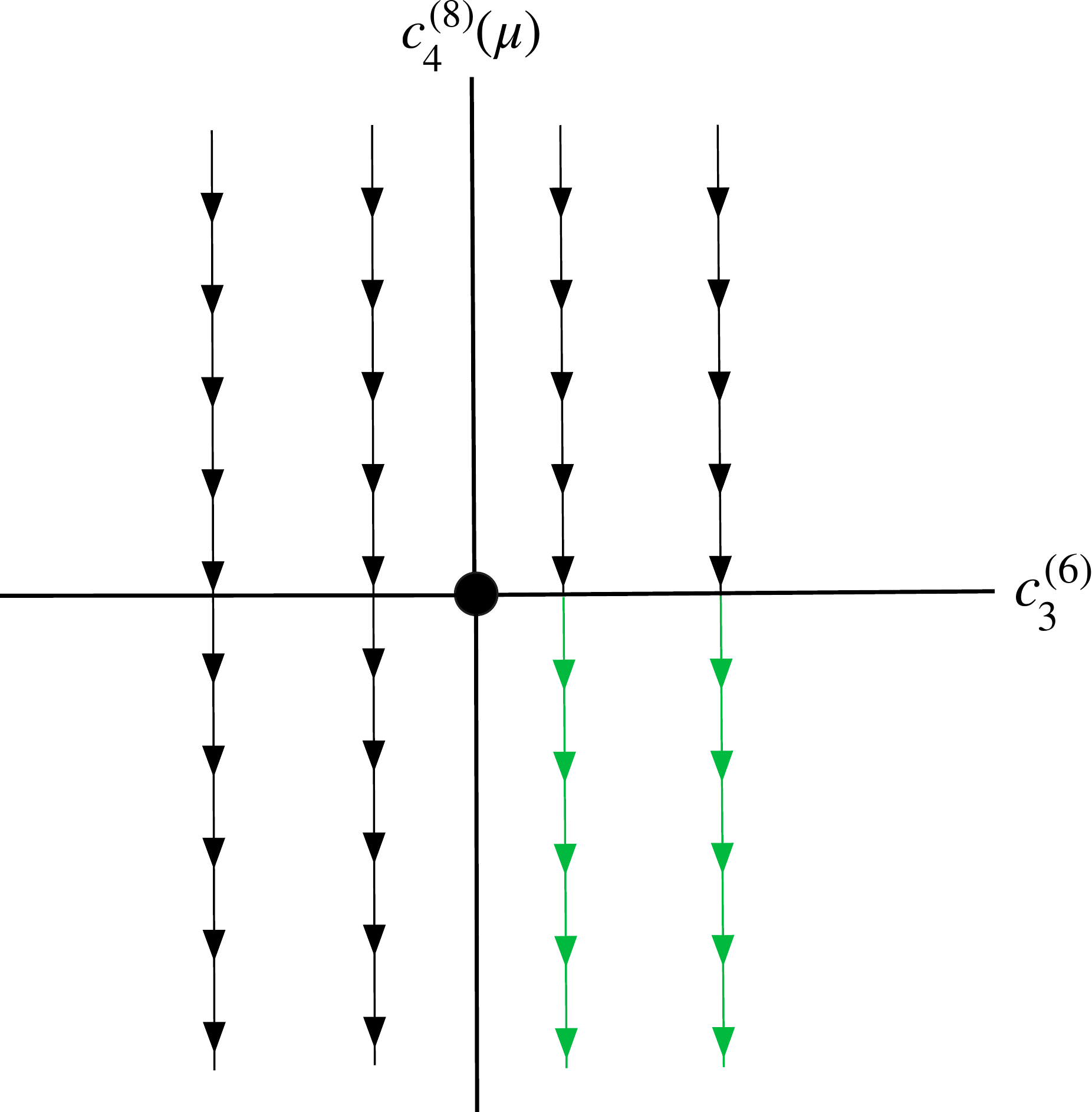}
\caption{\small The $4d$ RG flows for ${\cal L}$ in \eq{L2468} at $m^2 = \l =0$ limit. The flow is non-trivial only when $ c_3^{(6)} \ne 0$. The stable, broken branch of the flow (green lines) is for $0<\m<\m_L$ with $\m_L=\m_R \exp[ 16\pi^2 v^2 / 10m_\phi^2 ]$ and $\m_R<\m_L$
the renormalization scale.
\label{d4c6c81}}
\end{figure}

\begin{figure}[!htbp]
\begin{minipage}{.5\textwidth}
\includegraphics[width=8cm]{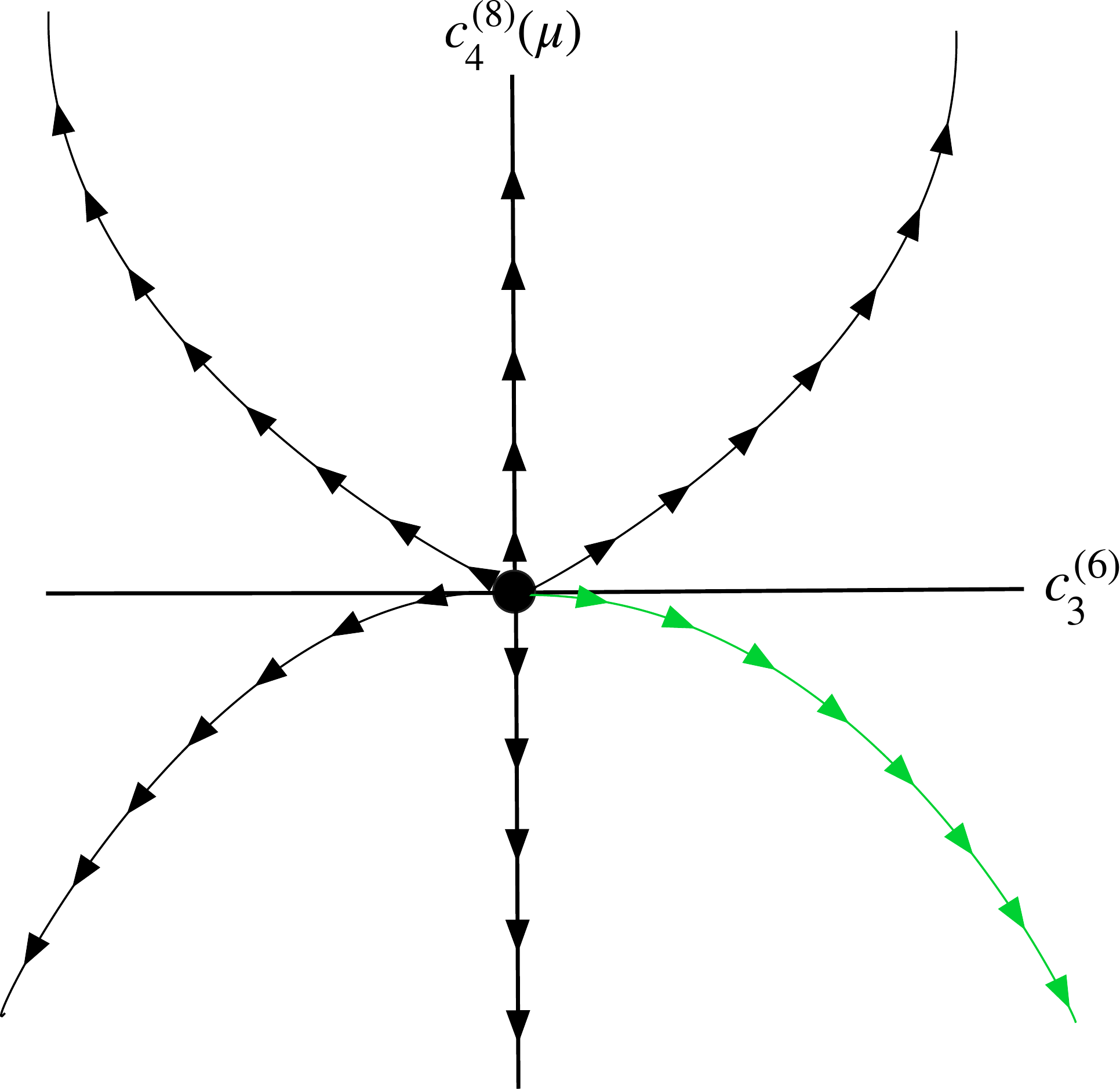}
\end{minipage}
\begin{minipage}{.5\textwidth}
\includegraphics[width=8cm]{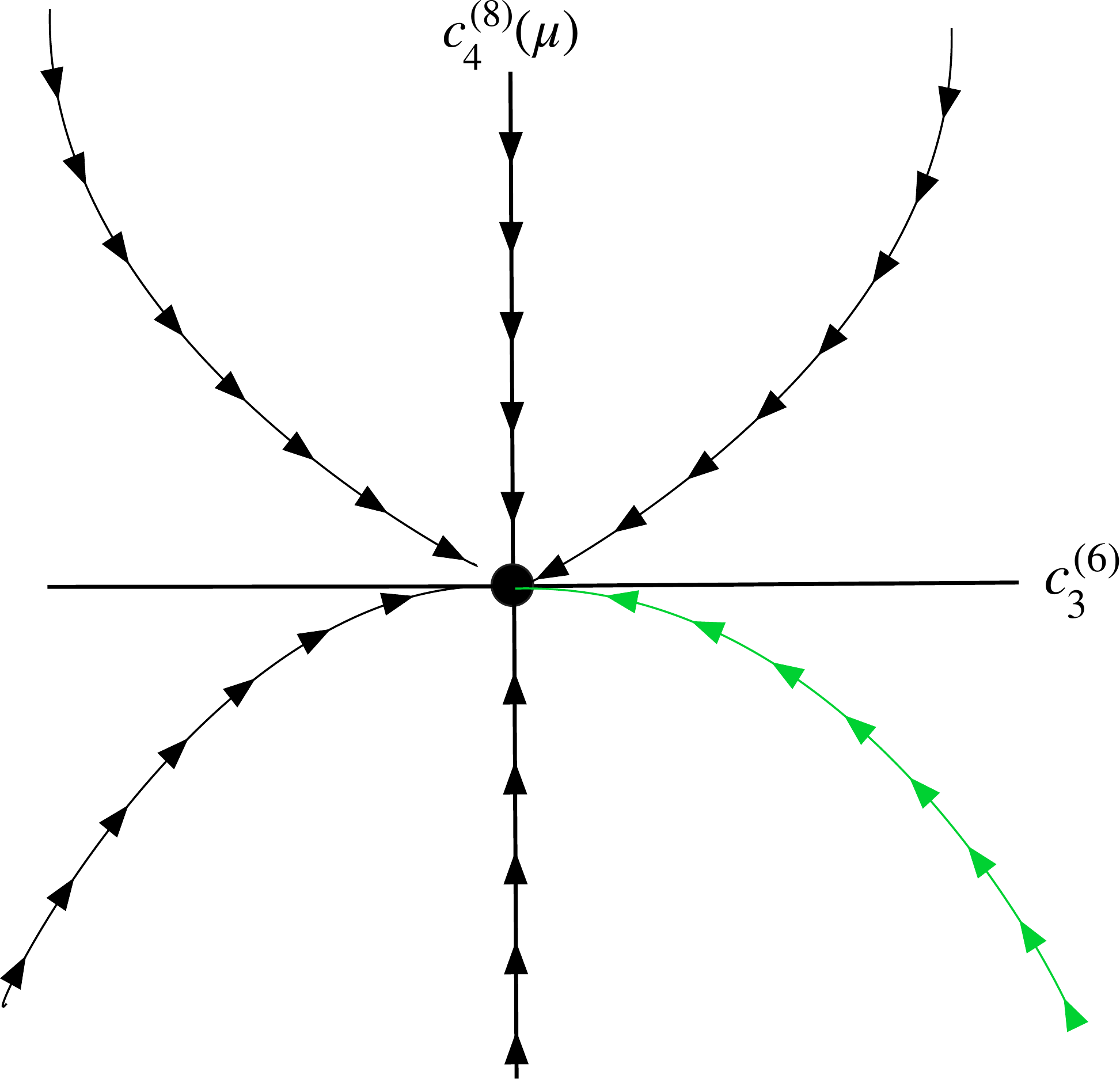}
\end{minipage}
\caption{\small Left: RG flows for ${\cal L}$ in \eq{L2468} at $m^2 = \l =0$ limit, in $d = 3$. The phase admits a Gaussian fixed point, which is UV, while the couplings both diverge in the IR.
Right: RG flows of \eq{L2468} at $m^2 = \l =0$ limit, in $d = 5$. The phase admits a Gaussian fixed point, which is IR, hence both couplings decrease for deceasing renormalization scale.
\label{PDc8d35}}
\end{figure}

In $d=4$ the $\b$-function of $c_3^{(6)}$ vanishes and the coupling does not run: $c_3^{(6)}(\m) \equiv  c_3^{(6)}$.
The other $\b$-function is
\be\label{c63muc63}
\b_{  c_4^{(8)}} =  \frac{7 ( c_3^{(6)})^2 }{16 \pi^2}\, .
\ee
This distinguishes two cases:
\begin{itemize}
\item $ c_3^{(6)} = 0$\\
For $\ve=0$, $\b_{  c_4^{(8)}}$
vanishes independently of $c_4^{(8)}$. Neither $ c_3^{(6)}$ nor $ c_4^{(8)}$ runs and the phase diagram has a WF-line 
with points labelled by $c_4^{(8)}$ ending on a Gaussian fixed point where ${c_{3\bullet}^{(6)}} = { c_{4\bullet}^{(8)}}=0$.
The potential in \eq{wtV1lc6c8} becomes
\be
V^{(8)} = - \frac{ c^{(8)}_4}{8! \L^4} \phi^8 \, \nonumber
\ee
and determines $\mathbb{Z}_2$-symmetric phase that is stable or unstable when $ c_4^{(8)} < 0$ or $ c_4^{(8)} > 0$ respectively.
\item $ c_3^{(6)} \ne 0$\\
$\b_{ c_4^{(8)}}$ is strictly non-zero. The RG flow does not move towards or away from a fixed point. 
Nevertheless, $ c_4^{(8)}$ has a non-trivial flow at a constant distance equal to $c_3^{(6)}$ from the WF-line
that forces it to vanish in the IR at some scale $\m_L$ and to diverge in the UV.
The corresponding RG equation is given by 
\be\label{rgc860}
c_4^{(8)}(\m) =  c_4^{(8)} + \frac{7 ( c_3^{(6)})^2 }{16 \pi^2} \ln \frac{\m}{\m_{\rm R}} \, 
\ee
and its running is depicted in \fig{c84mu1}. It vanishes at the scale 
\be
\m_L = e^{ -  \frac{16 \pi^2  c_4^{(8)} }{ 7 ( c_3^{(6)})^2 } }  
\nonumber\, ,
\ee
while at scales $\m <(>) \m_L$ it admits negative (positive) values.
\end{itemize}
We can now combine \eq{mphyc6c8} with \eq{c63muc63} and \eq{rgc860} 
to evaluate the running of $ m_\phi^2$ as a function of $\m$ in $d = 4$, to obtain
\be
m_\phi^2 (\m) = \frac{7}{10} \frac{( c_3^{(6)})^2 }{ | c_4^{(8)}(\m) | } v^2\, . \nonumber
\ee
This scale dependence is depicted in \fig{muc6c8d4}. The case here is particular since the behaviour of 
$m_\phi^2$ is the opposite to that of the mass in \fig{mulc6d4}. Recall that
%%%%%%%%%%%%%%%%%%%%%%%%%%%%%%%%%%%%%%%
%\FloatBarrier
%%%%%%%%%%%%%%%%%%%%%%%%%%%%%%%%%%%%%%%
%
in the latter case there was a high but not infinite scale (the UV Landau pole) where $\l(\m)$ and as a consequence the scalar mass, became infinite. 
Instead, \fig{c84mu1} shows that $ c_4^{(8)}(\m)$ decreases for decreasing scale and when $\m$ reaches a small but non-zero value the coupling vanishes
and the scalar mass blows up. Therefore here we have an "inverse" or IR Landau pole.
%%%%%%%%%%%%%%%%%%%%%%%%%%%%%%%%%%%%%%%
%%%%%%%%%%%%%%%%%%%%%%%%%%%%%%%%%%%%%%%
Let us now turn to $d \ne 4$ where $\b_{ c_3^{(6)}}$ is not necessarily zero.
Then $c_3^{(6)}$ is a function of $\m$ and the solution to \eq{bc63bc84} admits a non-trivial RG flow given by
\be\label{rgc68}
 c_3^{(6)}(\m) =  c_3^{(6)} \left( \frac{\m}{\m_{\rm R}}  \right)^{-2\ve}\, .
\ee
Notice that the above relation matches \eq{c6mue} in the limit $ \wt \l \to 0$.
For the other coupling, \eq{bc63bc84} suggests the RG flow 
\be\label{rgc86e1}
 c_4^{(8)}(\m) = c_4^{(8)}(\m_{\rm R}) \left( \frac{\m}{\m_{\rm R}}  \right)^{-3\ve} + \frac{7 (  c_3^{(6)})^2 }{48 \pi^2 \ve} \left[ 1 - \left( \frac{\m}{\m_{\rm R}}  \right)^{-3\ve} \right]
\ee
and choosing for simplicity the renormalization prescription $ c_4^{(8)}(\m_{\rm R}) =  c_4^{(8)} + \frac{7 (  c_3^{(6)})^2 }{48 \pi^2 \ve}$, we get 
\be\label{rgc86e2}
 c_4^{(8)}(\m) =  c_4^{(8)} \left( \frac{\m}{\m_{\rm R}}  \right)^{-3\ve} + \frac{7 ( c_3^{(6)})^2 }{48 \pi^2 \ve} \, .
\ee
Regarding the fixed points things are simpler here than in the previous case since, the only point that eliminates both $ \b_{ c_3^{(6)}} $ and $\b_{ c_4^{(8)}}$ simultaneously is
\be
c_{3\bu}^{(6)} = 0  \hskip .3cm {\rm and}  \hskip .3cm c_{4\bu}^{(8)} = 0 \, .
\ee    
This means that $O^{(6)}$-$O^{(8)}$ system admits only a Gaussian fixed point.
All the above information is depicted in \fig{d4c6c81} for $d=4$ and in \fig{PDc8d35} for $d=3,5$.
To complete the picture we look at the behaviour of the operators $O^{(6)}_3$ and $O^{(8)}_4$ with respect to the fixed points, as we move from UV to IR.
We need to know also the nature of the fixed points and this is given in the following: 
\be\label{wtc638cases}
\frac{\partial \beta_{ c_3^{(6)}}}{\partial c_3^{(6)}} = -2 \ve \, 
\begin{cases}
     \bullet : -2 \ve & \begin{cases}
                            d=4:0\,\, (\text{$ c_3^{(6)}$ does not run}) \\
                            d=3:-2 < 0 \rightarrow {\rm UV} \\
                            d=5: +2 >0 \rightarrow {\rm IR}
                            \end{cases}
                             \\
     \star : -2 \ve & \begin{cases}
                           d=4: - \\
                           d=3: - \\
                           d=5: - 
                          \end{cases}
   \end{cases}
\ee
%%%%%%%%%%%%
\be\label{wtc84cases}
\frac{\partial \beta_{ c_4^{(8)}}}{\partial c_4^{(8)}} = -3 \ve\,
\begin{cases}
     \bullet : -3 \ve & \begin{cases}
                            d=4:0\,\, \text{( G exists only for $ c_3^{(6)}, c_4^{(8)} = 0$. Otherwise no fixed-points)} \\
                            d=3:-3 < 0 \rightarrow {\rm UV} \\
                            d=5: +3 >0 \rightarrow {\rm IR}
                            \end{cases}
    \\
     \star : -3 \ve & \begin{cases}
                           d=4:-  \,\, \text{( WF-line only for $ c_3^{(6)} = 0$ and $ c_4^{(8)} \ne 0$. There is no flow)}\\
                           d=3: - \\
                           d=5: - 
                          \end{cases}
\end{cases}
\ee
As we have already mentioned, 
\be
\Delta_{O^{(6)}} = d_{O^{(6)}} = 4 -3\ve \,\,\,{\rm and} \,\,\, \Delta_{O^{(8)}} = d_{O^{(8)}} = 4 - 4\ve \, .
\ee
Therefore, we have for the 1-loop exponent that determines the nature of an operator:
%%%%%%%%%%%%   TABLE   %%%%%%%%%%
\hskip 2cm
\begin{center}
\begin{tabular}{|c|c|c|c|}
\hline 
$$ & $\Delta_O-d$ & $\bullet$ & $\star$ \\
\hline \hline   
%%%%
$O^{(6)}_3$  & $-2\ve $ &$-2 \ve$ & $-$\\ \hline
$O^{(8)}_4$  & $- 3\ve 
$ &$ - 3 \ve $ & $-$\\ \hline
\end{tabular}
%\center{TABLE 6. The 1-loop exponent that determines the nature of an operator following Table 2.}
\end{center}
%%%%%%%%%%%%%%%%%%%%%%%%%%%%
and this distinguishes the following cases:
%%%%%%%%%%%%   TABLE   %%%%%%%%%%
\hskip 2cm
\begin{center}
\begin{tabular}{|c|c|c|c|}
\hline 
$$ & $d=4$ & $d=3$ & $d=5$ \\
\hline \hline   
%%%%
$O^{(6)}_3$  & \text{truly marginal(for $ c_3^{(6)} \ne 0$)}  &$ \bullet: {\rm rel} \hskip .5cm  \vline \hskip .5cm \star: {-}$ & $ \bullet: {\rm irrel} \hskip .5cm  \vline \hskip .5cm \star: {-}$\\ \hline
$O^{(8)}_4$  & $ {\rm irrelevant} $(from \eq{rgc860}) & $\bullet: {\rm rel} \hskip .5cm  \vline \hskip .5cm \star: {-} $ & $ \bullet: {\rm irrel} \hskip .5cm  \vline \hskip .5cm \star: {-} $\\ \hline
\end{tabular}
%\center{TABLE 7. Classifications of the operators at the fixed points according to Table 5.}
\end{center}
%%%%%%%%%%%%%%%%%%%%%%%%%%%%
As a last comment, recall that unitarity sets energy bounds when a HDO is inserted in the Lagrangian. 
Such a constraint can be calculated following \cite{Luty} as in Sect. \ref{onlyphi} and for $n_1 = 8$ and $k_1 = \frac{n_1}{2} = 4$, the bound
\be\label{unE4}
E_{4} \le \frac{16\sqrt{3} \pi^{5/4}}{\left( c_4^{(8)} \right)^{1/4}} \L \approx 116 \frac{\L}{\left( c_4^{(8)} \right)^{1/4}} \, .
\ee 
is obtained.
%%%%%%%%%%%%%%%%%%%%%%%%%%%%%%%%%%%%%%%
%%%%%%%%%%%%%%%%%%%%%%%%%%%%%%%%%%%%%%%
%
%
%%%%%%%%%%%%%%%%%%%%%%%%%%%%%%%%%%%%%%%
\FloatBarrier
%%%%%%%%%%%%%%%%%%%%%%%%%%%%%%%%%%%%%%%
%

%---------------------------------------------------------------------------------------------------------------------------------------------------
\section{Conclusion}\label{Concl}
%---------------------------------------------------------------------------------------------------------------------------------------------------

We considered the example of a scalar Generalized Effective Field Theory, which is an Effective Field Theory where the couplings 
associated with Higher Dimensional Operators run with the renormalization scale in an independent fashion.
Such extensions are non-renormalizable, possess in general Ostrogradsky ghosts and superluminal propagation of information.
Despite of all these drawbacks they can be meaningful in certain regimes of their phase diagram. 
A necessary condition for this is that the infinite tower of HDO can be truncated. In order to do this in a systematic way we
parametrized them by inverse powers of an auxiliary scale $\L$ and truncated the series at fixed powers of $\L$.
In this context we showed how the GEFT can be made ghost-free and subsequently finite.
The couplings of the HDO contribute to the $\b$-function of the marginal coupling in GEFT which could be observable. 

An interesting corner of effective field theories is the massless limit, the reason being that to the extent that the breaking of scale invariance
induced by the presence of $\L$ is spontaneous, it is connected to the spontaneous breaking of the internal symmetry in the broken phase.
Then we could have a Coleman-Weinberg like mechanism \cite{Coleman} where instead of gauge fields, the breaking is triggered by the HDO.
Indeed, the phase diagrams that we have seen, do possess such phases.

We also extended the analysis to dimensions $d\ne 4$, exploiting the flexibility of the $\ve$-expansion.
In this case the phase diagram has more structure as Wilson-Fisher fixed points or lines start to appear where
the system becomes scale invariant or approximately scale invariant, given that our analysis was restricted to 1-loop.

%---------------------------------------------------------------------------------------------------------------------------------------------------
%\section{Acknowledgement}
%---------------------------------------------------------------------------------------------------------------------------------------------------
%We would like to thank .......
%---------------------------------------------------------------------------------------------------------------------------------------------------

%---------------------------------------------------------------------------------------------------------------------------------------------------
\begin{appendices}
%---------------------------------------------------------------------------------------------------------------------------------------------------

%---------------------------------------------------------------------------------------------------------------------------------------------------
\section{Scaleless integrals}\label{Sc.In.}
%---------------------------------------------------------------------------------------------------------------------------------------------------

Here the calculation of divergent, scaleless integrals is reviewed. 
Our interest is restricted to the scalar integrals $A_0$ and $B_0$. 

Let as start with $A_0$ which comes from Tadpole diagrams. As an example consider a two-point diagram of the form
%---------------------------------------------------------------------------------------------------------------------------------------------------
\vskip .5cm
\begin{center}
\begin{tikzpicture}[scale=0.7]
\draw [dashed] (0,0)--(1.8,0);
\draw [dashed] (0,0.9) circle [radius=0.9];
\draw [dashed] (-1.8,0)--(-0,0);
\node at (0,0.6) {$k$};
\draw [->]  (0.45,0.4) arc [start angle=-45, end angle=-135, radius=0.6cm];
\node at (-1.2,-0.4) {$p$};
\node at (1.2,-0.4) {$p$};
\draw [->]  (1.7,0.25)--(1,0.25);
\draw [->]  (-1.7,0.25)--(-1,0.25);
\node at (3.2,0) {$= i {\cal M}^1$};
\end{tikzpicture}
\end{center}%---------------------------------------------------------------------------------------------------------------------------------------------------
with only one momentum running the loop. In addition, we assume that there is a mass scale
$m_0^2$ that on-shell forces the external momentum to be $p^2 = m_0^2$. Then the above diagram in $d$-dimensions, evaluates to
\be\label{M1}
{\cal M}^1 \simeq \int{ \frac{d^d k}{(2 \pi)^d i}}  \frac{1 }{ k^2 - m_0^2} = A_0(m_0^2) \, ,
\ee
which is equal to the known result  in Dimensional Regularization
\be\label{M1.exp}
A_0(m_0^2) = m_0^2 \Bigl [  \frac{2}{\ve} + \ln \frac{\m^2}{m_0^2} + 1 \Bigr] \, .
\ee
Now the question that we want to address is how the integral of \eq{M1} is calculated in the absence of a scale or, in other words, when $m_0 \to 0$. 
The result is a scaleless integral of the form
\be\label{M1.0}
A_0(0) = \int{ \frac{d^d k}{(2 \pi)^d i}}  \frac{1 }{ k^2} \, .
\ee
Of course if we take the $m_0^2\to 0$ limit of \eq{M1.exp} the integral is zero. 
Nevertheless, let us look at the behaviour of $A_0(0)$ performing an explicit calculation of \eq{M1.0}.
The first step is to move to Euclidean space and remove $i$ from the above expression. Then, we get that
\bea
A_0(0) &=& \int{ \frac{d^d k_E}{(2 \pi)^d }}  \frac{1 }{ k_E^2}  \Rightarrow \nonumber\\ 
A_0(0) &\sim& \Omega_d \int_0^\infty d k_E \frac{k_E^{d-1}}{k_E^2} \, ,
\eea
where we have performed the integration in the $(d-1)$-spherical surface, extracting the solid angle
\be
\Omega_d = \frac{2 \pi^{\frac{d}{2}}}{\Gamma(\frac{d}{2})} \, . \nonumber
\ee
Now, this scaleless integral gives
\bea\label{M1.1}
A_0(0) &\sim&\Omega_d \int_0^\infty d k_E k_E^{d-3} \Rightarrow \nonumber\\ 
A_0(0) &\sim&\Omega_d \Bigl [ \int_0^M d k_E k_E^{d-3}  + \int_M^\infty d k_E k_E^{d-3}    \Bigr] \, ,
\eea
and notice that it does not converge in any dimension and for that reason we inserted an arbitrary scale $M$ to split the integral into its IR and UV parts.
The next step is to calculate \eq{M1.1} close to $4$-dimensions. To do so we take $d = 4 - \ve$.
Here we should be careful since the way we move away from $4-d$ is different for the IR and UV limits. 
For the former case we define $\ve_{\rm IR} <0$ with $d= 4 - \ve_{\rm IR}$
and for the latter $\ve \equiv \ve_{\rm UV} >0$ with $d= 4 - \ve_{\rm UV}$.
With these in mind we obtain
\bea
A_0(0) &\sim& \Omega_d \Bigl [ \frac{M^{2-\ve_{\rm IR}}}{2-\ve_{\rm IR}}  - \frac{M^{2-\ve_{\rm UV}}}{2-\ve_{\rm UV} }   \Bigr] \Rightarrow \nonumber\\
A_0(0) &\sim& M^2 \Omega_d \Bigl [ \frac{M^{-\ve_{\rm IR}}}{2-\ve_{\rm IR}}  - \frac{ M^{-\ve_{\rm UV}}}{2-\ve_{\rm UV}}    \Bigr]\, , \nonumber
\eea
and expanded for $\ve_{\rm IR},\ve_{\rm UV} <<1$ the above expression becomes
\be
A_0(0) \sim M^2 \Omega_d \Bigl [ \frac{\ve_{\rm IR}}{2} \Bigl (\frac{1}{2} - \ln M \Bigr ) -  \frac{\ve_{\rm UV}}{2} \Bigl (\frac{1}{2} - \ln M \Bigr )   \Bigr] \, . \nonumber
\ee  
It is easy to check that for $\ve_{\rm IR} \ne \ve_{\rm UV}$, $\ln M$ is not canceled between the two parts of the parenthesis. 
This indicates that $A_0$ depends on the way that we split the integrals into an IR and a UV part which should not be the case. Therefore the only solution is 
$\ve_{\rm IR} = \ve_{\rm UV}$ when the integral is identically zero. This shows that $A_0$ is both IR- and UV-divergent and as a consequence
\be\label{M1.f}
A_0(0) = 0 \, .
\ee

Let us now turn to the scaleless integral with two propagators, the $B_0$-integral. Consider the two-point function diagram
 %-------------------------------------
\vskip .5cm
\begin{center}
\begin{tikzpicture}[scale=0.7]
\draw [dashed] (1,0)--(3,0);
\draw [dashed] (0,0) circle [radius=1];
\draw [dashed] (-3,0)--(-1,0);
\node at (0,1.02) {$ > $};
\node at (0,-0.95) {$ < $};
\draw [->]  (-1.9,0.2)--(-1.2,0.2);
\node at (-1.7,-0.4) {$p$};
\node at (1.7,-0.4) {$p$};
\node at (0,0.5) {$k+p$};
\node at (0,-0.4) {$k$};
\draw [<-]  (1.9,0.2)--(1.2,0.2);
%\node at (2.6,0.2) {$-p$};
\node at (4.5,0) {$=\, \, i{\cal M}^2$};
\end{tikzpicture}
\end{center}
which in $d$-dimensions, when the masses are non-zero, reads 
\be\label{M2}
{\cal M}^2 \simeq \int{ \frac{d^d k}{(2 \pi)^d i}}  \frac{1 }{ k^2 - m_{1,0}^2}  \frac{1 }{ (k+p)^2 - m_{2,0}^2} = B_0(p,m_{1,0},m_{2,0}) \, .
\ee
Evaluating it in DR, the known result
\be\label{M2.0}
B_0(p,m_{1,0},m_{2,0}) = \frac{2}{\ve} + \int_0^1 dx \ln \frac{\m^2}{\Delta_B}\, ,
\ee
with $\Delta_B = - p^2 x (1-x) + m_{1,0}^2 + m_{2,0}^2$ is obtained.
Similarly to the previous case, we want to calculate this integral in the limit $m_{0,1},m_{0,2}^2 \to 0$.
Following the same steps with the $A_0$ calculation, we go to Euclidean space for general $d$ and then $B_0$ becomes
\bea
B_0(p,0,0) &=&  \int{ \frac{d^d k_E}{(2 \pi)^d }}  \frac{1 }{ k_E^4} \Rightarrow \nonumber\\
B_0(p,0,0) &\sim&  \Omega_d \int_0^\infty d k_E \frac{k_E^{d -1} }{ k_E^4} \Rightarrow \nonumber\\
B_0(p,0,0) &\sim&  \Omega_d \int^M_0 d k_E k_E^{d -5} + \Omega_d \int^\infty_M d k_E k_E^{d -5} \nonumber
\eea
where we have extracted the solid angle $\Omega_d$ and inserted an arbitrary scale $M$ to separate the IR from the UV divergent part. This integral does not converge in any dimension.\\
Close to $d=4$ demands that $d= 4 - \ve_{\rm IR}$ with $\ve_{\rm IR} <0$ and $d= 4 - \ve_{\rm UV}$ with $\ve_{\rm UV} >0$.
The integral then evaluates to
\bea
B_0(p,0,0) &\sim&   \Omega_d \Bigl [ - \frac{M^{-\ve_{\rm IR}}}{\ve_{\rm IR}}  + \frac{M^{-\ve_{\rm UV}}}{\ve_{\rm UV}}  \Bigr] \Rightarrow \nonumber\\
B_0(p,0,0) &\sim& \Omega_d \Bigl [ \ln M - \frac{1}{\ve_{\rm IR}} - \ln M + \frac{1}{\ve_{\rm UV}}  \Bigr] \Rightarrow \nonumber\\
B_0(p,0,0) &\sim& \Omega_d \Bigl [ \frac{1}{\ve_{\rm UV}}  - \frac{1}{\ve_{\rm IR}}  \Bigr]\, , \nonumber
\eea
which is independent of the arbitrary scale $M$. This shows that for the above integral, IR and UV divergences can be split from each other. 
This is summarized in the following:
\be\label{M2.f}
\Bigl [ B_0(p,0,0) \Bigr]_{\rm IR} = \frac{1}{16 \pi^2} \frac{2}{\ve_{\rm IR}} \hskip .3cm {\rm and} \hskip .3cm \Bigl [ B_0(p,0,0) \Bigr]_{\rm UV} = \frac{1}{16 \pi^2} \frac{2}{\ve_{\rm UV}} \, .
\ee
We finish by a discussion of the role of these scaleless integrals in our computations.
First of all, they appear in the computations involved in the operator insertion method because we chose to consider
any operator, including the mass operator, as a deformation of the kinetic term. As a result the propagators in the loops are
massless and the resulting integrals scaleless. Now according to our discussion about operator insertions, the method is 
sensitive to particular UV divergences. This means that when a scaleless integral is encountered during operator insertion operations, 
if possible, its UV divergent part must be extracted. 
We saw above that this is not possible for $A_0(0)$ which is always identically zero, but it is possible for $B_0$
that has a non-trivial UV divergent part. 
On the other hand when we renormalize \eq{LHDO} directly as in Sect. \ref{gairo} there is no reason to split the IR from the UV part
which means that in such a case we must use the exact result, that is $B_0(p,0,0)=0$.

%---------------------------------------------------------------------------------------------------------------------------------------------------
\section{Review of the ${\cal L}^{(2)}+{\cal L}^{(4)}$ (usual $\phi^4$) theory}\label{2-4}
%---------------------------------------------------------------------------------------------------------------------------------------------------

In this section we review the renormalizable $\phi^4$-theory, rederiving textbook results \cite{Peskin}, but
using for renormalization the operator-insertion method.
The Lagrangian is 
\be\label{L4fi0}
 {\cal L} = -\frac{1}{2} \phi \Box \phi  - c^{(2)} O^{(2)} - c^{(4)} O^{(4)} \, ,
\ee
where $ O^{(2)}  = \frac{\phi^2}{2}$ with coupling $ c^{(2)} = m^2 $ and $ O^{(4)}  = \frac{\phi^4}{4!}$ with coupling $ c^{(4)} = \l $.
The classical dimension of $\phi$ in $d=4-\ve$ dimensions is $[\phi]=d_\phi=\frac{d-2}{2}$.
The classical dimensions of the inserted operators are
\be\label{cdO2O4}
d_{O^{(2)}} = 2 - \ve \,\,\, {\rm and} \,\,\, d_{O^{(4)}} = 4 - 2\ve \, .
\ee
For this model we face two $\b$-functions which, according to \eq{totalbeta}, are given by
\be\label{bc2a}
\b_{c^{(2)}} =  -2 c^{(2)} + \G_{O^{(2)}} \, c^{(2)} +  \G_{2 2 2} \, c^{(2)} c^{(2)} + \G_{2 4 4} \, c^{(4)} c^{(4)}
\ee
for $l = 2$ and by 
\be\label{bc4a}
\b_{c^{(4)}} =  -\ve c^{(4)}  + \frac{ m^2}{2} \G_m^{6} + \G_{O^{(4)}} \, c^{(4)} +  \G_{4 2 2} \, c^{(2)} c^{(2)} + \G_{4 4 4} \, c^{(4)} c^{(4)}
\ee
for $l=4$.
The anomalous dimensions in the above $\b$-functions are evaluated through the procedure described in Sect. \ref{sobf}.
As a first step we consider \eq{GnGnk.m.01} for $n=1$ and the insertion of only one field, $\phi(y)$. Then renormalization gives
\be
\langle \phi(x_1) \phi(y) \rangle_0 = Z_\phi \langle \phi(x_1) \phi(y) \rangle \, ,
\ee
where ${\cal L}_{\rm int,0} = - c_0^{(2)} O_0^{(2)} - c_0^{(4)} O_0^{(4)}$ and $Z_\phi$ defined in \eq{Phiclr}. 
Performing the expansion of the exponential, at order ${\cal O}(c_0^{(2)}, c_0^{(4)})$, we get
\be\label{f1f1}
\langle \phi(x_1) \phi(y) \frac{-i m^2 \phi^2(z)}{2} \rangle_0 + \langle \phi(x_1) \phi(y) \frac{-i \l \phi^4(z)}{4!}  \rangle_0 = Z_\phi \langle \phi(x_1) \phi(y) \rangle \, .
\ee
The contractions on the first term on the l.h.s result in a finite propagator from $x_1$ to $y$. 
The second term on the same side gives the divergent diagram
\be\label{M00}
%-------------------------------------
\begin{tikzpicture} [scale=0.9]
\draw [dashed] (6,0)--(7.6,0);
\draw [dashed] (7.6,0.5) circle [radius=0.5];
\draw [dashed] (7.6,0)--(9.1,0);
\node at (7.6,0.04) {$\times $};
\node at (7.6,0.04) {$\times $};
%\draw [thick] [fill=black] (3.7,0) circle [radius=0.1];
\end{tikzpicture}
%------------------------------------- 
\ee
where the mark on the loop indicates the spot where the operator is inserted.
This is the two-point Tadpole integral $A_0$ which here is scaleless, so according to Appendix \ref{Sc.In.} it is equal to zero.
Since also the r.h.s of \eq{f1f1} gives a finite propagator from $x_1$ to $y$, equivalence holds if $Z_\phi = 1$ which gives the known result, $\d_\phi = 0$.

Next we focus on $\b_{c^{(2)}}$. For its linear part the relevant Green's function is given by \eq{flOl} for $n=l=2$:
\be
\langle \phi(x_1) \phi(x_2) O^{(2)} (y) \rangle_0 = Z_{ O^{(2)} } \langle \phi(x_1) \phi(x_2) O^{(2)}(y) \rangle \, ,
\ee
with $Z_{ O^{(2)} }$ given by \eq{ZOdo1}. It can be rewritten as
\be\label{GZ1112}
\langle \phi(x_1) \phi(x_2) O^{(2)} (y) \rangle_0 - \d_{ O^{(2)} } \langle \phi(x_1) \phi(x_2) O^{(2)}(y) \rangle =  \langle \phi(x_1) \phi(x_2) O^{(2)}(y) \rangle \, .
\ee
According to the discussion below \eq{GnGnk.m1}, ${\cal L}_{\rm int,0} = - c_0^{(4)} O_0^{(4)}$, therefore the l.h.s is
\be\label{GZ11}
 {\cal N}^{-1}  \langle 0 | T[ \phi_0(x_1) \phi_0(x_2) \frac{\phi_0^2(y)}{2}  e^{i\int d^d z \frac{-\l_0}{4!} \phi_0^4(z)} ] | 0 \rangle - \d_{ O^{(2)} } \langle \phi(x_1) \phi(x_2) \frac{\phi_0^2(y)}{2} \rangle \, .
\ee 
Expanding the exponential to ${\cal O}(\l_0)$ and performing contractions we get, at 1-loop, the diagrams
\be
%-------------------------------------
\begin{tikzpicture} [scale=0.9]
\draw [dashed] (6,0)--(7.6,0);
\draw [dashed] (7.6,0.5) circle [radius=0.5];
\draw [dashed] (7.6,0)--(9.1,0);
\node at (7.6,1) {$\times $};
%\draw [thick] [fill=black] (3.7,0) circle [radius=0.1];
\draw [dashed] (10,0)--(11.5,0);
\node at (9.7,0) {$-$};
%\node at (13.8,0) {$ \cdot ( \d^{11}_O + \d_\phi) $};
\node at (10.7,0.01) {$\times $};
\node at (10.7,0.01) {$\times $};
\end{tikzpicture}
%------------------------------------- 
\nonumber
\ee
The first diagram looks like a usual scalar tadpole but it is in fact a three-point function contributing to $G^{(2)}(x_1,x_2)$. 
The second diagram corresponds to a counter-term insertion. 
Again, the mark on the loop indicates the spot where the operator is inserted.
Now recall that the propagators in the loop are massless, so the first diagram is a scaleless $B_0$-integral given by \eq{M2.f}. Extracting only its UV part, finiteness requires 
\be\label{dO11}
\d_{ O^{(2)} }  = - \frac{\l}{16 \pi^2} \frac{1}{\ve} \, ,
\ee
while using \eq{An.d.m1} or its 1-loop version below, we get
\be\label{GaO11}
\G_{ O^{(2)} }  = \frac{\l}{16 \pi^2}  \, .
\ee
For the non-linear part of \eq{bc2a} there are two possible contributions. 
The first will determine $\G_{222}$ and corresponds to $l=r=s=2$ in \eq{flOlOl}. The second gives $\G_{244}$ using \eq{fnOlOl} for $n = 2$ and $k'=r=s=4$ with $q_{(2,4)}=2$. Both anomalous dimensions vanish in this case.
With each part of $\b_{c^{(2)}}$ determined, \eq{bc2a} takes the form
\be\label{bc2b}
\b_{c^{(2)}} = -2 c^{(2)} + \frac{\l}{16 \pi^2} c^{(2)} \equiv  -2 m^2 + \frac{\l m^2}{16 \pi^2} \, .
\ee

\begin{figure}[!htbp]
\centering
\includegraphics[width=8cm]{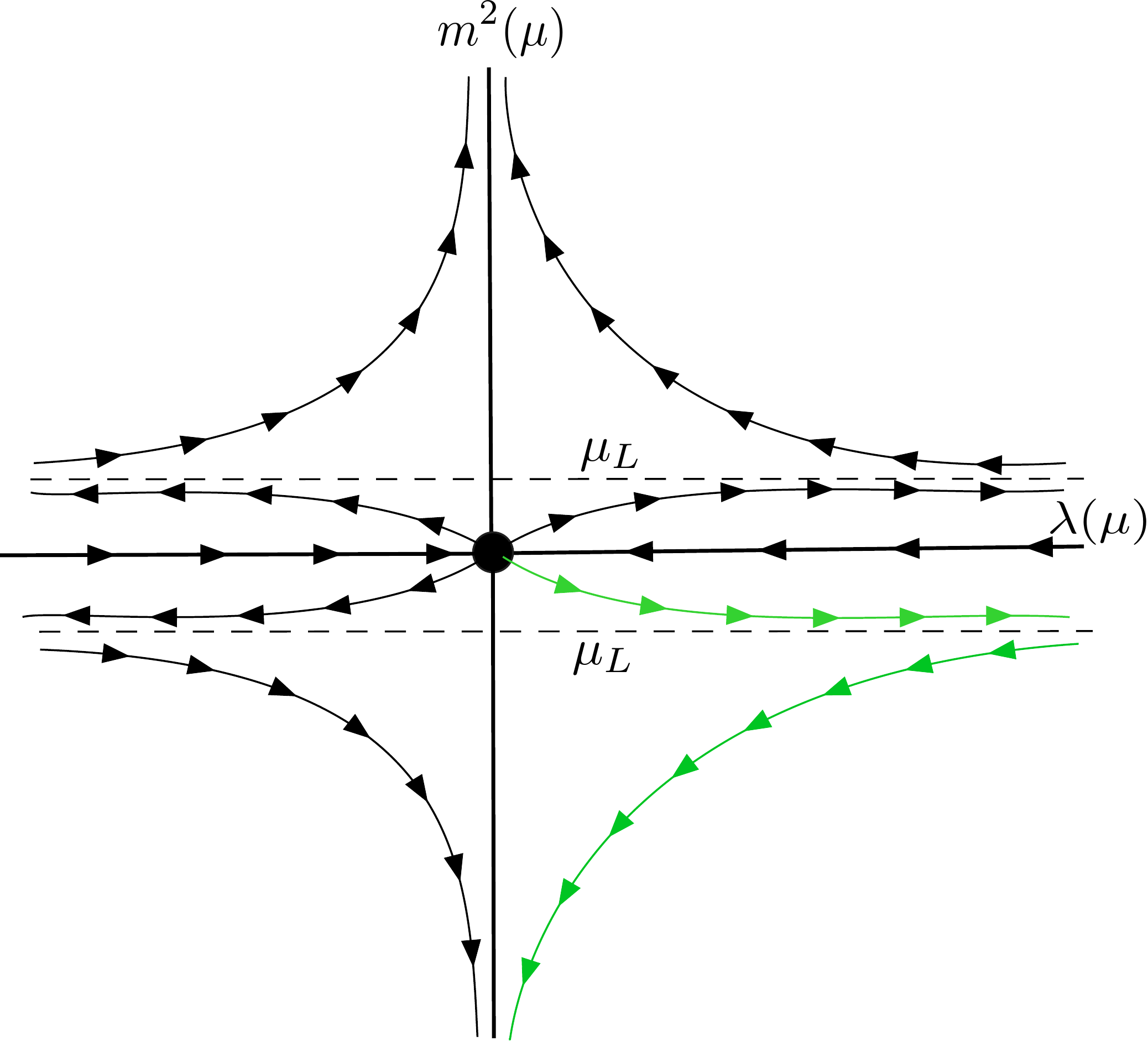}
\caption{\small RG flows for \eq{L4fi0}, in $d=4$. The $ \l >0 , m^2 < 0$ (green) flow corresponds to the $\mathbb Z_2$ broken phase.
\label{Ph.D.e0}}
\end{figure}

\begin{figure}[!htbp]
\begin{minipage}{.5\textwidth}
\includegraphics[width=8cm]{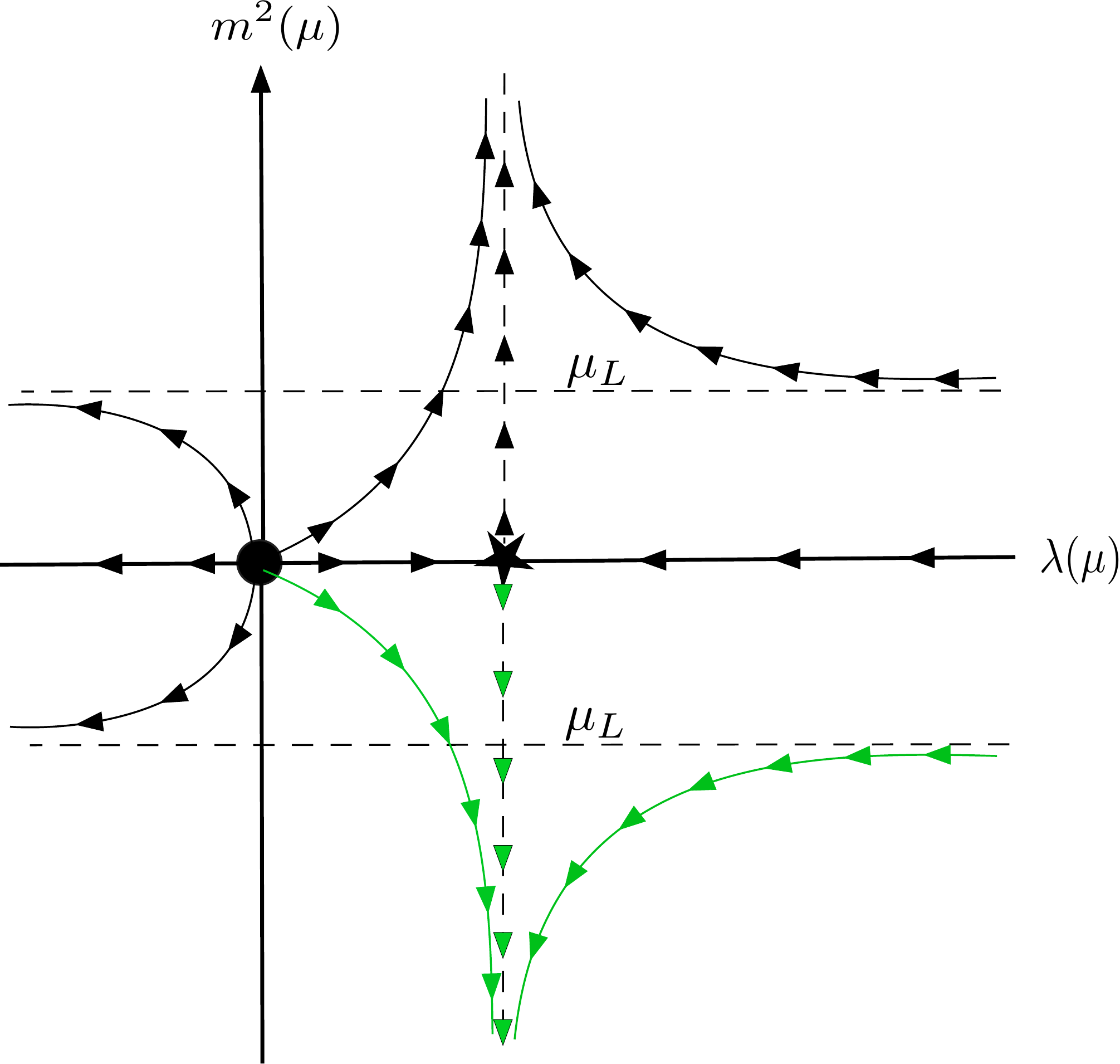}
\end{minipage}
\begin{minipage}{.5\textwidth}
\includegraphics[width=8cm]{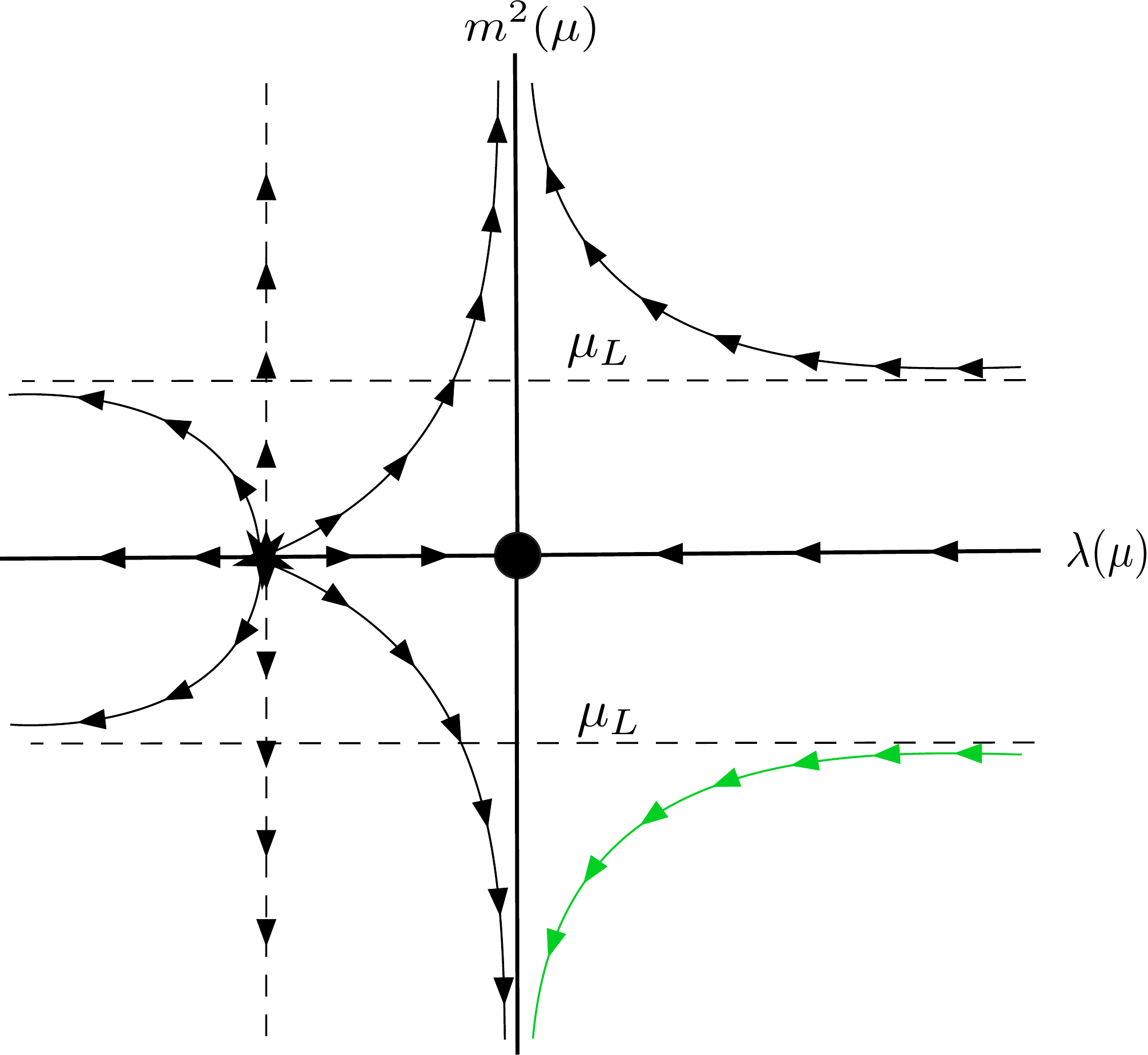}
\end{minipage}
\caption{\small Left: RG flows for \eq{L4fi0}, in $d=3$. The $ \l >0 , m^2 < 0$ (green) flow corresponds to the broken phase. 
Here, $\star$ denotes the WF fixed point where scale invariance is restored while the system is still interacting.
Right: RG flows for \eq{L4fi0}, in $d=5$. The broken phase has a Landau pole brach where $\l$ 
diverges as $\m$ reaches $\m_L$, similarly to the $\ve=0$ case. 
\label{PDepos}}
\end{figure}

Let us now move on to $\b_{c^{(4)}}$.
Since the Lagrangian does not include a dim-6 operator, the second term of \eq{bc4a} is absent.
The linear part on the other hand, is evaluated through \eq{flOl} for $n=l=4$ and the corresponding expectation value is
\be
\langle \phi(x_1) \phi(x_2) \phi(x_3) \phi(x_4) O^{(4)}(y) \rangle_0 = Z_{ O^{(4)} } \langle \phi(x_1) \phi(x_2) \phi(x_3)  \phi(x_4) O^{(4)}(y) \rangle  \, , 
\ee
or using \eq{ZOdo1},
\be\label{GZ21}
\langle \phi(x_1) \phi(x_2) \phi(x_3) \phi(x_4) O^{(4)}(y) \rangle_0 - \d_{ O^{(4)} } \langle \phi(x_1) \cdots  \phi(x_4) O^{(4)}(y) \rangle =  \langle \phi(x_1) \cdots  \phi(x_4) O^{(4)}(y) \rangle  \, .
\ee
Now the bare part of the above expression involves ${\cal L}_{\rm int,0} = - c_0^{(2)} O_0^{(2)} $.
Then the l.h.s gives
\be
{\cal N}^{-1}  \langle 0 | T[ \phi_0(x_1) \cdot\cdot\cdot \phi_0(x_4) \frac{\phi_0^4(y)}{4!} e^{i\int d^d z \frac{-m_0^2}{2} \phi_0^2(z)} ] | 0 \rangle - \d_{ O^{(4)} } \langle \phi(x_1) \cdots  \phi(x_4) \frac{\phi^4(y)}{4!} \rangle  \nonumber
\ee
with the first term contributing only bubble and disconnected diagrams and the second giving a 4-point vertex.
Thus, at 1-loop, there is no non-vanishing contribution to $Z_{ O^{(4)} }$ which means that $\d_{ O^{(4)} } = 0$. Using \eq{An.d.m1} the anomalous dimension of $O^{(4)}$ is
\be\label{GaO22}
\G_{ O^{(4)} }  = 0  \, .
\ee
The last step here is the evaluation of the non-linear part in \eq{bc4a}.
The first of the two non-linear terms, $\G_{422}$, is determined using the formula of \eq{fnOlOl} for $n=4$ and $k'=r=s=2$ along with $q_{(4,2)}=4$. This admits only disconnected propagators and
then, $Z_{422}$ is trivial and as a consequence $\G_{422}$ vanishes. 
For the the second non-linear term, $\G_{444}$, we have a non-vanishing contribution. The relevant expectation value follows from \eq{flOlOl} for $l=r=s=4$:
\be
\langle \phi(x_1) \phi(x_2) \phi(x_3) \phi(x_4) O^{(4)}(y) O^{(4)}(z) \rangle_0 = Z_{444} \langle \phi(x_1) \phi(x_2) \phi(x_3) \phi(x_4) O^{(4)}(y) O^{(4)}(z) \rangle  \, .
\ee
With the level-2 inserted operator $O^{(4)} = \phi^4/4!$, this is
 \be\label{Z444}
\langle \phi(x_1) \cdots \phi(x_4) \frac{\phi^4(y)}{4!} \frac{\phi^4(z)}{4!} \rangle_0 \, - \, 
\d_{444} \langle \phi(x_1) \cdots \phi(x_4)\frac{\phi^4(y)}{4!} \frac{\phi^4(z)}{4!} \rangle = \langle \phi(x_1) \cdots \phi(x_4)\frac{\phi^4(y)}{4!} \frac{\phi^4(z)}{4!} \rangle  \, .
\ee
The bare part after the contractions gives the known 1-loop diagram 
\be
%-------------------------------------
\begin{tikzpicture} [scale=0.9]
\draw [dashed] (6.4,0)--(7.4,0);
\draw [dashed] (6.4,1)--(7.4,1);
\draw [dashed] (7.6,0.5) circle [radius=0.5];
\draw [dashed] (7.9,0)--(9,0);
\draw [dashed] (7.9,1)--(9,1);
\node at (7.6,1) {$\times $};
\node at (7.6,1) {$\times $};
\node at (7.6,0) {$\times $};
\node at (7.6,0) {$\times $};
%%%%
%\draw [dashed] (11,1)--(11.7,0);
%\draw [dashed] (11.7,0)--(12.4,1);
%\draw [dashed] (11,0)--(12.5,0);
%\node at (10,0) {$- $};
\node at (10,0.5) {$= {\cal M}^{44}$};
%\node at (11.7,0.01) {$\times $};
%\node at (11.7,0.01) {$\times $};
\end{tikzpicture}
%------------------------------------- 
\nonumber
\ee
without any coupling included.
The diagram has three possible channels, $s$, $t$ and $u$, given by 
\be\label{3channels}
s=(p_1+p_2)^2 \hskip 1cm t=(p_1+p_3)^2 \hskip 1cm u=(p_1+p_4)^2\, ,
\ee
with each channel contributing a scaleless $B_0$ integral. 
Separating only the UV part of \eq{M2.f} and adding the three channels we get 
\be
{\cal M}^{44} = (-i) \frac{3}{16 \pi^2} \frac{1}{\ve}+ {\rm finite}\, ,
\ee
The renormalized term on the l.h.s of \eq{Z444} gives exactly the same diagram, multiplied by $\d_{444}$, but now this is finite.
Notice that both sides of \eq{Z444} include the extra $(-i)$ factor.
According to the discussion below \eq{fnOlOl}, such expectation values should be considered as finite vertices 
so that the corresponding counter-term absorbs any remnant divergence.
In other words finiteness requires $\d_{444} = \frac{3}{16 \pi^2} \frac{1}{\ve}$ determining 
\be
\G_{444} = \frac{3}{16 \pi^2}\, .
\ee 
Putting all the pieces together we can reconstruct $\b_{c^{(4)}}$:
\be\label{bc4b}
\b_{c^{(4)}} = - \ve c^{(4)} + \frac{3}{16 \pi^2} ( c^{(4)} )^2 \equiv - \ve \l + \frac{3}{16 \pi^2} \l^2\, .
\ee
With \eq{bc2b} and \eq{bc4b} in hand, the quantum parts of the $\b$-functions are
\be\label{qpbc}
\vec{\b}^q_{c^{(l)}} = \begin{bmatrix}
   \b^q_{c^{(2)}}      \\
   \b^q_{c^{(4)}}    \\
\end{bmatrix}   = \begin{bmatrix}
    \frac{\l m^2}{16 \pi^2}      \\
    \frac{3\l^2}{16 \pi^2}     \\
\end{bmatrix}
\ee
and the full 1-loop $\b$-functions are
\[\vec{\b}_{c^{(l)}}  = \begin{bmatrix}
  -2 m^2 + \frac{\l m^2}{16 \pi^2}      \\
  - \ve \l + \frac{3\l^2}{16 \pi^2}     \\
\end{bmatrix} \, . \]
As mentioned, these are the right expressions expected from standard treatments. 
Based on these $\b$-functions we review the properties of the phase diagrams for $d=4,3,5$. 
On Figs. \ref{Ph.D.e0} and \ref{PDepos} we plot the corresponding renormalization group (RG) flows according to the following conventions:
\begin{itemize}
\item We denote by $\bullet$ the Gaussian fixed points and by $\star$ the Wilson-Fisher fixed points.
\item $\m_L$ is the Landau pole.
\item The arrow on an RG flow line points towards the IR. This means that it points to the direction of an increasing mass, when the mass is non-zero.
For a flow line where the mass remains constant, the arrow indicates the flow of $\l$ towards the IR. 
\item We keep in mind that when $\l < 0$, an instability eventually develops for large field values.
\item We also consider flows with $\l > |{\l_{\star}}|$ and $\l < - |{\l_{\star}}|$ even though they may lie beyond the validity of the $\ve$ expansion.
\item The green flows are the regions on the phase diagram where a stable mechanism of spontaneous breaking of $\mathbb Z_2$ is at work.
\end{itemize}
%
%%%%%%%%%%%%%%%%%%%%%%%%%%%%%%%%%%%%%%%
\FloatBarrier
%%%%%%%%%%%%%%%%%%%%%%%%%%%%%%%%%%%%%%%
%
%
%%%%%%%%%%%%%%%%%%%%%%%%%%%%%%%%%%%%%%%
%%%%%%%%%%%%%%%%%%%%%%%%%%%%%%%%%%%%%%%
%
Let us summarize the features of the phase diagrams.
%, using the language described in Appendix \ref{FixedPoints}. 
Each fixed point is seen by the couplings as either a UV fixed point or an IR fixed point. 
This is decided by the derivative of the $\b$-function evaluated at the fixed point.
In the case where the two couplings featuring in the model are $m^2$ and $\l$, we have (we consider only $\ve =0,\pm 1$):

\be\label{mcases}
\frac{\partial \beta_{m^2}}{\partial m^2} = -2 + \frac{\l}{16 \pi^2} 
\begin{cases}
\bullet: -2 < 0 \,\, \text{for any $d$} \rightarrow {\rm UV} \\
\star :  \begin{cases} 
                                                d = 4 : - \\
                                                -2 + \frac{\ve}{3} < 0 \,\, \text{for any $d \ne 4 $} \rightarrow {\rm UV}
                                                  \end{cases}
\end{cases}
\ee

\be\label{lcases}
   \frac{\partial \beta_{\l}}{\partial \l} = -\ve + \frac{6}{16 \pi^2} \l =
   \begin{cases}
     \bullet : -\ve & \begin{cases}
                            d=4:0\,\, \rightarrow {\rm IR}  \\
                            d=3:-1 < 0 \rightarrow {\rm UV} \\
                            d=5: +1 >0 \rightarrow {\rm IR}
                            \end{cases}
     \\
     \star : +\ve & \begin{cases}
                           d=4: - \\
                           d=3: +1 >0 \rightarrow {\rm IR}\\
                           d=5: -1 < 0 \rightarrow {\rm UV} 
                          \end{cases}
   \end{cases}
 \ee
From \eq{mcases} we see that the mass feels both fixed points as UV. 

The nature of an operator is decided by the quantity $\Delta_{O^{(l)}}-d$ according to:
\be\label{Oprule}
 \Delta_{O^{(l)}}-d :
\begin{cases}
< 0 \rightarrow {\rm relevant}\\
> 0 \rightarrow {\rm irrelevant}\\
=0 \rightarrow {\rm RG\, equation}
\end{cases}
\ee
An appropriate definition for the scaling dimension $\Delta_{O^{(l)}}$ is needed. In general it is defined as
\be\label{sc.d.Ol}
\Delta_{O^{(l)}} = d_{O^{(l)}} + \g_{O^{(l)}}
\ee
with $d_{O^{(l)}}$ defined in \eq{dim.an.Oc}.
The second term on the r.h.s refers to the full anomalous dimension of the associated operator and
it should not be confused with $\G_{O^{(l)}}$ of \eq{An.d.m1}, which comes from just the linear part of $\b_{c^{(l)}}$.
Taking into account the full quantum part of the $\b$-function, we define \cite{Hollowood}
\be\label{an.d.1}
\g_{O^{(l)}} = \frac{\partial \b^q_{c^{(l)}} (c^{(l)})}{\partial c^{(l)} } \Bigg |_{c^{(l)} \to c^{(l)}_\star} \, .
\ee
Note that $\g_{O^{(l)}}$ contributes to \eq{sc.d.Ol} only for $d\ne4$ where a $c^{(l)}_\star$ exists,
so any error in the characterization of the operators using $\g_{O^{(l)}}$ vs $\G_{O^{(l)}}$ in the definition of the scaling dimension is invisible in $d=4$.
These results are summarized in the following tables. The nature of the fixed points in various dimensions, as determined by $\frac{\partial \b^q_\l}{\partial \l}\vert_{\bullet, \star}$ are:
%%%%%%%%%%%%   TABLE   %%%%%%%%%%
\hskip 2cm
\begin{center}
\begin{tabular}{|c|c|c|c|}
\hline 
$d$ & $\bullet$ & $\star$\\
\hline \hline   
%%%%
$4$  & $\rm{trivial\, (UV\, when \, \l=0)}$ & $-$ \\ \hline
$3$  & $\rm{UV}$ & $\rm{IR}$ \\ \hline
$5$  & $\rm{IR}$ & $\rm{UV}$ \\ \hline
\end{tabular}
%\center{TABLE 1. The nature of the fixed points in various dimensions, as determined by $\frac{\partial \b^q_\l}{\partial \l}\vert_{\bullet, \star}$.}
\end{center}
%%%%%%%%%%%%%%%%%%%%%%%%%%%%
The 1-loop exponent is determined by the nature of an operator, according to \eq{sc.d.Ol} using \eq{cdO2O4} and 
Eq. \ref{qpbc} combined with \eq{an.d.1}. With respect to the IR, when it is negative the operator is relevant, 
if it is positive then the operator is irrelevant and if it is zero it is marginal and the RG running must decide (when it applies):
%%%%%%%%%%%%   TABLE   %%%%%%%%%%
\hskip 2cm
\begin{center}
\begin{tabular}{|c|c|c|c|}
\hline 
$$ & $\Delta_{O^{(l)}}-d$ & $\bullet$ & $\star$ \\
\hline \hline   
%%%%
$O^{(2)}$  & $-2 + \frac{\l}{16 \pi^2}$ &$-2$ & $-2+\frac{\ve}{3}$\\ \hline
$O^{(4)}$  & $-\ve + 2 \frac{3\l}{16 \pi^2}$ &$-\ve\,\, (\rm irrel.\, in\, d=4)$ & $\ve$\\ \hline
%$O^3$  & $- 1 - \frac{\ve}{2} + \frac{3\l}{16 \pi^2}$ &$- 1 - \frac{\ve}{2}$ & $- 1 + \frac{\ve}{2}$\\ \hline
\end{tabular}
%\center{TABLE 2. The 1-loop exponent that determines the nature of an operator, according to \eq{sc.d.Ol} using 
%\eq{cdO2O4} and Eq. \ref{qpbc} combined with \eq{an.d.1}. With respect to the IR, when it is negative the operator is relevant, 
%if it is positive then the operator is irrelevant and if it is zero it is marginal and the RG running must decide (when it applies). %With respect to the UV, the 
%operator is irrelevant when it is negative and relevant when it is positive.}
\end{center}
%%%%%%%%%%%%%%%%%%%%%%%%%%%%
Finally, the chracterization of operators at the fixed points according to the above is:
%%%%%%%%%%%%   TABLE   %%%%%%%%%%
\hskip 2cm
\begin{center}
\begin{tabular}{|c|c|c|c|}
\hline 
$$ & $d=4$ & $d=3$ & $d=5$ \\
\hline \hline   
%%%%
$O^{(2)}$  & $\bullet \, (\l=0): {\rm rel}\hskip .5cm  \vline \hskip .5cm \star: -$ & $ \bullet: {\rm rel} \hskip .5cm  \vline \hskip .5cm \star: {\rm rel}$ & $\bullet: {\rm rel} \hskip .5cm  \vline \hskip .5cm \star: {\rm rel}$\\ \hline
$O^{(4)}$  & $ \bullet: {\rm marg. \, irrel} \hskip .5cm  \vline \hskip .5cm \star: {-}$ &$ \bullet: {\rm rel} \hskip .5cm  \vline \hskip .5cm \star: {\rm irrel}$ & $ \bullet: {\rm irrel} \hskip .5cm  \vline \hskip .5cm \star: {\rm rel}$\\ \hline
%$O^3$  & $ \bullet \, (\l=0): {\rm rel}\hskip .5cm  \vline \hskip .5cm \star: - $ & $\bullet: {\rm rel} \hskip .5cm  \vline \hskip .5cm \star: {\rm rel} $ & $ \bullet: {\rm rel} \hskip .5cm  \vline \hskip .5cm \star: {\rm rel} $\\ \hline
\end{tabular}
%\center{TABLE 3. Classification of operators at the fixed points according to TABLE 2.}
\end{center}
%%%%%%%%%%%%%%%%%%%%%%%%%%%%

\end{appendices}
%---------------------------------------------------------------------------------------------------------------------------------------------------

%\pagebreak

%---------------------------------------------------------------------------------------------------------------------------------------------------


\begin{thebibliography}{9}

\bibitem{Warsaw} 
W. Buchmuller, D. Wyler, Nucl. Phys. B {\bf 268} (1986) 621-653.
B. Grzadkowski, M. Iskrzynski, M. Misiak, J. Rosiek,
JHEP 10 (2010) 085.

\bibitem{Einhorn} 
M. B. Einhorn and J. Wudka, JHEP {\bf 08}, (2001), 025

\bibitem{Colangelo}
M. Buchler and G. Colangelo, Eur. Phys. J. C {\bf 32},(2004), 427-442.
E.E. Jenkins, A.V. Manohar, M. Trott JHEP {\bf 10}, (2013), 087.
E. E. Jenkins, A. V. Manohar, and M. Trott, [hep-ph/1310.4838].
R. Alonso, E.E. Jenkins, A.V. Manohar, M. Trott JHEP {\bf 04}, (2014), 159.
I. Brivio and M. Trott, JHEP {\bf 07}, (2017), 148.
A. Dedes, W. Materkowska, M. Paraskevas, J. Rosiek, K. Suxho JHEP {\bf 06}, (2017), 143.
A. Dedes, M. Paraskevas, J. Rosiek, K. Suxho and L. Trifyllis, JHEP {\bf 08}, (2018), 103.
A. Dedes, K. Suxho and L. Trifyllis, JHEP {\bf 06}, (2019), 115

\bibitem{FotisLetter}
N. Irges and F. Koutroulis, [hep-th/1905.05506].

\bibitem{Manohar} 
see for example A. V. Manohar, 
%\textit{Introduction to Effective Field Theories},  
 [hep-ph/1804.05863].

\bibitem{Neubert}
see for example M. Neubert, 
%\textit{Les Houches Lectures on Renormalization Theory and Effective Field Theories}, 
 [hep-ph/1901.06573v1].

\bibitem{Peskin}
M. E. Peskin and D. V. Schroeder,
%\textit{An Introduction to Quantum Field Theory},
Perseus Books Publishing, 1995.

\bibitem{Hollowood}
T. J. Hollowood,
%\textit{Renormalization Group and Fixed Points in Quantum Field Theory}, 
Springer Briefs in Physics, (2013).

\bibitem{Serone}
A. Bilal,
%\textit{Advanced Quantum Field Theory: Renormalization, Non-Abelian Gauge Theories and Anomalies},
Lecture notes, Brussels (2014).
M. Serone,
%\textit{Notes on Quantum Field Theory},
SISSA via Bonomea 265, I-34136, Trieste, Italy (2017).

\bibitem{phi6quantum}
G. 't Hooft, M.J.G. Veltman, NATO Sci.Ser. B{\bf 4} (1974), 177-322.
A. R. Chowdhury and T. Roy, Phys. Rev. D{\bf 15} (1977) no8, 2186.
K. Hagiwara, S. Ishihara, R. Szalapski, D. Zeppenfeld, Phys. Rev. D{\bf 48} (1993) 2182-2203.
C. Arzt, Phys. Lett. B {\bf 342} (1995) 189.
F. Goertz, Phys. Rev. D{\bf 94}, (2016) 015013.
S. Nagy, J. Polonyi and I. Steib, Phys. Rev. D{\bf 97} (2018), 085002.
M. Safari and G. P. Vacca, Phys. Rev. D{\bf 97} (2018) no4, 041701.
G. Passarino, [hep-ph/1901.04177].
J. C. Criado, M. Perez-Victoria, JHEP {\bf 1903}, (2019) 038.
R. Trinchero, [hep-th/1904.01616].

\bibitem{Nicolis}
A. Adams, N. A-Hamed, S. Dubovsky, A. Nicolis, R. Rattazzi,
JHEP {\bf 0610}, (2006) 014.

\bibitem{Richard}
M. Ostrogradsky, Mem. Ac. St. Petersbourg {\bf VI 4} (1850) 385.
R. P. Woodard, Scholarpedia 10 (2015) no.8, 32243.

\bibitem{Anselmi}
D. Anselmi,
%\textit{A General Field-Covariant Formulation Of Quantum Field Theory}, 
Eur. Phys. J. C {\bf 73}, (2013) 2338.

\bibitem{Luty}
S. Chang and M. A. Luty,
%\textit{The Higgs Trilinear Coupling and the Scale of New Physics},
[hep-ph/1902.05556v1].

\bibitem{Coleman}
S. R. Coleman and E. J. Weinberg,
%\textit{Radiative corrections as the origin of spontaneous symmetry breaking},
Phys. Rev. D{\bf 7}, (1973) 1888.


\end{thebibliography}
\end{document}